\documentclass[11pt]{article}
\pdfoutput=1 

\usepackage{jheppub} 
                     
\usepackage{braket,slashed,bm}
\usepackage{array,multirow}

\usepackage{amsmath,amssymb}

\usepackage[T1]{fontenc} 

\usepackage{xparse}
\usepackage{etoolbox}

\usepackage{booktabs}

\usepackage{float}

\usepackage{hypcap}
\usepackage{adjustbox}

\usepackage{dsfont}

\usepackage[all]{nowidow}

\usepackage{feynmp-auto-mod}

\newcommand{\nn}{\nonumber\\}

\renewcommand{\L}{\mathcal{L}}
\renewcommand{\O}{\mathcal{O}}

\newcommand{\p}{\partial}

\renewcommand{\O}{\mathcal{O}}

\newcommand{\tr}{\mathrm{Tr}}
\newcommand{\T}{\mathcal{T}}
\newcommand{\N}{\mathcal{N}}
\newcommand{\F}{\mathcal{F}}

\newcommand{\E}{\mathcal{E}}
\newcommand{\msbar}{$\overline{\text{MS}}$}
\newcommand{\rismom}{\mbox{RI-$\tilde{\text{S}}$MOM}}
\newcommand{\M}{\mathcal{M}}

\newcommand{\<}{\langle}
\renewcommand{\>}{\rangle}

\newcommand{\mytag}{\\[-\baselineskip] \stepcounter{equation}\tag{\theequation}}

\allowdisplaybreaks

\title{\boldmath Non-perturbative renormalization scheme for the $CP$-odd three-gluon operator}

\author[a]{Vincenzo Cirigliano,}

\author[a]{Emanuele Mereghetti,}

\author[b]{Peter Stoffer}

\affiliation[a]{Theoretical Division, Los Alamos National Laboratory, Los Alamos, NM 87545, USA}
\affiliation[b]{Department of Physics, University of California at San Diego, 9500 Gilman Drive,\\ La Jolla, CA 92093-0319, USA}

\emailAdd{cirigliano@lanl.gov}
\emailAdd{emereghetti@lanl.gov}
\emailAdd{pstoffer@ucsd.edu}

\abstract{
We define a regularization-independent momentum-subtraction scheme for the $CP$-odd three-gluon operator at dimension six. This operator appears in effective field theories for heavy physics beyond the Standard Model, describing the indirect effect of new sources of $CP$-violation at low energies. In a hadronic context, it induces permanent electric dipole moments. The hadronic matrix elements of the three-gluon operator are non-perturbative objects that should ideally be evaluated with lattice QCD. We define a non-perturbative renormalization scheme that can be implemented on the lattice and we compute the scheme transformation to \msbar{} at one loop. Our calculation can be used as an interface to future lattice-QCD calculations of the matrix elements of the three-gluon operator, in order to obtain theoretically robust constraints on physics beyond the Standard Model from measurements of the neutron electric dipole moment.
}

\begin{document}

\mbox{}

\vspace{-1.75cm}
\hfill{}\begin{minipage}[t][0cm][t]{5cm}
\raggedleft
\small
LA-UR-20-20500
\end{minipage}
\vspace{0.25cm}

\bigskip

\maketitle


\section{Introduction}

Permanent electric dipole moments (EDMs) of non-degenerate systems break the symmetries of parity ($P$) and time reversal ($T$), and
consequently, in Lorentz-invariant theories, the combination of charge conjugation and parity ($CP$). 
While $C$ and $P$ are separately maximally broken in the Standard Model (SM) of particle physics by the weak interaction,
$CP$ is broken in a much more subtle fashion: in the SM with three generations of quarks, $CP$ is
broken by the phase of the Cabibbo--Kobayashi--Maskawa (CKM) quark-mixing matrix and the QCD $\theta$ term. 
So far, $CP$ violation has been observed in kaon~\cite{Christenson:1964fg,Batley:2002gn,Abouzaid:2010ny}, $D$-meson~\cite{Aaij:2019kcg},
and $B$-meson~\cite{Abe:2001xe,Aubert:2001nu} decays, and it is compatible with the CKM mechanism. 
On the other hand, SM $CP$ violation is insufficient to explain the observed matter-antimatter asymmetry in the universe~\cite{Gavela:1993ts,Huet:1994jb}. 
EDMs of the nucleon, light nuclei, diamagnetic and paramagnetic atoms, and molecules offer an important window into non-SM $CP$ violation,
by combining extremely high experimental sensitivies with unobservably small CKM backgrounds, see~\cite{Chupp:2017rkp} for a review.
Currently, the strongest bounds are those on the electron EDM, $|d_e| < 1.1 \times 10^{-16}$~$e$~fm (at the 90\% confidence level), inferred from experiments with the ThO and HfF molecules 
\cite{Andreev:2018ayy,Cairncross:2017fip,Baron:2013eja}, on the neutron EDM, $|d_n | < 1.8 \times 10^{-13}$~$e$~fm~\cite{Baker:2006ts,Afach:2015sja,Abel:2020gbr}, and
on the EDM of the $^{199}$Hg atom, $|d_{\rm Hg}| < 7.4 \times 10^{-17}$~$e$~fm~\cite{Graner:2016ses}. In all three cases, the CKM background is 
several orders of magnitude smaller than current and future sensitivies \cite{Khriplovich:1981ca,Pospelov:1991zt,Booth:1993af,Czarnecki:1997bu,Pospelov:2013sca,Seng:2014lea}.
The present generation of EDM experiments is already putting severe constraints on models of physics beyond the SM 
and on electroweak baryogenesis scenarios. These constraints will become even more stringent in the next generation of experiments, which aims at improving the 
electron and neutron EDM sensitivities by one or two orders of magnitude, respectively, and the $^{225}$Ra EDM sensitivity by four orders of magnitude \cite{Bishof:2016uqx}.
In addition, forthcoming experiments will for the first time investigate the EDMs of the proton and light ions~\cite{Anastassopoulos:2015ura,Abusaif:2018oly}
and the EDMs of unstable particles, like the $\tau$, $\Lambda$, and the $\Lambda_{c,s}$ baryons~\cite{Fu:2019utm,Botella:2016ksl}.

While the observation of a non-zero EDM in any of these experiments will be a clear indication of new physics, connecting a 
nuclear or atomic EDM with the fundamental, high-energy mechanism of $CP$ violation requires gaining control over hadronic and nuclear uncertainties. 
At the quark level, flavor-diagonal $CP$ violation can be model-independently described by 
extending the SM Lagrangian with gauge-invariant higher-dimensional operators to an effective field theory (SMEFT)~\cite{Buchmuller:1985jz,Grzadkowski:2010es}, which parametrizes the indirect effects of physics at scales $\Lambda \gg v$, where $v=246$ GeV is the Higgs vacuum expectation value.
Heavy SM degrees of freedom can then be integrated out by matching the SMEFT Lagrangian onto an $SU(3)_c \times U(1)_{\rm em}$-invariant low-energy effective field theory (LEFT)~\cite{Jenkins:2017jig,Jenkins:2017dyc}. The complete one-loop matching was carried out recently in~\cite{Dekens:2019ept}.
At the hadronic scale, the LEFT Lagrangian includes the dimension-four QCD $\bar\theta$ term, the dimension-five electric and chromo-electric dipole moments (CEDMs) of 
the $u$, $d$, and $s$ quarks (which arise from dimension-six SMEFT operators at the electroweak scale), and, at dimension six, the $CP$-odd three-gluon operator and several four-quark operators.   
The quark-level operators induce $CP$-violating hadronic interactions, such as the EDMs of the neutron and proton and $CP$-violating pion-nucleon and nucleon-nucleon couplings,
which then feed into the calculations of the EDM and Schiff moments of light and heavy nuclei.
As QCD is nonperturbative at the hadronic scale, the hadronic matrix elements required to match the quark-level and hadronic/nuclear EFTs 
need to be evaluated via nonperturbative methods. In particular, lattice QCD (LQCD) has emerged as a powerful tool to compute hadronic matrix elements, in which  all  sources  of  systematic uncertainty can be quantified, controlled, and improved. This has led to the first LQCD calculations of the nucleon EDM from 
the $u$- and $d$-quark EDMs~\cite{Gupta:2018lvp}, with few-percent uncertainties, and to the first estimates of the nucleon EDM induced by 
the QCD $\bar\theta$ term~\cite{Shintani:2005xg,Shintani:2006xr,Shintani:2008nt,Shintani:2015vsx,Shindler:2015aqa,Guo:2015tla,Abramczyk:2017oxr,Dragos:2019oxn} and by the quark CEDM~\cite{Bhattacharya:2018qat,Syritsyn:2018mon}.
These estimates, though preliminary and still affected by large uncertainties, promise to deliver controlled EDM calculations for the next generation of experiments.

An important issue to be addressed in the interpretation of LQCD results is the mixing between different $CP$-violating operators. The lattice spacing in LQCD 
effectively works as a gauge-invariant cut-off, causing mixing between operators of different dimension, in addition to the familiar logarithmic mixing of dimensional regularization. 
To unambigously identify the effects of various LEFT operators, it is then necessary to define a renormalization scheme. 
This scheme needs to be interfaced with the \msbar{} scheme, which is employed for the calculation of the operator mixing and running between the weak and the hadronic scale in the LEFT~\cite{Jenkins:2017dyc} and above the weak scale in the SMEFT~\cite{Alonso:2013hga,Jenkins:2013wua,Jenkins:2013zja}.
Such a matching calculation has been performed in~\cite{Bhattacharya:2015rsa} for the dimension-five quark CEDM operator.
In this paper we focus on the $CP$-odd three-gluon operator~\cite{Buchmuller:1985jz,Weinberg:1989dx}\footnote{In the literature, this operator is often called ``Weinberg operator'' in attribution to~\cite{Weinberg:1989dx}, although it was already listed as part of the general set of gauge-invariant dimension-six operators in~\cite{Buchmuller:1985jz}.}
\begin{equation}
	\label{eq:Definition3GluonOperator}
	\O_1^{(6)}  = i g \tr[G_{\mu\nu} {G^\mu}_\lambda \widetilde G^{\nu\lambda}] \, ,
\end{equation}
where  $G_{\mu\nu}$ is the gluon field strength, the trace is taken in color space, and $\widetilde G^{\mu \nu} = \epsilon^{\mu \nu \alpha \beta} G_{\alpha \beta}/2$ is the dual field-strength tensor.
The three-gluon operator induces a non-zero gluon chromo-electric dipole moment (gCEDM), and we will thus also denote  $\O_1^{(6)}$ by gCEDM. 
The perturbative renormalization of the gCEDM has been calculated at one loop in~\cite{Braaten:1990gq,Braaten:1990zt,Chang:1990dja}. The anomalous dimension has recently been calculated to two and three loops in~\cite{deVries:2019nsu}.
A first study of the renormalization of ${\cal O}_1^{(6)} $ using the gradient flow has been presented in~\cite{Rizik:2018lrz,Rizik:2020naq}.
In this work we define the three-gluon operator in a regularization-independent (RI) momentum-subtraction (MOM) scheme, construct the complete basis of gauge-invariant and nuisance operators
needed to carry out the nonperturbative renormalization, and  calculate the matching between the \msbar{} scheme and the momentum-subtraction scheme at one loop.

The paper is organized as follows. In Sect.~\ref{sec:Mixing}, we discuss mixing and different types of operators that need to be taken into account in our scheme. In Sect.~\ref{sec:Basis}, we discuss the construction of the operators and present the resulting basis of operators that mix with the gCEDM. In Sect.~\ref{sec:Scheme}, we define a regularization-independent renormalization scheme and in Sect.~\ref{sec:OneLoopMatching} we present the results for the matching between the RI and the \msbar{} scheme at one loop, before we conclude in Sect.~\ref{sec:Conclusions}. More details on the operator basis construction are provided in the appendices.


\section{Operator mixing}
\label{sec:Mixing}

In this paper we define the $CP$-odd three-gluon operator \eqref{eq:Definition3GluonOperator} in a regularization-independent momentum-subtraction scheme \cite{Martinelli:1994ty}.
In this scheme, the renormalization conditions are imposed on quark, gluon, and photon Green's functions, computed in a fixed gauge, with off-shell external states of large space-like virtualities. 
The renormalized operators $\O_j^\text{RI}$ thus defined are independent of the ultraviolet regulator, and, since the renormalization conditions  can be implemented both on the lattice and in perturbation theory, one can convert them into the \msbar{} scheme
\begin{align}
	\O_i^\text{\msbar} = C_{ij} \O_j^\text{RI} ,
\end{align}
with  matching coefficients $C_{ij}$ computed in perturbation theory.
The implementation of the RI-MOM schemes requires  working off-shell in a fixed gauge. In this case a given gauge-invariant operator mixes with two classes of operators of the same or lower dimension 
\cite{Dixon:1974ss,KlubergStern:1975hc,Joglekar:1975nu,Deans:1978wn,Collins:1984xc}:
\begin{itemize}
	\item[I.] gauge-invariant and ghost-free operators that do not vanish by equations of motion (EOM) and have the same properties as $\O^{(6)}_1$ under Lorentz, chiral, and discrete symmetries ($C$, $P$, and $CP$),
	\item[II.] ``nuisance'' operators, which we denote by $\mathcal N$. These operators are allowed by the solution of the Ward identities associated with BRST invariance. They do not need to be gauge invariant, they vanish by the EOM and can be constructed as off-shell BRST variations of operators with ghost number $-1$, with otherwise the same properties as $\O_1^{(6)}$.
\end{itemize}
We will discuss the construction of operators in class I and II in Sect.~\ref{sec:Basis}, where we will further divide the operators in class II into
\begin{itemize}
	\item[IIa.] gauge-invariant operators that vanish by EOM,
	\item[IIb.] gauge-variant operators.
\end{itemize}
The mixing with gauge-variant operators (class IIb) can be avoided by working in background-field gauge~\cite{Abbott:1980hw}. 

In lattice calculations, the traditional RI-MOM scheme~\cite{Martinelli:1994ty} suffers from unwanted infrared effects, which can be suppressed by choosing subtraction points with non-exceptional kinematics as in the RI-SMOM prescription~\cite{Aoki:2007xm,Sturm:2009kb}.
In our scheme, we will impose the renormalization conditions at non-exceptional but asymmetric kinematic points (dubbed \rismom{} scheme in~\cite{Bhattacharya:2015rsa}). As the scheme involves momentum insertion into the operator, we also need to take into account mixing with operators that are total derivatives.

We now establish the conventions used throughout the paper.
If we consider only single-operator insertions, the relation between bare and renormalized operators in any scheme is linear:
\begin{align}
	\label{eq:OperatorMixing}
	\O_i^{(0)} = Z_{ij} \O_j \, ,
\end{align}
where the superscript $^{(0)}$ denotes bare operators.
By general consideration, it can be proved that the renormalization matrix has triangular structure \cite{Dixon:1974ss,KlubergStern:1975hc,Joglekar:1975nu,Deans:1978wn,Collins:1984xc} 
\begin{equation}
\left(\begin{array}{c} \mathcal O^{(0)} \\ \mathcal N^{(0)} 
\end{array}\right)
= 
\left(\begin{array}{cc}
Z_O  &  Z_{ON}\\
0 & Z_N 
\end{array}\right)
\, 
\left(\begin{array}{c}
\mathcal O \\ \mathcal  N
\end{array}\right) \, .
\label{eq:Zstructure}
\end{equation}
The matching coefficients for the translation between the \msbar{} and RI-MOM schemes are therefore given by
\begin{align}\label{eq:Cij}
  C_{ij} = \big(Z^\text{\msbar}\big)_{ik}^{-1} Z_{kj}^\text{RI} \, .
\end{align}
In this paper we consider the matching at one-loop, where~\eqref{eq:Cij} simplifies to
\begin{align}\label{eq:Zdef}
	Z_{ij} = \mathds{1}_{ij} + \Delta_{ij} \, , \quad C_{ij} = \mathds{1}_{ij} - \Delta_{ij}^\text{\msbar} + \Delta_{ij}^\text{RI} \, .
\end{align}
Since the matrix elements of nuisance operators vanish between physical states \cite{Deans:1978wn,Collins:1984xc},
when computing hadronic matrix elements we can neglect nuisance operators. In particular, the contribution to the neutron EDM will be extracted from
\begin{align}
	\label{eq:physicalME}
	\< N | \O_i^\text{\msbar} | \gamma^* N \> = \Big( \mathds{1}_{ij} - \Delta_{ij}^\text{\msbar} + \Delta_{ij}^\text{RI} \Big) \< N | \O_j^\text{RI} | \gamma^* N \> \, ,
\end{align}
where the operator $\O_i^\text{\msbar}$ arises from the insertion of the effective Lagrangian, which carries no external momentum. Hence, in~\eqref{eq:physicalME} the summed index $j$ only runs over gauge-invariant, physical operators that are not total derivatives. Note, however, that in order to determine the factors $\Delta_{ij}$ in~\eqref{eq:physicalME}, either perturbatively or nonperturbatively, and in particular the renormalized operators $\O_j^\text{RI}$,
one is forced to also determine the mixing with nuisance operators. 

We need to calculate the mixing matrices in the two schemes, which can be obtained by considering the insertions of bare operators into amputated $n$-point Green's functions:
\begin{align}\label{eq:npoint}
	\< \psi_1^{(0)} \ldots \psi_n^{(0)} \O_i^{(0)} \>^\text{amp} = Z_\psi^{-n/2} Z_{ij} \< \psi_1 \ldots \psi_n \O_j \>^\text{amp} \, ,
\end{align}
where $\psi$ denotes a generic field and $\sqrt{Z_\psi}$ its field-renormalization factor.


\section{Construction of the operator basis}
\label{sec:Basis}

The dimension-four QCD Lagrangian is given by
\begin{align}
	\label{eq:LQCD}
	\L_\text{QCD} = \bar q( i \gamma^\mu D_\mu - \M ) q - \frac{1}{4} G^{\mu\nu}_a G_{\mu\nu}^a - \bar \theta_\text{QCD} \frac{g^2}{32\pi^2} G^{\mu\nu}_a \widetilde G^a_{\mu\nu} \, ,
\end{align}
where the quark field includes the three light quarks, $q := (u, d, s)^T$. We define quark-mass and charge matrices as
\begin{align}
	\label{eq:MassChargeMatrices}
	\M = \text{diag}(m_u, m_d, m_s) \, , \quad Q = \text{diag}\Big(\frac{2}{3}, -\frac{1}{3}, -\frac{1}{3}\Big) \, .
\end{align}
In the LEFT, the Lagrangian~\eqref{eq:LQCD} is supplemented with QED as well as a tower of effective operators~\cite{Jenkins:2017jig}. Here, we will be interested in the extraction of the neutron EDM from the matrix element~\eqref{eq:physicalME} with the insertion of the dimension-six three-gluon operator~\eqref{eq:Definition3GluonOperator}. We consider the matrix element at $\O(e)$ and we will neglect higher-order QED corrections. This allows us to disregard the photon kinetic term and instead treat the photon field as an external source~\cite{Gasser:1983yg,Gasser:1984gg}. In this case, the covariant derivative is given by
\begin{align}
	D_\mu = \partial_\mu - i g t^a G^a_\mu - i v_\mu - i \gamma_5 a_\mu = \partial_\mu - i g t^a G^a_\mu - i l_\mu P_L - i r_\mu P_R
\end{align}
where $t^a = \lambda^a/2$ and $\lambda^a$ are the Gell-Mann matrices in color space, and $v_\mu$, $a_\mu$, $l_\mu$, and $r_\mu$ are traceless, Hermitian $3\times3$ matrices in flavor space, fulfilling
\begin{align}
	l_\mu = v_\mu - a_\mu \, , \quad r_\mu = v_\mu + a_\mu
\end{align}
and taking the physical values
\begin{align}
	l_\mu = r_\mu = v_\mu = e Q A_\mu \, , \quad a_\mu = 0 \, .
\end{align}

In the following, we will construct the basis of operators that are needed to renormalize the $CP$-odd three-gluon operator.
The symmetries of the Lagrangian strongly constrain the possible mixings. Neglecting the QCD $\bar\theta$ term, the leading-order Lagrangian is $P$- and $CP$-even, 
implying that we only need to consider $CP$-odd operators as possible counterterms to the three-gluon operator. In addition, 
in the limit  $\mathcal M \rightarrow 0$ and $e Q \rightarrow 0$, the Lagrangian has  
an $SU(3)_L \times SU(3)_R$ chiral symmetry, i.e., it is invariant under  the transformation
\begin{align}
	q_{L,R} \rightarrow U_{L,R}\, q_{L,R} \, , \quad U_{L,R} \in SU(3)_{L,R} \, ,
\end{align}
where $q_{L,R} = P_{L,R} q$ and the chiral projectors are $P_L = (1-\gamma_5)/2$, $P_R = (1+\gamma_5)/2$. While chiral symmetry is broken by quark masses and charges, one can formally recover chiral invariance by assigning spurion transformation properties 
to the mass matrix and external fields. Since the three-gluon operator is chirally invariant,
it can mix only with operators that are chirally invariant in the spurion sense.   
Chiral symmetry applies to the continuum theory. If the lattice regularization breaks chiral symmetry, additional spurions are present in the effective Lagrangian,
which can induce more mixings of the three-gluon operator. We will restrict our analysis to the case where chiral symmetry is preserved by the lattice regulator.  

In Sect.~\ref{sec:GaugeInvariant}, we briefly describe the construction of the relevant set of gauge-invariant \mbox{class-I} operators, while in Sect.~\ref{sec:Nuisance} we explain how we construct the class-II nuisance operators.  More details on the construction of the operator basis are provided in App.~\ref{sec:BasisConstructionI} and~\ref{sec:BasisConstructionII}. In Sect.~\ref{sec:Operators}, we present the complete basis of operators that are needed to renormalize the gCEDM. Based on general considerations, we discuss the structure of the mixing matrix in Sect.~\ref{sec:MixingStructure}.

\subsection{Gauge-invariant operators}
\label{sec:GaugeInvariant}

We construct the basis of operators up to dimension six that renormalize the $CP$-odd three-gluon operator.
The dynamical degrees of freedom that we need to consider are the gluon field and quark fields.

In order to implement the constraints of chiral symmetry, we rewrite the non-gauge part of the leading-order Lagrangian as
\begin{align}
	\L = \frac{i}{2} \bar q_L \overleftrightarrow{\slashed D} q_L +  \frac{i}{2} \bar q_R \overleftrightarrow{\slashed D} q_R - \bar q_L \M q_R  - \bar q_R \M^\dagger q_L \, ,
\end{align}
where $\overleftrightarrow{\slashed D} := \slashed D - \overleftarrow{\slashed D}$, and $\slashed D := \gamma^\mu D_\mu$, $\overleftarrow{\slashed D} := \gamma^\mu \overleftarrow{D}_\mu$. The left-acting covariant derivative is
\begin{align}
	\overleftarrow{D}_\mu = \overleftarrow{\partial}_\mu + i g t^a G^a_\mu + i l_\mu P_L + i r_\mu P_R \, .
\end{align}
The mass matrix is promoted to a spurion field and the transformations
\begin{align}
	\label{eq:SpurionTransformations}
	\M &\stackrel{\chi\;}{\mapsto} U_L \M U_R^\dagger \, , \quad \M^\dagger \stackrel{\chi\;}{\mapsto} U_R \M^\dagger U_L^\dagger \, , \nn
	l_\mu &\stackrel{\chi\;}{\mapsto} U_L l_\mu U_L^\dagger + i U_L \p_\mu U_L^\dagger \, , \quad
	r_\mu \stackrel{\chi\;}{\mapsto} U_R r_\mu U_R^\dagger + i U_R \p_\mu U_R^\dagger
\end{align}
formally make the leading-order Lagrangian invariant under chiral transformations. The field-strength tensors associated with the external fields are
\begin{align}
	F_{\mu\nu}^L := \p_\mu l_\nu - \p_\nu l_\mu - i [ l_\mu, l_\nu ] \, , \quad F_{\mu\nu}^R := \p_\mu r_\nu - \p_\nu r_\mu - i [ r_\mu, r_\nu ] \, ,
\end{align}
with the physical values
\begin{align}
	F_{\mu\nu}^{L,R} = e Q F_{\mu\nu} \, , \quad F_{\mu\nu} = \p_\mu A_\nu - \p_\nu A_\mu \, .
\end{align}
Since we are interested only in effects of $\O(e)$, we will consistently restrict the basis to operators that are at most linear in the external photon field.
Gauge-invariant operators will be expressed in terms of covariant derivatives, the external field-strength tensors, the gluon field-strength tensor
\begin{align}
	G^{a}_{\mu \nu} &= \partial_\mu G^a_\nu - \partial_\nu G^a_\mu + g f^{abc} G^b_\mu G^c_\nu \, ,
\end{align}
and the dual field strengths
\begin{align}
	\widetilde G^{a\, \mu\nu} = \frac{1}{2} \epsilon^{\mu\nu\lambda\sigma} G_{\lambda\sigma}^a \, , \qquad 
	\widetilde F^{\mu\nu}_{L,R}   = \frac{1}{2} \epsilon^{\mu\nu\lambda\sigma} F_{\lambda\sigma}^{L,R} \, , 
\end{align}
with $\epsilon^{0123} = + 1$.

The commutator of the covariant derivative is related to the field-strength tensors by
\begin{align}
	\label{eq:CovariantDerivativeCommutator}
	[ D_\mu, D_\nu ] = - i g G_{\mu\nu} - i F_{\mu\nu}^L P_L - i F_{\mu\nu}^R P_R \, , \quad G_{\mu\nu} = t^a G_{\mu\nu}^a \, .
\end{align}
The covariant derivative in the adjoint representation is defined by
\begin{align}
	D_\mu (\cdot) = \p_\mu (\cdot) - i g [ G_\mu, \;\cdot{}\; ] - i [ l_\mu, \;\cdot{}\; ] P_L - i [ r_\mu, \;\cdot{}\; ] P_R \, ,
\end{align}
i.e., the covariant derivatives of the field-strength tensors are
\begin{align}
	D_\rho F^L_{\mu\nu} &= (\p_\rho F^L_{\mu\nu} - i [l_\rho, F_{\mu\nu}^L] ) \, , \quad
	D_\rho F^R_{\mu\nu} = (\p_\rho F^R_{\mu\nu} - i [r_\rho, F_{\mu\nu}^R] ) \, , \nn
	D_\rho G_{\mu\nu} &= \p_\rho G_{\mu\nu} - i g [ G_\rho, G_{\mu\nu} ] \, , \;\; \text{or} \;\; (D_\rho G_{\mu\nu})^a = \p_\rho G_{\mu\nu}^a + g f^{abc} G_\rho^b G_{\mu\nu}^c \, .
\end{align}
We use the same symbol $D_\mu$ for the covariant derivative in different representations. Note that the covariant derivative fulfills the Jacobi identity
\begin{align}
	[D_\mu, [D_\nu, D_\lambda ] ] + [D_\lambda, [D_\mu, D_\nu ] ] + [D_\nu, [D_\lambda, D_\mu ] ] = 0
\end{align}
as well as the Leibniz rule
\begin{align}
	\label{eq:Leibniz}
	D_\mu ( A B ) = (D_\mu A) B + A (D_\mu B) \, ,
\end{align}
where each $D_\mu$ denotes the proper covariant derivative belonging to the representation of the object that it acts upon. The Jacobi identity and Leibniz rule imply the Bianchi identity
\begin{align}
	\label{eq:BianchiIdentity}
	(D_\mu [D_\nu, D_\lambda ] ) + (D_\lambda &[D_\mu, D_\nu ] ) + (D_\nu [D_\lambda, D_\mu ] ) \nn
		&= -i g \left( D_\mu G_{\nu\lambda} + D_\lambda G_{\mu\nu} + D_\nu G_{\lambda\mu} \right) \nn
			&\quad - i \left( D_\mu F_{\nu\lambda}^L + D_\lambda F_{\mu\nu}^L + D_\nu F_{\lambda\mu}^L \right) P_L \nn
			&\quad - i \left( D_\mu F_{\nu\lambda}^R + D_\lambda F_{\mu\nu}^R + D_\nu F_{\lambda\mu}^R \right) P_R = 0 \, .
\end{align}
All three brackets have to vanish separately. By contracting the Bianchi identity with the Levi-Civita tensor, one obtains
\begin{align}
	D_\mu \widetilde G^{\mu\nu} = 0 \, , \quad D_\mu \widetilde F_L^{\mu\nu} = 0 \, , \quad D_\mu \widetilde F_R^{\mu\nu} = 0 \, .
\end{align}
These identities play an important role in identifying the minimal set of operators that mix with the three-gluon operator.

The gauge-invariant operators mixing with the gCEDM are obtained by constructing an exhaustive list of operators that are Lorentz scalars, chirally invariant, $P$-odd, and $CP$-odd, using as building blocks the fields and spurions
\begin{align}
	\label{eq:BuildingBlocks}
	\bar q_{L,R} \, , \quad q_{L,R} \, , \quad G_{\mu\nu} \, , \quad F_{\mu\nu}^{L,R} \, , \quad D_\mu \, , \quad \M \, , \quad \M^\dagger \, ,
\end{align}
and subsequently removing all redundancies. Details on this construction are given in App.~\ref{sec:BasisConstructionI}. The counting of operators can be automatized using Hilbert series techniques~\cite{Lehman:2015via,Henning:2015daa,Lehman:2015coa,Henning:2015alf,Henning:2017fpj}. We use these methods as a cross-check to count the number of operators that are invariant under the Lorentz group (which is isomorphic to $SU(2)_L \times SU(2)_R$), global $SU(3)_c \times U(1)_\mathrm{em}$, and the chiral group $\chi = SU(3)_L \times SU(3)_R$. From the complete list including total derivatives and EOM operators, we select the operators that are $P$-odd and $CP$-odd.

\subsection{Nuisance operators}
\label{sec:Nuisance}

Because of gauge fixing and the peculiar nature of BRST symmetry, by which BRST variations of elementary fields are composite operators,
gauge-invariant operators can mix with non-invariant operators~\cite{Dixon:1974ss,KlubergStern:1975hc,Joglekar:1975nu,Deans:1978wn,Collins:1984xc}.
The form of the operators is dictated by the Ward--Slavnov--Taylor identities associated with BRST symmetry, and we follow here the construction of~\cite{Deans:1978wn}.

The construction relies on the fact that the nuisance operators can be written as BRST variations of ``seed operators'' with ghost number $-1$. The seed operators need not be gauge invariant. Their building blocks consist of the dynamical fields, the ghost fields, the spurions, as well as external sources for the fields, which are set to zero after applying the BRST variation. Details on the derivation are given in App.~\ref{sec:BasisConstructionII}. The construction provides us with a list of nuisance operators of both classes IIa and IIb. The class-IIa operators (gauge-invariant operators that vanish by the EOM) are linear combinations of gauge-invariant operators constructed in Sect.~\ref{sec:GaugeInvariant}.\footnote{These operators correspond to the EOM redundancies that are usually removed from the set of operators in the construction of EFT Lagrangians through field redefinitions.}
They can be presented in a compact form by introducing the fields
\begin{align}
	\label{eq:EOMField}
	q_E := (i \slashed D - \M) q \, , \qquad \bar q_E = - \bar q (i \overleftarrow{\slashed D} + \M) \, .
\end{align}
The complete list of operators is provided in Sect.~\ref{sec:Operators}.

\subsection{Operator basis}
\label{sec:Operators}

The final matrix element~\eqref{eq:physicalME} that is needed to extract the neutron EDM contains an external photon state. The photon is allowed to couple either to the electromagnetic current or directly to an effective operator. As we are working at leading order in the QED coupling, we disregard operators containing more than one photon field.

In a cut-off scheme, the gCEDM mixes with $CP$-odd operators of dimension six or lower. 
There are no $CP$-odd, chirally invariant operators with dimension smaller than four. 

In the following, we present the complete basis of operators that renormalize the gCEDM operator at leading order in the QED coupling.
In order to make the operators manifestly Hermitian in $D$ dimensions, we introduce the following symbols:\footnote{The definition of $\tilde\sigma^{\mu\nu}$ differs from the one in~\cite{Bhattacharya:2015rsa} by an evanescent term.}
\begin{align}
	\label{eq:GammaSigmaTilde}
	\tilde\gamma^\mu &:= \frac{i}{3!} \epsilon^{\alpha\beta\gamma\mu} \gamma_\alpha \gamma_\beta \gamma_\gamma = \frac{1}{2} [ \gamma^\mu, \gamma_5 ] = P_L \gamma^\mu P_R - P_R \gamma^\mu P_L \, , \nn
	\tilde\sigma^{\mu\nu} &:= \frac{i}{2} \epsilon^{\mu\nu\alpha\beta} \sigma_{\alpha\beta} \, .
\end{align}

\paragraph{Dimension four}
At dimension four, we have two physical and one nuisance operator:
\begin{align*}
	\label{eq:FinalOperatorsDim4}
	\arraycolsep=1.4pt
	\begin{array}{rl}
		\O_1^{(4)} &= \tr[G_{\mu\nu} \widetilde G^{\mu\nu}] \, , \\
		\O_2^{(4)} &= \p_\mu (\bar q \tilde\gamma^\mu q) \, , \\
		\N_1^{(4)} &= i ( \bar q_E \gamma_5 q + \bar q \gamma_5 q_E ) \, .
	\end{array}
	\mytag
\end{align*}
The operator $\O_1^{(4)}$ is the QCD $\bar\theta$ term, the operator $\O_2^{(4)}$ is a total derivative and contributes due to momentum insertion. The nuisance operator belongs to class IIa.

\paragraph{Dimension five}
At dimension five, there is a single chiral invariant operator:
\begin{align}
	\label{eq:FinalOperatorDim5}
	\O_1^{(5)} &= \epsilon_{ijk} \epsilon_{lmn} \M_{mj} \M_{nk} \bar q^i i \gamma_5 q^l \, .
\end{align}
For a diagonal mass matrix, the following relation holds~\cite{Kaplan:1986ru,Leutwyler:1989pn}:
\begin{align}
	\O_1^{(5)} = 2 \det(\M) \bar q i \gamma_5 \M^{-1} q \, .
\end{align}

\paragraph{Dimension six}

At dimension six, we find the following operator basis:
\begin{align}
	\label{eq:FinalOperatorsDim6}
	\O_1^{(6)} &= i g \tr[G_{\mu\nu} {G^\mu}_\lambda \widetilde G^{\nu\lambda}] \, , \nn
	\O_2^{(6)} &= i g ( \bar q \tilde\sigma^{\mu\nu} \M t^a q ) G_{\mu\nu}^a \, , \nn
	\O_3^{(6)} &= i e ( \bar q \tilde\sigma^{\mu\nu}  \M Q q ) F_{\mu\nu} \, , \nn
	\O_4^{(6)} &= \tr[\M^2] \tr[G_{\mu\nu} \widetilde G^{\mu\nu}] \, ,  \nn
	\O_5^{(6)} &= \p_\nu \tr[ (D^\mu G_{\mu\lambda}) \widetilde G^{\nu\lambda} ] \, , \nn
	\O_6^{(6)} &= \p_\mu \left( \bar q \tilde\gamma^\mu \M^2 q \right) - \frac{1}{N_f} \O_7^{(6)} \, , \nn
	\O_7^{(6)} &= \tr[\M^2] \p_\mu \left( \bar q \tilde\gamma^\mu q \right) \, , \nn
	\O_8^{(6)} &= e \p_\mu \big( \bar q \gamma_\nu Q q \widetilde F^{\mu\nu} \big) \, , \nn
	\O_9^{(6)} &= \Box \, \tr[G_{\mu\nu} \widetilde G^{\mu\nu}] \, , \nn
	\O_{10}^{(6)} &= \Box \p_\mu( \bar q \tilde\gamma^\mu q ) \, ,
\end{align}
where $N_f = 3$ is the number of quark flavors.
The basis in~\eqref{eq:FinalOperatorsDim6} contains the gCEDM and three additional purely gluonic operators, $\mathcal O_4^{(6)}$, a mass correction to the QCD $\bar\theta$ term, and
$\mathcal O_{5,9}^{(6)}$, which are total derivatives.  $\mathcal O^{(6)}_{2}$ and $\mathcal O^{(6)}_{3}$ are the quark CEDM and EDM, respectively. Due to the different chiral properties, 
the gCEDM can mix into them only via insertions of $\mathcal M$ and $\mathcal M Q$. $\mathcal O^{(6)}_{6,7,10}$ are derivatives of the axial current,
with the appropriate number of mass insertions required for chiral invariance.
An important result of our construction is that there are no $SU(3)$ chirally invariant, $CP$-odd four-quark operators. 

In addition to the physical operators, we find a set of 20 nuisance operators at dimension six. 
There are 10 gauge-invariant operators that vanish by EOM  (class IIa):
\begin{align}
	\label{eq:FinalOperatorsDim6IIa}
	\N_1^{(6)} &=  i g ( \bar q_E \tilde\sigma^{\mu\nu} t^a q + \bar q \tilde\sigma^{\mu\nu} t^a q_E) G_{\mu\nu}^a \, ,  \nn
	\N_2^{(6)} &=  i e ( \bar q_E \tilde\sigma^{\mu\nu} Q q + \bar q \tilde\sigma^{\mu\nu} Q q_E ) F_{\mu\nu} \, , \nn
	\N_3^{(6)} &=  ( \bar q_E \M \tilde\gamma^\mu D_\mu q + \bar q \overleftarrow{D}_\mu \tilde\gamma^\mu \M q_E ) \, , \nn
	\N_4^{(6)} &=  i \left( \bar q_E \M^2 \gamma_5 q + \bar q \M^2 \gamma_5 q_E \right) - \frac{1}{N_f} \N_5^{(6)} \, , \nn
	\N_5^{(6)} &=  \tr[\M^2] i ( \bar q_E \gamma_5 q + \bar q \gamma_5  q_E ) \, , \nn
	\N_6^{(6)} &= i \p_\mu \Big( \bar q_E \gamma_5 D^\mu q + \bar q \overleftarrow{D}^\mu \gamma_5 q_E \Big) \, , \nn
	\N_7^{(6)} &=  \p_\mu ( \bar q_E \tilde\sigma^{\mu\nu} D_\nu q - \bar q \overleftarrow{D}_\nu \tilde\sigma^{\mu\nu} q_E ) \, , \nn
	\N_8^{(6)} &= \p_\lambda \left( G_{\mu\nu}^a \Big( D^\rho G_{\rho\sigma}^a + g \bar q t^a \gamma_\sigma q \Big) \right) \epsilon^{\mu\nu\lambda\sigma} \, , \nn
	\N_9^{(6)} &= \p_\mu \Big( \bar q_E \M \tilde\gamma^\mu q + \bar q \M \tilde\gamma^\mu q_E \Big) \, , \nn
	\N_{10}^{(6)} &= i \Box \Big( \bar q_E \gamma_5 q + \bar q \gamma_5 q_E \Big) \, .
\end{align}
Finally, we find another 10 gauge-variant operators (class IIb):
\begin{align}
	\label{eq:FinalOperatorsDim6IIb}
	\N_{11}^{(6)} &= G_{\mu\nu}^a \left(\p_\lambda \Big( D^\rho G_{\rho\sigma}^a + g \bar q t^a \gamma_\sigma q - g f^{abc} (\p_\sigma \bar c^b) c^c \Big) \right)  \epsilon^{\mu\nu\lambda\sigma} \, , \nn
	\N_{12}^{(6)} &= i g^2 ( \bar q_E \gamma_5 q + \bar q \gamma_5 q_E) G_\mu^a G^\mu_a \, , \nn
	\N_{13}^{(6)} &= i g^2 ( \bar q_E \gamma_5 t^a q + \bar q \gamma_5 t^a q_E) G_\mu^b G^\mu_c d^{abc} \, , \nn
	\N_{14}^{(6)} &= g ( \bar q_E \gamma_5 t^a q - \bar q \gamma_5 t^a q_E ) \p_\mu G^\mu_a \, , \nn
	\N_{15}^{(6)} &= g ( \bar q_E \gamma_5 t^a D_\mu q - \bar q \overleftarrow{D}_\mu \gamma_5 t^a q_E ) G^\mu_a \, , \nn
	\N_{16}^{(6)} &= i g ( \bar q_E \tilde\sigma^{\mu\nu} t^a q + \bar q \tilde\sigma^{\mu\nu} t^a q_E ) \p_\mu G_\nu^a \, , \nn
	\N_{17}^{(6)} &= i g ( \bar q_E \M \tilde\gamma_\mu t^a q - \bar q \M \tilde\gamma_\mu t^a q_E ) G_a^\mu \, , \nn
	\N_{18}^{(6)} &= \p_\lambda \left( (\p_\mu G_\nu^a) \Big( D^\rho G_{\rho\sigma}^a + g \bar q t^a \gamma_\sigma q - g f^{abc} (\p_\sigma \bar c^b) c^c \Big) \right) \epsilon^{\mu\nu\lambda\sigma} \, , \nn
	\N_{19}^{(6)} &= \p_\mu \Big( g ( \bar q_E \gamma_5 t^a q - \bar q \gamma_5  t^a q_E ) G^\mu_a \Big) \, , \nn
	\N_{20}^{(6)} &= i \p_\mu \Big( g ( \bar q_E \tilde\sigma^{\mu\nu} t^a q + \bar q \tilde\sigma^{\mu\nu} t^a q_E ) G_\nu^a \Big) \, .
\end{align}
The operators that are written in terms of the EOM quark fields $\bar q_E$ and $q_E$ obviously vanish by the quark EOM. A subtlety arises in connection with the operators involving pure gauge-field terms: $\N_8^{(6)}$ vanishes by the ``naive'' classical EOM, i.e., by the EOM without gauge-fixing and ghost terms. On the other hand, the class-IIb operators $\N_{11}^{(6)}$ and $\N_{18}^{(6)}$ vanish by the full EOM including ghost (and auxiliary-field) terms. Nevertheless, neither class-IIa nor class-IIb operators contribute to physical matrix elements~\cite{KlubergStern:1975hc,Deans:1978wn,Collins:1984xc}, as all of them are given by BRST variations. In particular, the class-IIa operator can be obtained as the BRST variation of a seed operator with ghost number $-1$,
\begin{align}
	\label{eq:N8Seed}
	\N_8^{(6)} = \hat W\big[ \p_\lambda( G_{\mu\nu}^a \hat J_\sigma^a ) \epsilon^{\mu\nu\lambda\sigma} \big] \, ,
\end{align}
where the BRST operator $\hat W$ is defined in~\eqref{eq:WHatOperator} and given explicitly in~\eqref{eq:WHatOperatorSimp}, and where $\hat J_\sigma^a$ is the source for a composite BRST variation, see App.~\ref{sec:SeedOperators}. The seed operator in~\eqref{eq:N8Seed} is a linear combination of the seed operators~\eqref{eq:PureGaugeSeedOperators}.

As discussed in~\cite{Simma:1993ky}, the difference between an operator that vanishes by the naive classical EOM (i.e., the EOM without ghost and gauge-fixing terms) and the corresponding operator vanishing by the full classical EOM is again a BRST variation. Therefore, this difference is a class-IIb operator that does not affect physical matrix elements. In the case of $\N_8^{(6)}$, this difference would be given by the BRST variation
\begin{align}
	\hat W \big[ \partial_\lambda ( G_{\mu\nu}^a \partial_\sigma \bar c^a ) \epsilon^{\mu\nu\lambda\sigma} \big] \, .
\end{align}
The fact that this term does not appear in our basis is due to the ghost quantum EOM~\eqref{eq:GhostQEOM}, which is a constraint on the effective action beyond the classical level, see App.~\ref{sec:STI}.

\subsection{Mixing structure}
\label{sec:MixingStructure}

\begin{table}[t]
\begin{align*}
	\scriptsize
	\scalebox{0.65}{$
	\def\arraystretch{1.9}
	\setlength{\arraycolsep}{0.01cm}
	\begin{array}{l |l cccccccccc |l cccccccccc |l cccccccccc |l c |l cc |l c }
		\toprule
			&\;\;& \O_{1}^{(6)} 
			& \O_{2}^{(6)} 
			& \O_{3}^{(6)} 
			& \O_{4}^{(6)} 
			& \O_{5}^{(6)} 
			& \O_{6}^{(6)} 
			& \O_{7}^{(6)} 
			& \O_{8}^{(6)} 
			& \O_{9}^{(6)} 
			& \O_{10}^{(6)} 
			&\;\;
			& \N_{1}^{(6)} 
			& \N_{2}^{(6)} 
			& \N_{3}^{(6)} 
			& \N_{4}^{(6)} 
			& \N_{5}^{(6)} 
			& \N_{6}^{(6)} 
			& \N_{7}^{(6)} 
			& \N_{8}^{(6)} 
			& \N_{9}^{(6)} 
			& \N_{10}^{(6)} 
			&\;\;
			& \N_{11}^{(6)} 
			& \N_{12}^{(6)} 
			& \N_{13}^{(6)} 
			& \N_{14}^{(6)} 
			& \N_{15}^{(6)} 
			& \N_{16}^{(6)} 
			& \N_{17}^{(6)} 
			& \N_{18}^{(6)} 
			& \N_{19}^{(6)} 
			& \N_{20}^{(6)} 
			&\;\;
			& \O_{1}^{(5)}
			&\;\;
			& \O_{1}^{(4)}
			& \O_{2}^{(4)}
			&\;\;
			& \N_{1}^{(4)} \\[0.03cm]
		\hline
		\O_{1}^{(6)}	&&	\times &\times &\times &\times &\times &\times &\times &\times &\times &\times 
					&&	\times &\times &\times &\times &\times &\times &\times &\times &\times &\times
					&&	\times &\times &\times &\times &\times &\times &\times &\times &\times &\times
					&&	\times
					&&	\times &\times
					&&	\times  \\
		\O_{2}^{(6)}	&&	&\times&\times&\times&&\times&\times&&&\times
					&&	&&\times&\times&\times&&&&\times&\times
					&&	&&&&&&\times&&&
					&&	\times
					&&	& \times
					&&	\times  \\
		\O_{3}^{(6)}	&&	&&\times&&&&&&&
					&&	&&&&&&&&&
					&&	&&&&&&&&&
					&&	
					&&	&
					&&	\\
		\O_{4}^{(6)}	&&	&&&\times&&&\times&&&
					&&	&&&&\times&&&&&
					&&	&&&&&&&&&
					&&	
					&&	&
					&&	\\
		\O_{5}^{(6)}	&&	&&&&\times&\times&\times&\times&\times&\times
					&&	&&&&&\times&\times&\times&\times&\times
					&&	&&&&&&&\times&\times&\times
					&&	
					&&	&\times
					&&	\\
		\O_{6}^{(6)}	&&	&&&&&\times&&&&
					&&	&&&&&&&&&
					&&	&&&&&&&&&
					&&	
					&&	&
					&&	\\
		\O_{7}^{(6)}	&&	&&&&&&\times&&&
					&&	&&&&&&&&&
					&&	&&&&&&&&&
					&&	
					&&	&
					&&	\\
		\O_{8}^{(6)}	&&	&&&&&&&\times&&
					&&	&&&&&&&&&
					&&	&&&&&&&&&
					&&	
					&&	&
					&&	\\
		\O_{9}^{(6)}	&&	&&&&&&&&\times&\times
					&&	&&&&&&&&&\times
					&&	&&&&&&&&&
					&&	
					&&	&
					&&	\\
		\O_{10}^{(6)}	&&	&&&&&&&&&\times
					&&	&&&&&&&&&
					&&	&&&&&&&&&
					&&	
					&&	&
					&&	\\
		\midrule
		\N_{1}^{(6)}	&&	&&&&&&&&&
					&&	\times&\times&\times&\times&\times&\times&\times&\times&\times&\times
					&&	\times&\times&\times&\times&\times&\times&\times&\times&\times&\times
					&&	
					&&	&
					&&	\times	\\
		\N_{2}^{(6)}	&&	&&&&&&&&&
					&&	&\times&&&&&&&&
					&&	&&&&&&&&&
					&&	
					&&	&
					&&	\\
		\N_{3}^{(6)}	&&	&&&&&&&&&
					&&	&&\times&\times&\times&&&&\times&
					&&	&&&&&&\times&&&
					&&	
					&&	&
					&&	\\
		\N_{4}^{(6)}	&&	&&&&&&&&&
					&&	&&&\times&&&&&&
					&&	&&&&&&&&&
					&&	
					&&	&
					&&	\\
		\N_{5}^{(6)}	&&	&&&&&&&&&
					&&	&&&&\times&&&&&
					&&	&&&&&&&&&
					&&	
					&&	&
					&&	\\
		\N_{6}^{(6)}	&&	&&&&&&&&&
					&&	&&&&&\times&\times&\times&\times&\times
					&&	&&&&&&&\times&\times&\times
					&&	
					&&	&
					&&	\\
		\N_{7}^{(6)}	&&	&&&&&&&&&
					&&	&&&&&\times&\times&\times&\times&\times
					&&	&&&&&&&\times&\times&\times
					&&	
					&&	&
					&&	\\
		\N_{8}^{(6)}	&&	&&&&&&&&&
					&&	&&&&&\times&\times&\times&\times&\times
					&&	&&&&&&&\times&\times&\times
					&&	
					&&	&
					&&	\\
		\N_{9}^{(6)}	&&	&&&&&&&&&
					&&	&&&&&&&&\times&
					&&	&&&&&&&&&
					&&	
					&&	&
					&&	\\
		\N_{10}^{(6)}	&&	&&&&&&&&&
					&&	&&&&&&&&&\times
					&&	&&&&&&&&&
					&&	
					&&	&
					&&	\\
		\midrule
		\N_{11}^{(6)}	&&	&&&&&&&&&
					&&	\times&\times&\times&\times&\times&\times&\times&\times&\times&\times
					&&	\times&\times&\times&\times&\times&\times&\times&\times&\times&\times
					&&	
					&&	&
					&&	\times	\\
		\N_{12}^{(6)}	&&	&&&&&&&&&
					&&	\times&\times&\times&\times&\times&\times&\times&\times&\times&\times
					&&	\times&\times&\times&\times&\times&\times&\times&\times&\times&\times
					&&	
					&&	&
					&&	\times	\\
		\N_{13}^{(6)}	&&	&&&&&&&&&
					&&	\times&\times&\times&\times&\times&\times&\times&\times&\times&\times
					&&	\times&\times&\times&\times&\times&\times&\times&\times&\times&\times
					&&	
					&&	&
					&&	\times	\\
		\N_{14}^{(6)}	&&	&&&&&&&&&
					&&	\times&\times&\times&\times&\times&\times&\times&\times&\times&\times
					&&	\times&\times&\times&\times&\times&\times&\times&\times&\times&\times
					&&	
					&&	&
					&&	\times	\\
		\N_{15}^{(6)}	&&	&&&&&&&&&
					&&	\times&\times&\times&\times&\times&\times&\times&\times&\times&\times
					&&	\times&\times&\times&\times&\times&\times&\times&\times&\times&\times
					&&	
					&&	&
					&&	\times	\\
		\N_{16}^{(6)}	&&	&&&&&&&&&
					&&	\times&\times&\times&\times&\times&\times&\times&\times&\times&\times
					&&	\times&\times&\times&\times&\times&\times&\times&\times&\times&\times
					&&	
					&&	&
					&&	\times	\\
		\N_{17}^{(6)}	&&	&&&&&&&&&
					&&	&&\times&\times&\times&&&&\times&
					&&	&&&&&&\times&&&
					&&	
					&&	&
					&&	\\
		\N_{18}^{(6)}	&&	&&&&&&&&&
					&&	&&&&&\times&\times&\times&\times&\times
					&&	&&&&&&&\times&\times&\times
					&&	
					&&	&
					&&	\\
		\N_{19}^{(6)}	&&	&&&&&&&&&
					&&	&&&&&\times&\times&\times&\times&\times
					&&	&&&&&&&\times&\times&\times
					&&	
					&&	&
					&&	\\
		\N_{20}^{(6)}	&&	&&&&&&&&&
					&&	&&&&&\times&\times&\times&\times&\times
					&&	&&&&&&&\times&\times&\times
					&&	
					&&	&
					&&	\\
		\midrule
		\O_{1}^{(5)}	&&	&&&&&&&&&
					&&	&&&&&&&&&
					&&	&&&&&&&&&
					&&	\times 	
					&&	&
					&&	\\
		\midrule
		\O_{1}^{(4)}	&&	&&&&&&&&&
					&&	&&&&&&&&&
					&&	&&&&&&&&&
					&&	
					&&	\times & \times 	
					&&	\times 	\\
		\O_{2}^{(4)}	&&	&&&&&&&&&
					&&	&&&&&&&&&
					&&	&&&&&&&&&
					&&	
					&&	 & \times
					&&	\\
		\midrule
		\N_{1}^{(4)}	&&	&&&&&&&&&
					&&	&&&&&&&&&
					&&	&&&&&&&&&
					&&	
					&&	&
					&&	\times 	\\
		\bottomrule
	\end{array}$}%
\end{align*}%
	\caption{Structure of the inverse mixing matrix $Z^{-1}$, defined in~\eqref{eq:OperatorMixing}. The operators are defined in~\eqref{eq:FinalOperatorsDim4}, \eqref{eq:FinalOperatorDim5}, \eqref{eq:FinalOperatorsDim6}, \eqref{eq:FinalOperatorsDim6IIa}, and~\eqref{eq:FinalOperatorsDim6IIb}. The operator $\O_1^{(6)}$ is the $CP$-odd three-gluon operator.}
	\label{tab:MixingStructure}
\end{table}

In Table~\ref{tab:MixingStructure}, we give the structure of the mixing matrix at leading order in the QED coupling. The structure is determined according to the following rules.
\begin{enumerate}
	\item Dimensional argument: operators only mix into operators of the same or lower mass dimension.
	\item Operators containing the mass matrix only mix into operators with at least the same power of the mass matrix.
	\item Total derivative operators only mix into operators with at least the same structure of total derivatives.
	\item Nuisance operators do not mix into class-I operators~\cite{Dixon:1974ss,KlubergStern:1975hc,Joglekar:1975nu,Deans:1978wn,Collins:1984xc}.\footnote{This holds for regularization schemes where the path-integral measure is invariant under chiral rotations and the anomaly is due to evanescent terms, such as dimensional regularization or Wilson fermions on the lattice. In schemes where the Jacobian of chiral rotations is not unity, an additional finite renormalization and potentially the subtraction of power-divergent terms are required, see~\cite{Espriu:1983zz}.}
	\item At leading order in the QED coupling, photon operators only mix into photon operators.
\end{enumerate}
Due to the choice of the operator basis, the second rule involves a subtlety: because of~\eqref{eq:EOMField}, operators proportional to the mass matrix can be obtained from linear combinations of EOM operators with class-I operators without mass matrices. E.g., the relations
\begin{align}
	\label{eq:PseudoscalarDensityRelations}
	2 \bar q i \gamma_5 \M q &= \O_2^{(4)} - \N_1^{(4)} \, , \nn
	2 \Box \left( \bar q i \gamma_5 \M q \right) &= \O_{10}^{(6)} - \N_{10}^{(6)} \, ,
\end{align}
imply that the qCEDM operator $\O_2^{(6)}$, which contains a mass matrix, mixes into $\O_2^{(4)} - \N_1^{(4)}$ and $\O_{10}^{(6)} - \N_{10}^{(6)}$.

A mixing of gauge-variant nuisance operators (class-IIb) into gauge-invariant nuisance operators that vanish by the EOM (class-IIa) is not excluded.
Note that in the \msbar{} scheme, mixing only happens between operators of the same dimension, because mass insertions are explicitly treated as part of the operators.

The divergence of the axial current, $\O_2^{(4)}$ does not mix into the QCD $\bar\theta$ term $\O_1^{(4)}$~\cite{Espriu:1982bw,Breitenlohner:1983pi,Kaplan:1988ku,Larin:1993tq}: although $\O_1^{(4)}$ is a total divergence, it is the divergence of a gauge-variant current. In background-field gauge, the axial current cannot mix into this gauge-variant current and the same is applies for their divergences. Since $\O_{1,2}^{(4)}$ are gauge invariant, the same conclusion holds in any gauge. The argument of course applies as well to the dimension-six operators involving $G \widetilde G$.


\section{Renormalization scheme}
\label{sec:Scheme}

In order to calculate the matrix element of the \msbar{} three-gluon operator in~\eqref{eq:physicalME}, we need the first row of the conversion matrix between \msbar{} and \rismom{} schemes, $C_{1j}$, as well as a definition of the renormalized physical \rismom{} operators---nuisance operators as well as total-derivative operators (apart from the topological $\bar\theta$ term) do not contribute to the physical matrix element. In this section, we formulate renormalization conditions in an \rismom{} scheme, which can be implemented in lattice QCD. They define the renormalized physical operators in terms of bare operators, hence the renormalization conditions need to determine the entries of the mixing matrix,\footnote{In general one expects mixing  of the gCEDM and other physical operators with evanescent operators.   
In    App.~\ref{sec:Evanescent}
   we  specify a set of evanescent operators that defines our minimal scheme, and which allows us to effectively ignore the evanescent operators in the one-loop matching calculation.}
\begin{align}
	\O_i^\text{RI} = (Z^\text{RI})_{ij}^{-1} \O_j^{(0)} \, .
\end{align}
Table~\ref{tab:MixingStructure} shows the structure of the full renormalization matrix of all the operators that potentially mix with the three-gluon operator at leading order in the QED coupling.

Only those entries of the inverse mixing matrix $(Z^\text{RI})_{ij}^{-1}$ are needed, where both $i$ and $j$ run over physical operators. However, in order to determine these entries, one still has to impose $n$ conditions if the operator $\O_i$ mixes with $n$ operators $\O_j$, $j=1,\ldots,n$. Let us denote the renormalization condition for insertions of the operator $\O_i$ into $m$-point Green's functions by
\begin{align}
	\label{eq:RIrenormalizationConditions}
	R_k[\O_i] := \< \psi_1 \ldots \psi_{m_k} \O_i^\text{RI} \>^\text{amp}\Big|_{S_k} = \sum_{j=1}^n (Z^\text{RI})_{ij}^{-1} Z_\psi^{m_k/2} \< \psi_1^{(0)} \ldots \psi_{m_k}^{(0)} \O_j^{(0)} \>^\text{amp} \Big|_{S_k} \, ,
\end{align}
where $|_{S_k}$ denotes the evaluation at a certain kinematic point and appropriate contractions in Lorentz and Dirac space defined by condition $k$, with $k = 1, \ldots, n$. The desired renormalization factors are then obtained by inversion of an $n\times n$ matrix $A$
\begin{align}
	(Z^\text{RI})_{ij}^{-1} = \sum_{k=1}^n [ A^{-1} ]_{jk} R_k[\O_i] \, , \quad [A]_{kj} = Z_\psi^{m_k/2} \< \psi_1^{(0)} \ldots \psi_{m_k}^{(0)} \O_j^{(0)} \>^\text{amp} \Big|_{S_k} \, .
\end{align}
In the \msbar{} scheme, a relation similar to~\eqref{eq:RIrenormalizationConditions} holds, with the renormalization matrix $(Z^\text{\msbar})_{ij}^{-1}$ chosen to cancel only the dimensionally regulated poles.

In Sect.~\ref{sec:VertexRules}, we compute the counterterm vertex rules for the insertions of all operators of the basis, which allows us to determine the number of independent renormalization constraints that can be obtained from a particular Green's function. The explicit renormalization conditions $R_k\big[\O_1^{(6)}\big]$ on the Green's functions with insertions of the three-gluon operator are formulated in Sect.~\ref{sec:RenormalizationConditions}. To carry out the full renormalization program, we also need to impose renormalization conditions $R_k[\O_i]$ on the Green's functions with insertions of the other physical operators $\O_i$ that mix with the three-gluon operator. They can be chosen as a subset of the conditions used for the gCEDM. Alternative conditions for these operators could be obtained in a straightforward way from~\cite{Bhattacharya:2015rsa}.

\subsection{Counterterm vertex rules}
\label{sec:VertexRules}

\begin{figure}[t]
	\centering
	\scalebox{0.75}{
	\setlength{\unitlength}{0.9cm}
	\begin{picture}(4,3.5)
		\put(2.25,3){$q$}
		\put(0.85,0){$p_a$}
		\put(2.85,0){$p_b$}
		\put(-0.7,0){$a,\alpha$}
		\put(4,0){$b,\beta$}
		\includegraphics[width=3.6cm]{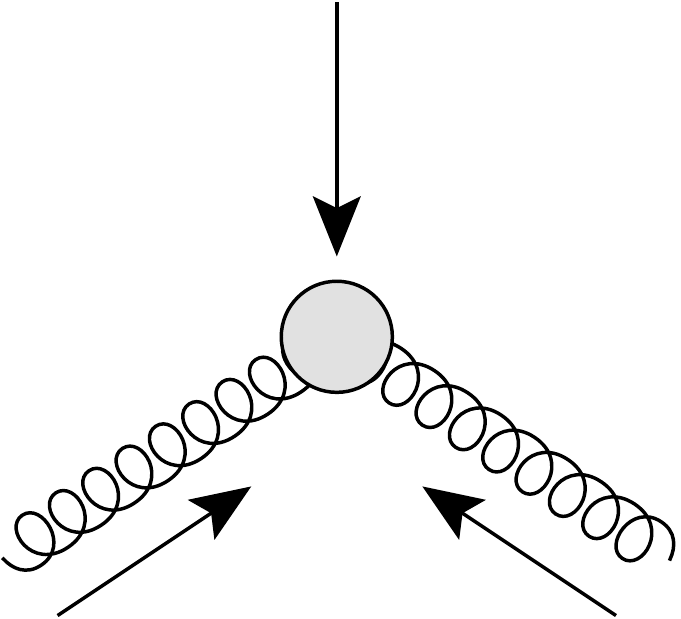}
	\end{picture}
	\hspace{1.75cm}
	\begin{picture}(4,3.5)
		\put(3.0,2.75){$q$}
		\put(0.85,0){$p_a$}
		\put(2.85,0){$p_b$}
		\put(0.85,3.25){$p_c$}
		\put(-0.7,0){$a,\alpha$}
		\put(4,0){$b,\beta$}
		\put(-0.7,3){$c,\gamma$}
		\includegraphics[width=3.6cm]{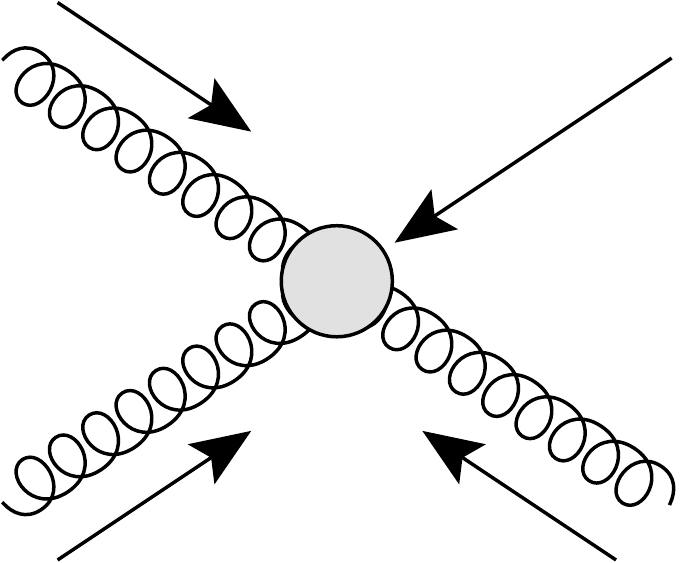}
	\end{picture}
	\hspace{1.75cm}
	\begin{picture}(4,3.5)
		\put(3.0,2.75){$q$}
		\put(0.85,0){$p_a$}
		\put(2.85,0){$p_b$}
		\put(0.85,3.25){$p_c$}
		\put(-0.35,0){$a$}
		\put(4,0){$b$}
		\put(-0.7,3){$c,\gamma$}
		\includegraphics[width=3.6cm]{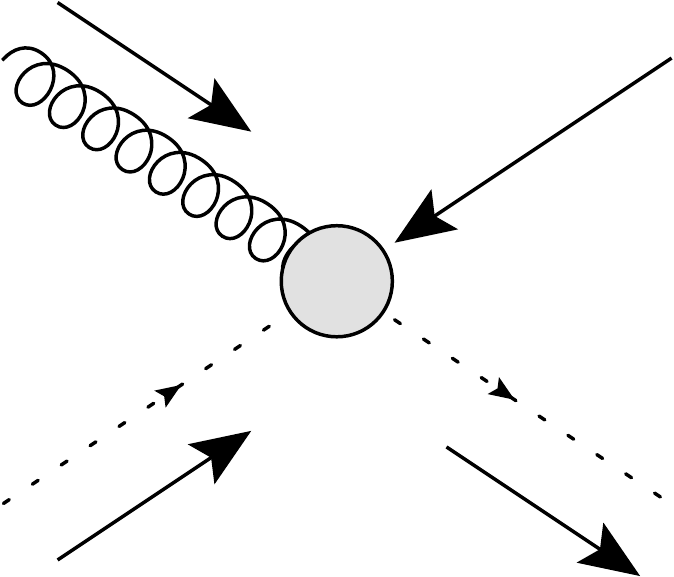}
	\end{picture}}%
	\caption{Gluon two-point, gluon three-point, and ghost-gluon three-point functions with momentum insertion into the operator.}
	\label{img:gluons}
\end{figure}
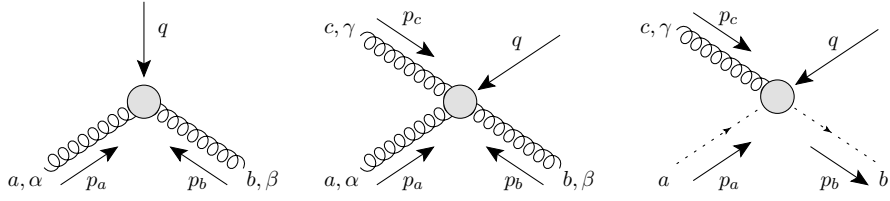

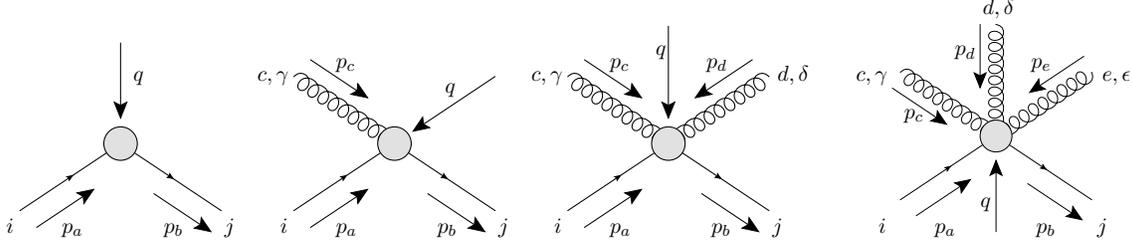
\begin{figure}[t]
	\centering
	\scalebox{0.75}{
	\setlength{\unitlength}{0.9cm}
	\begin{picture}(4,3.5)
		\put(2.25,3){$q$}
		\put(0.85,0){$p_a$}
		\put(2.85,0){$p_b$}
		\put(-0.25,0){$i$}
		\put(4.05,0){$j$}
		\includegraphics[width=3.6cm]{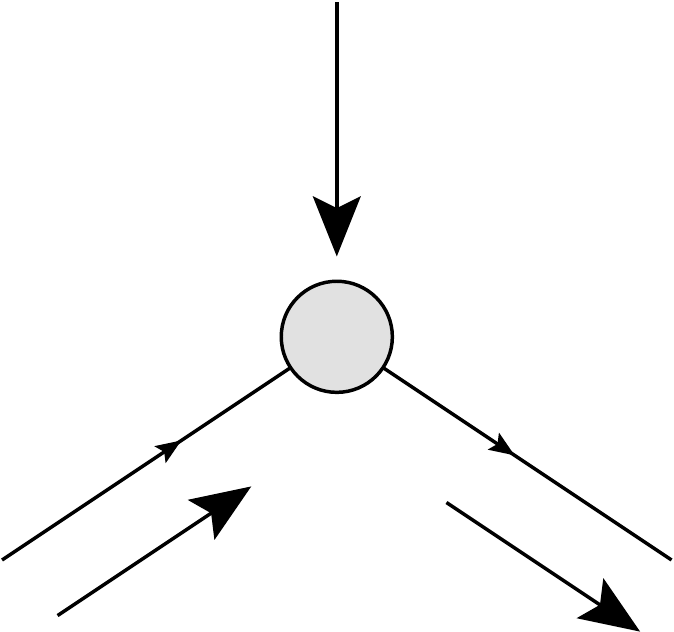}
	\end{picture}
	\hspace{1cm}
	\begin{picture}(4,3.5)
		\put(3.0,2.75){$q$}
		\put(0.85,0){$p_a$}
		\put(2.85,0){$p_b$}
		\put(-0.25,0){$i$}
		\put(4.05,0){$j$}
		\put(0.85,3.25){$p_c$}
		\put(-0.7,3){$c,\gamma$}
		\includegraphics[width=3.6cm]{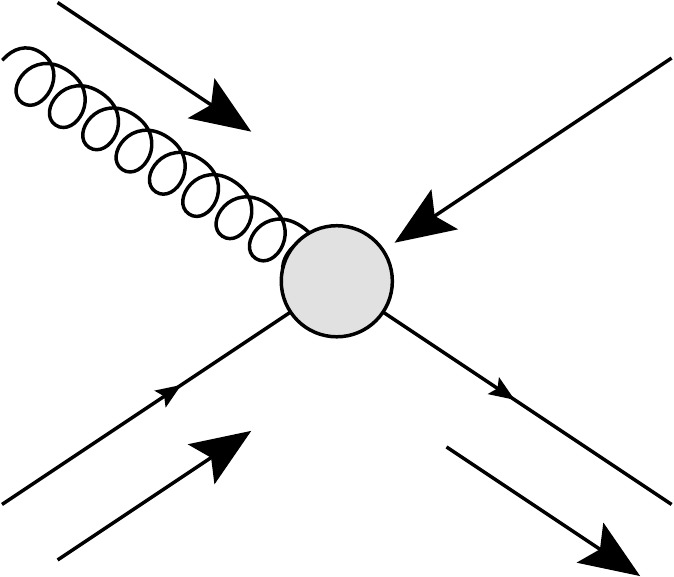}
	\end{picture}
	\hspace{1cm}
	\begin{picture}(5,4)
		\put(1.75,3.5){$q$}
		\put(0.85,0){$p_a$}
		\put(2.85,0){$p_b$}
		\put(-0.25,0){$i$}
		\put(4.05,0){$j$}
		\put(0.85,3.25){$p_c$}
		\put(2.75,3.25){$p_d$}
		\put(-0.7,3){$c,\gamma$}
		\put(4.15,3){$d,\delta$}
		\includegraphics[width=3.6cm]{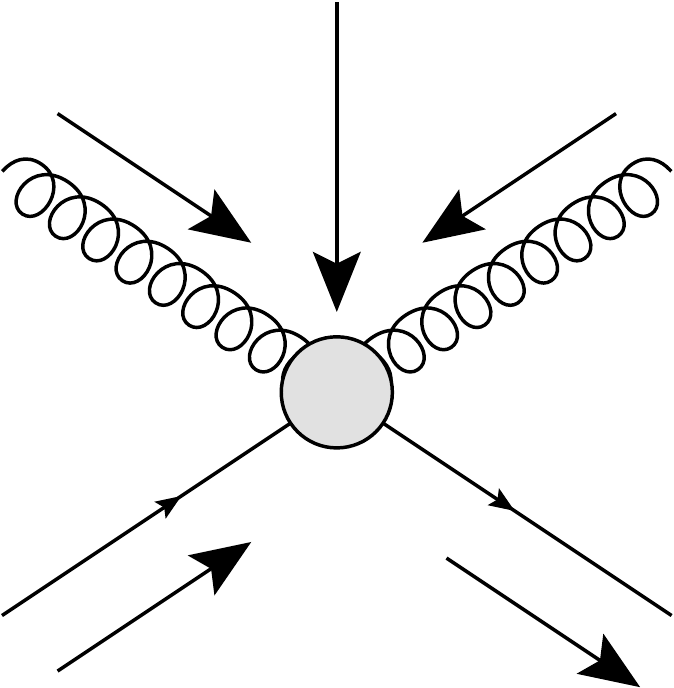}
	\end{picture}
	\hspace{1cm}
	\begin{picture}(5,4)
		\put(-0.25,0){$i$}
		\put(4.05,0){$j$}
		\put(0.85,0){$p_a$}
		\put(2.85,0){$p_b$}
		\put(0.25,2.25){$p_c$}
		\put(1.25,3.5){$p_d$}
		\put(2.75,3.25){$p_e$}
		\put(1.75,0.5){$q$}
		\put(-0.7,3){$c,\gamma$}
		\put(1.8,4.3){$d,\delta$}
		\put(4.15,3){$e,\epsilon$}
		\includegraphics[width=3.6cm]{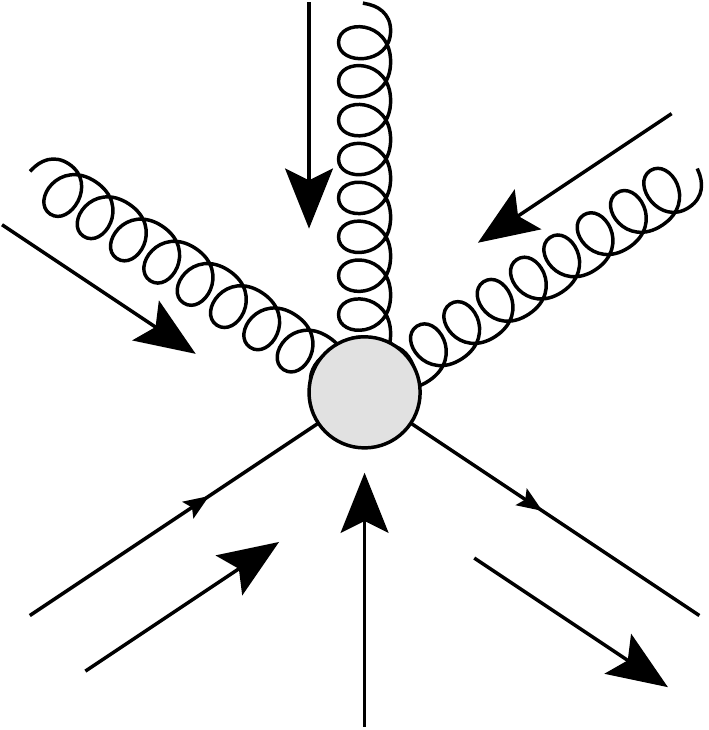}
	\end{picture}}%
	\caption{Quark two-point and quark-gluon three-, four-, and five-point functions with momentum insertion into the operator.}
	\label{img:quarksgluons}
\end{figure}

\begin{figure}[t]
	\centering
	\scalebox{0.75}{
	\setlength{\unitlength}{0.9cm}
	\begin{picture}(4,3.5)
		\put(3.0,2.75){$q$}
		\put(0.85,0){$p_a$}
		\put(2.85,0){$p_b$}
		\put(-0.25,0){$i$}
		\put(4.05,0){$j$}
		\put(0.85,3.25){$p_c$}
		\put(-0.35,3){$\gamma$}
		\includegraphics[width=3.6cm]{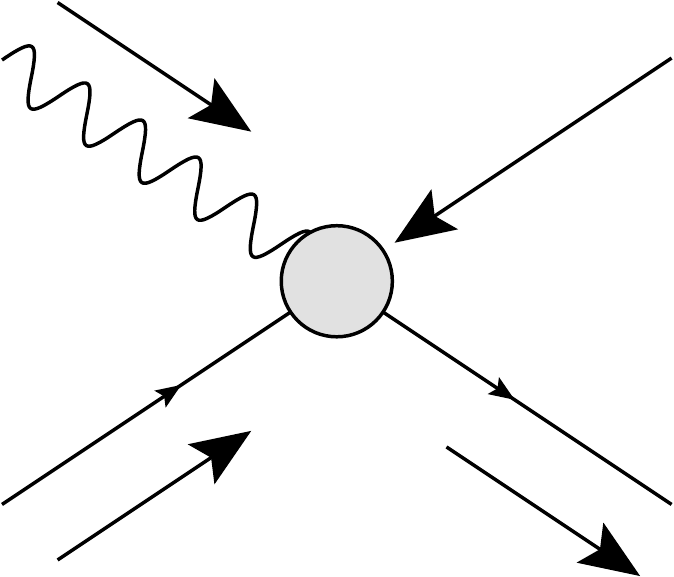}
	\end{picture}
	\hspace{1.5cm}
	\begin{picture}(4,3.5)
		\put(1.75,3.5){$q$}
		\put(0.85,0){$p_a$}
		\put(2.85,0){$p_b$}
		\put(-0.25,0){$i$}
		\put(4.05,0){$j$}
		\put(0.85,3.25){$p_c$}
		\put(2.75,3.25){$p_d$}
		\put(-0.7,3){$c,\gamma$}
		\put(4.15,3){$\delta$}
		\includegraphics[width=3.6cm]{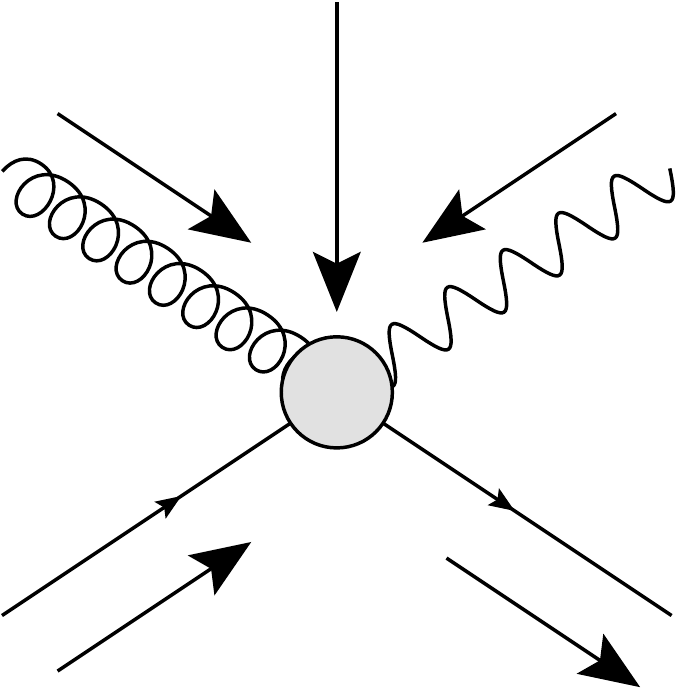}
	\end{picture}
	\hspace{1.5cm}
	\begin{picture}(5,4)
		\put(-0.25,0){$i$}
		\put(4.05,0){$j$}
		\put(0.85,0){$p_a$}
		\put(2.85,0){$p_b$}
		\put(0.25,2.25){$p_c$}
		\put(1.25,3.5){$p_d$}
		\put(2.75,3.25){$p_e$}
		\put(1.75,0.5){$q$}
		\put(-0.7,3){$c,\gamma$}
		\put(1.8,4.3){$d,\delta$}
		\put(4.15,3){$\epsilon$}
		\includegraphics[width=3.6cm]{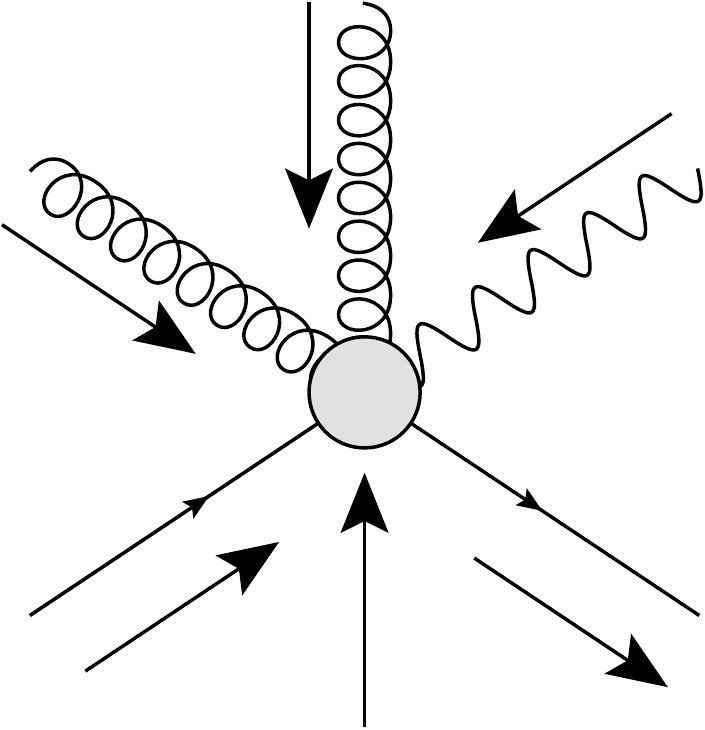}
	\end{picture}}%
	\caption{Quark-photon three-point and quark-gluon-photon four- and five-point functions with momentum insertion into the operator.}
	\label{img:photons}
\end{figure}
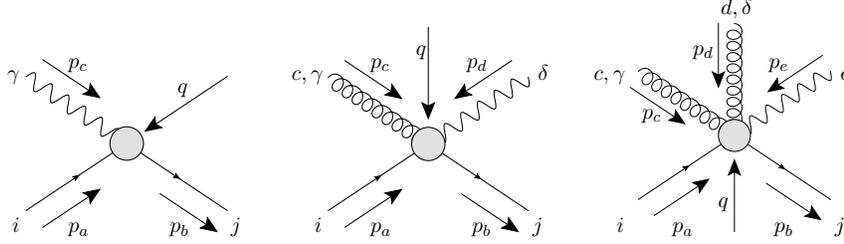

The renormalization conditions for the \rismom{} scheme need to render all renormalized operators finite and determine the finite contributions to the mixing matrix. We will formulate the conditions as the requirement that at certain kinematic points the renormalized amputated Green's functions agree with their tree-level expressions. In order to determine the number of independent conditions that can be obtained from each Green's function, in the following we calculate the $n$-point vertex rules for all the operators of the basis~\eqref{eq:FinalOperatorsDim4}, \eqref{eq:FinalOperatorDim5}, \eqref{eq:FinalOperatorsDim6}, \eqref{eq:FinalOperatorsDim6IIa}, and \eqref{eq:FinalOperatorsDim6IIb}. We insert momentum $q$ into the operator. The convention for signs and factors of $i$ is given by
\begin{align}
	i (2\pi)^4 \delta^{(4)}(q + p_i - p_f) \T_{fi} &:= \sum_{a,n} i \int d^4x e^{-i q \cdot x} \< f | c_a^{(n)} \O_a^{(n)}(x) | i \> \nn
		&\quad +  \sum_{a,n} i \int d^4x e^{-i q \cdot x} \< f | n_a^{(n)} \N_a^{(n)}(x) | i \> \, .
\end{align}
We only list the contact terms. We define kinematics and indices for all the necessary Green's functions as in Figs.~\ref{img:gluons}, \ref{img:quarksgluons}, and \ref{img:photons}. Lorentz indices are denoted by Greek letters $\alpha$, $\beta$, \ldots, color indices by $a$, $b$, \ldots, and quark-flavor indices by $i$, $j$. All the gluon and photon momenta are incoming, while for the quark and ghost lines the momentum flow is in the direction of the fermion- and ghost-number flow. 

\paragraph{Gluon two-point function}

The gluon two-point function is given by
\begin{align*}
	\T_{ab} &= \epsilon_\alpha(p_a) \epsilon_\beta(p_b) \Pi^{\alpha\beta}_{ab} \, , \\*
	\Pi^{\alpha\beta}_{ab} &= \delta_{ab} i \epsilon^{\alpha\beta\mu\nu} {p_a}_\mu {p_b}_\nu \begin{aligned}[t]
		& \bigg\{ 2 c_1^{(4)} + 2 \tr[\M^2] c_4^{(6)} - 2 q^2 c_9^{(6)} \\
		& - (p_a^2 + p_b^2) \left( \frac{1}{2} c_5^{(6)} + 2 n_8^{(6)} + 2 n_{11}^{(6)} + n_{18}^{(6)} \right) \bigg\} \, . \end{aligned}
	\mytag
\end{align*}
It provides a constraint on the dimension-four operator coefficient $c_1^{(4)}$ and three independent constraints on dimension-six operator coefficients.

\paragraph{Gluon three-point function}

The gluon three-point function is given by
\begin{align*}
	\T_{abc} &= \epsilon_\alpha(p_a) \epsilon_\beta(p_b) \epsilon_\gamma(p_c) \Gamma^{\alpha\beta\gamma}_{abc} \, , \\
	\Gamma^{\alpha\beta\gamma}_{abc} &= g f^{abc} \epsilon^{\alpha\beta\gamma\mu} q_\mu \bigg\{ 2 c_1^{(4)} + 2 \tr[\M^2] c_4^{(6)} - 2 q^2 c_9^{(6)} - (p_a^2+p_b^2+p_c^2) \left( \frac{1}{2} c_5^{(6)} + 2 n_8^{(6)} \right) \bigg\} \\
		&\quad + g f^{abc} \epsilon^{\alpha\beta\gamma\mu} \left( {p_a}_\mu p_b \cdot p_c + {p_b}_\mu p_c \cdot p_a + {p_c}_\mu p_a \cdot p_b \right) \bigg\{ -\frac{1}{2} c_1^{(6)} \bigg\} \\
		&\quad + g f^{abc} \epsilon^{\alpha\beta\gamma\mu} \left( {p_a}_\mu p_a^2 + {p_b}_\mu p_b^2 + {p_c}_\mu p_c^2 \right) \bigg\{ 2 n_{11}^{(6)} \bigg\} \\
		&\quad + g f^{abc} \begin{aligned}[t]
			& \bigg[ \epsilon^{\alpha\beta\mu\nu} \left( p_a^\gamma {p_b}_\mu {p_c}_\nu + p_b^\gamma {p_a}_\mu {p_c}_\nu \right) \\
				& + \epsilon^{\beta\gamma\mu\nu} \left( p_b^\alpha {p_c}_\mu {p_a}_\nu + p_c^\alpha {p_b}_\mu {p_a}_\nu \right) \\
				& + \epsilon^{\gamma\alpha\mu\nu} \left( p_c^\beta {p_a}_\mu {p_b}_\nu + p_a^\beta {p_c}_\mu {p_b}_\nu \right) \bigg] \bigg\{ - \frac{1}{2} c_1^{(6)} - c_5^{(6)} - 4 n_8^{(6)} - 4 n_{11}^{(6)} - 2 n_{18}^{(6)} \bigg\} \end{aligned} \\
		&\quad + g f^{abc} \begin{aligned}[t]
			& \bigg[ \epsilon^{\alpha\beta\mu\nu} {p_a}_\mu {p_b}_\nu \left( p_a^\gamma - p_b^\gamma \right) \\
				& + \epsilon^{\beta\gamma\mu\nu} {p_b}_\mu {p_c}_\nu \left( p_b^\alpha - p_c^\alpha \right) \\
				& + \epsilon^{\gamma\alpha\mu\nu} {p_c}_\mu {p_a}_\nu \left( p_c^\beta - p_a^\beta \right) \bigg] \bigg\{ c_5^{(6)} + 4 n_8^{(6)} + 4 n_{11}^{(6)} + 2 n_{18}^{(6)} \bigg\} \end{aligned} \\
		&\quad + g f^{abc} \begin{aligned}[t]
			& \left( \epsilon^{\alpha\beta\mu\nu} p_c^\gamma {p_c}_\mu + \epsilon^{\beta\gamma\mu\nu} p_a^\alpha {p_a}_\mu + \epsilon^{\gamma\alpha\mu\nu} p_b^\beta {p_b}_\mu \right) q_\nu \bigg\{ -2 n_{11}^{(6)} - n_{18}^{(6)} \bigg\} \end{aligned} \\
		&\quad + g f^{abc} \begin{aligned}[t]
			& \bigg[ \epsilon_{\mu\nu\lambda\sigma} \left( g^{\alpha\beta} g^{\gamma\sigma} + g^{\beta\gamma} g^{\alpha\sigma} + g^{\gamma\alpha} g^{\beta\sigma} \right) p_a^\mu p_b^\nu p_c^\lambda \bigg] \\
			&\qquad \times \bigg\{ \frac{1}{2} c_1^{(6)} + c_5^{(6)} + 4 n_8^{(6)} + 4 n_{11}^{(6)} + 2 n_{18}^{(6)} \bigg\} \, . \end{aligned}
	\mytag
\end{align*}
It again provides a constraint on the dimension-four-operator coefficient $c_1^{(4)}$ (which can also be fixed from the two-point function) as well as six independent constraints on dimension-six-operator coefficients. Three of them are linearly dependent with the constraints from the two-point function.

\paragraph{Ghost-gluon three-point function}

The ghost-gluon three-point function is given by
\begin{align*}
	\T_{abc} &= \epsilon_\gamma(p_c) \Gamma^{\gamma}_{abc} \, , \\
	\Gamma^{\gamma}_{abc} &= g f^{abc} \epsilon^{\gamma\mu\nu\lambda} {p_a}_\mu {p_b}_\nu {p_c}_\lambda \bigg\{ -2 n_{11}^{(6)} - n_{18}^{(6)} \bigg\} \, .
	\mytag
\end{align*}
It does not provide a new linear combination of coefficients, i.e., by renormalizing the gluon two- and three-point functions, the ghost-gluon three-point function will be automatically finite.

\paragraph{Quark two-point function}

The quark two-point function is given by
\begin{align*}
	\T_{ji} &= \bar u_j(p_b) \Gamma_{ji} u_i(p_a) \, , \\
	\Gamma_{ji} &= \det(\M) (\M^{-1})_{ji}  \gamma_5 \bigg\{ -2 c_1^{(5)} \bigg\} \\
		&\quad + \M_{ji}  \gamma_5 \begin{aligned}[t]
			& \bigg\{ 2 n_1^{(4)} + (p_a^2 + p_b^2) n_3^{(6)} + q^2 \left( - n_3^{(6)} - n_6^{(6)} - n_{9}^{(6)} - 2 n_{10}^{(6)} \right) \\
			& - 2 \tr[\M^2] \left( \frac{1}{N_f} n_4^{(6)} - n_5^{(6)} \right) \bigg\} \end{aligned} \\
		&\quad + \M_{ji} i \tilde\sigma_{\mu\nu} p_a^\mu p_b^\nu \bigg\{ 2 \left( n_3^{(6)} + n_7^{(6)} + n_{9}^{(6)} \right) \bigg\} \\
		&\quad + (\M^2)_{ji} (\slashed p_a - \slashed p_b) \gamma_5 \bigg\{ c_{6}^{(6)} - n_3^{(6)} + n_4^{(6)} - 2 n_{9}^{(6)} \bigg\} \\
		&\quad + (\M^3)_{ji}  \gamma_5 \bigg\{ 2 n_4^{(6)} \bigg\} \\
		&\quad + \delta_{ji} (\slashed p_a - \slashed p_b) \gamma_5 \begin{aligned}[t]
			& \bigg\{ c_{2}^{(4)} + n_1^{(4)} - \tr[\M^2] \Big( \frac{1}{N_f} c_6^{(6)} - c_7^{(6)} + \frac{1}{N_f} n_4^{(6)} - n_5^{(6)} \Big) \\
				& - q^2 \Big( c_{10}^{(6)} + \frac{1}{2} n_6^{(6)} - \frac{1}{2} n_7^{(6)} + n_{10}^{(6)} \Big) - (p_a^2 + p_b^2) n_7^{(6)}  \bigg\} \end{aligned} \\
		&\quad + \delta_{ji} (\slashed p_a + \slashed p_b) \gamma_5 \bigg\{ \frac{1}{2} (p_a^2 - p_b^2) \Big( n_6^{(6)} + n_7^{(6)} \Big) \bigg\}
		 \, .
	\mytag
\end{align*}
It provides the two missing constraints on dimension-four coefficients, a constraint on the dimension-five coefficient, as well as 10 constraints on dimension-six coefficients.

\paragraph{Quark-gluon three-point function}

The quark-gluon three-point function is given by
\begin{align*}
	\T_{ji,c} &= \epsilon_\gamma(p_c) \bar u_j(p_b) \Gamma_{ji,c}^\gamma u_i(p_a) \, , \\
	\Gamma_{ji,c}^\gamma &= i g \M_{ji} t^c \tilde\sigma^{\gamma\mu} {p_c}_\mu \bigg\{ -2 c_2^{(6)} + 4 n_1^{(6)} - 2 n_7^{(6)} - 2 n_{9}^{(6)} + 2 n_{16}^{(6)} + 2 n_{20}^{(6)} \bigg\} \\
		&\quad + i g \M_{ji} t^c \tilde\sigma^{\gamma\mu} ({p_a}_\mu - {p_b}_\mu) \bigg\{ -2 n_3^{(6)} - 2 n_7^{(6)} - 2 n_{9}^{(6)} + n_{17}^{(6)} + 2 n_{20}^{(6)} \bigg\} \\
		&\quad + g \M_{ji} t^c \gamma_5 (p_a^\gamma + p_b^\gamma) \bigg\{ 2 n_3^{(6)} - n_{15}^{(6)} - n_{17}^{(6)} \bigg\} \\
		&\quad + g \delta_{ji} t^c \gamma^\gamma \gamma_5 \begin{aligned}[t]
			& \bigg\{ (p_a^2 - p_b^2) \left( n_6^{(6)} + n_7^{(6)} - n_{20}^{(6)} \right) \\
			& + (p_a \cdot p_c + p_b \cdot p_c) \left( - 2 n_1^{(6)} + n_6^{(6)} + n_7^{(6)} - n_{16}^{(6)} - n_{20}^{(6)} \right) \bigg\} \end{aligned} \\
		&\quad + g \delta_{ji} t^c \gamma_\mu \gamma_5 (p_a^\gamma p_b^\mu - p_b^\gamma p_a^\mu) \bigg\{ n_6^{(6)} + 3 n_7^{(6)} + n_{15}^{(6)} + n_{19}^{(6)} - n_{20}^{(6)} \bigg\} \\
		&\quad + g \delta_{ji} t^c \gamma_\mu \gamma_5 (p_a^\gamma p_a^\mu - p_b^\gamma p_b^\mu) \bigg\{ n_6^{(6)} - n_7^{(6)} + n_{19}^{(6)} + n_{20}^{(6)} \bigg\} \\
		&\quad + g \delta_{ji} t^c \gamma_\mu \gamma_5 (p_a^\mu + p_b^\mu) p_c^\gamma \bigg\{ n_6^{(6)} + n_7^{(6)} + n_{14}^{(6)} + n_{19}^{(6)} \bigg\} \\
		&\quad + g \delta_{ji} t^c \gamma_\mu \gamma_5 (p_a^\gamma + p_b^\gamma) p_c^\mu \bigg\{ 2 n_1^{(6)} - 2 n_7^{(6)} + n_{16}^{(6)} + n_{20}^{(6)} \bigg\} \\
		&\quad + i g \delta_{ji} t^c \epsilon^{\gamma\mu\nu\lambda} \gamma_\mu ({p_a}_\nu - {p_b}_\nu) {p_c}_\lambda \bigg\{ 2 n_1^{(6)} - 2 n_7^{(6)} - 2 n_8^{(6)} - 2 n_{11}^{(6)} + n_{16}^{(6)} - n_{18}^{(6)} + n_{20}^{(6)} \bigg\}
		\, . \\
	\mytag
\end{align*}
It provides 10 linearly independent constraints on dimension-six coefficients. Two of them are linearly dependent with the constraints from previously listed $n$-point functions.

\paragraph{Quark-gluon four-point function}

The quark-gluon four-point function is given by
\begin{align*}
	\T_{ji,cd} &= \epsilon_\gamma(p_c) \epsilon_\delta(p_d) \bar u_j(p_b) \Gamma_{ji,cd}^{\gamma\delta} u_i(p_a) \, , \\
	\Gamma_{ji,cd}^{\gamma\delta} &= g^2 \M_{ji} f^{cda} t^a \tilde\sigma^{\gamma\delta} \bigg\{ -2 c_2^{(6)} + 4 n_1^{(6)} + 2 n_3^{(6)} - 2 n_{17}^{(6)} \bigg\} \\
		&\quad + g^2 \M_{ji} \{ t^c, t^d\} \gamma_5 g^{\gamma\delta} \bigg\{  2 n_3^{(6)} + 4 n_{13}^{(6)} - 2 n_{15}^{(6)} - 2 n_{17}^{(6)} \bigg\} \\
		&\quad + g^2 \M_{ji} \delta^{cd} \gamma_5 g^{\gamma\delta} \bigg\{ 4 n_{12}^{(6)} - \frac{4}{3} n_{13}^{(6)} \bigg\} \\
		&\quad + i g^2 \delta_{ji} f^{cda} t^a \gamma_\mu \gamma_5 \left( (p_a+p_b)^\gamma g^{\mu\delta} - (p_a+p_b)^\delta g^{\mu\gamma} \right) \bigg\{ 2 n_1^{(6)} + \frac{1}{2} n_{15}^{(6)} \bigg\} \\
		&\quad + g^2 \delta_{ji} \{ t^c, t^d\} \gamma_\mu \gamma_5 \left( (p_a-p_b)^\gamma g^{\mu\delta} + (p_a-p_b)^\delta g^{\mu\gamma} \right) \\
			&\hspace{3cm} \times \bigg\{ n_6^{(6)} + n_7^{(6)} + \frac{1}{2} n_{15}^{(6)} + n_{19}^{(6)} - n_{20}^{(6)} \bigg\} \\
		&\quad + g^2 \delta_{ji} \{ t^c, t^d\} \gamma_\mu \gamma_5 \left( p_c^\gamma g^{\mu\delta} + p_d^\delta g^{\mu\gamma} \right) \bigg\{ n_6^{(6)} + n_7^{(6)} + n_{14}^{(6)} + n_{19}^{(6)} - n_{20}^{(6)} \bigg\} \\
		&\quad + g^2 \delta_{ji} \{ t^c, t^d\} \gamma_\mu \gamma_5 \left( p_d^\gamma g^{\mu\delta} + p_c^\delta g^{\mu\gamma} \right) \bigg\{ -2 n_1^{(6)} + n_6^{(6)} + n_7^{(6)} - n_{16}^{(6)} + n_{19}^{(6)} - n_{20}^{(6)} \bigg\} \\
		&\quad + g^2 \delta_{ji} \{ t^c, t^d\} \gamma_\mu \gamma_5 g^{\gamma\delta} (p_a-p_b)^\mu \bigg\{ - 2 n_7^{(6)} + 2 n_{13}^{(6)} - n_{15}^{(6)} + 2 n_{20}^{(6)} \bigg\} \\
		&\quad + g^2 \delta_{ji} \{ t^c, t^d\} \gamma_\mu \gamma_5 g^{\gamma\delta} (p_c+p_d)^\mu \bigg\{ 2 n_1^{(6)} - 2 n_7^{(6)} + n_{16}^{(6)} + 2 n_{20}^{(6)} \bigg\} \\
		&\quad + g^2 \delta_{ji} f^{cda} t^a \epsilon^{\gamma\delta\mu\nu} \gamma_\mu (p_a-p_b)_\nu \bigg\{ 2 n_1^{(6)} - 2 n_7^{(6)} - 2 n_8^{(6)}  - 2 n_{11}^{(6)} + 2 n_{20}^{(6)} \bigg\} \\
		&\quad + g^2 \delta_{ji} f^{cda} t^a \epsilon^{\gamma\delta\mu\nu} \gamma_\mu (p_c+p_d)_\nu \bigg\{ 2 n_1^{(6)} - 2 n_7^{(6)} - 2 n_8^{(6)}  + n_{16}^{(6)} + 2 n_{20}^{(6)} \bigg\} \\
		&\quad + g^2 \delta_{ji} \delta^{cd} \gamma_\mu \gamma_5 g^{\gamma\delta} (p_a-p_b)^\mu \bigg\{ 2 n_{12}^{(6)} - \frac{2}{3} n_{13}^{(6)} \bigg\}
		\, .
	\mytag
\end{align*}
Out of 11 constraints on dimension-six coefficients, three are linearly independent of the constraints from two- and three-point functions.

\paragraph{Quark-gluon five-point function}

The quark-gluon five-point function is given by
\begin{align*}
	\T_{ji,cde} &= \epsilon_\gamma(p_c) \epsilon_\delta(p_d) \epsilon_\epsilon(p_e) \bar u_j(p_b) \Gamma_{ji,cde}^{\gamma\delta\epsilon} u_i(p_a) \, , \\
	\Gamma_{ji,cde}^{\gamma\delta\epsilon} &= i g^3 \delta_{ji} t^a \Big( g^{\gamma\delta} \gamma^\epsilon d^{cdb} f^{eab} + g^{\delta\epsilon} \gamma^\gamma d^{deb} f^{cab} + g^{\epsilon\gamma} \gamma^\delta d^{ecb} f^{dab} \Big) \gamma_5 \\
		&\quad \times \bigg\{ - 2 n_1^{(6)} + 2 n_{13}^{(6)} - n_{15}^{(6)} \bigg\}
		\, .
	\mytag
\end{align*}
The condition that can be obtained from it is linearly dependent with the previously listed ones.

\paragraph{Quark-photon three-point function}

The quark-photon three-point function is given by
\begin{align*}
	\T_{ji} &= \epsilon_\gamma(p_c) \bar u_j(p_b) \Gamma_{ji}^\gamma u_i(p_a) \, , \\
	\Gamma_{ji}^\gamma &= i e (\M Q)_{ji} \tilde\sigma^{\gamma\mu} {p_c}_\mu \bigg\{ -2 c_3^{(6)} + 4 n_2^{(6)} - 2 n_7^{(6)} - 2 n_{9}^{(6)} \bigg\} \\
		&\quad + i e (\M Q)_{ji} \tilde\sigma^{\gamma\mu} ({p_a}_\mu - {p_b}_\mu) \bigg\{ - 2 n_3^{(6)} - 2 n_7^{(6)} - 2 n_{9}^{(6)} \bigg\} \\
		&\quad + e (\M Q)_{ji} \gamma_5 (p_a^\gamma + p_b^\gamma) \bigg\{ 2 n_3^{(6)} \bigg\} \\
		&\quad + e Q_{ji} \gamma^\gamma \gamma_5 \begin{aligned}[t]
			& \bigg\{ (p_a^2 - p_b^2) \left( n_6^{(6)} + n_7^{(6)} \right) \\
			& + (p_a \cdot p_c + p_b \cdot p_c) \left( - 2 n_2^{(6)} + n_6^{(6)} + n_7^{(6)} \right) \bigg\} \end{aligned} \\
		&\quad + e Q_{ji} \gamma_\mu \gamma_5 (p_a^\gamma p_b^\mu - p_b^\gamma p_a^\mu) \bigg\{ n_6^{(6)} + 3 n_7^{(6)} \bigg\} \\
		&\quad + e Q_{ji} \gamma_\mu \gamma_5 (p_a^\gamma p_a^\mu - p_b^\gamma p_b^\mu) \bigg\{ n_6^{(6)} - n_7^{(6)} \bigg\} \\
		&\quad + e Q_{ji} \gamma_\mu \gamma_5 (p_a^\mu + p_b^\mu) p_c^\gamma \bigg\{ n_6^{(6)} + n_7^{(6)} \bigg\} \\
		&\quad + e Q_{ji} \gamma_\mu \gamma_5 (p_a^\gamma + p_b^\gamma) p_c^\mu \bigg\{ 2 n_2^{(6)} - 2 n_7^{(6)} \bigg\} \\
		&\quad + i e Q_{ji} \epsilon^{\gamma\mu\nu\lambda} \gamma_\mu ({p_a}_\nu - {p_b}_\nu) {p_c}_\lambda \bigg\{ - c_8^{(6)} + 2 n_2^{(6)} - 2 n_7^{(6)} \bigg\}
		\, .
	\mytag
\end{align*}
Out of 7 conditions on the dimension-six coefficients, 3 are linearly independent of the previously listed ones.

\paragraph{Quark-gluon-photon four-point function}

The quark-gluon-photon four-point function is given by
\begin{align*}
	\T_{ji,c} &= \epsilon_\gamma(p_c) \epsilon_\delta(p_d) \bar u_j(p_b) \Gamma_{ji,c}^{\gamma\delta} u_i(p_a) \, , \\
	\Gamma_{ji,c}^{\gamma\delta} &= 
		e g (\M Q)_{ji} t^c \gamma_5 g^{\gamma\delta} \bigg\{ 4 n_3^{(6)} - 2 n_{15}^{(6)} - 2 n_{17}^{(6)} \bigg\} \\
		&\quad + e g Q_{ji} t^c \gamma_\mu \gamma_5 \left( (p_a-p_b)^\delta g^{\mu\gamma} \right) \bigg\{ 2 n_6^{(6)} + 2 n_7^{(6)} - 2 n_{20}^{(6)} \bigg\} \\
		&\quad + e g Q_{ji} t^c \gamma_\mu \gamma_5 \left( (p_a-p_b)^\gamma g^{\mu\delta} \right) \bigg\{ 2 n_6^{(6)} + 2 n_7^{(6)} + n_{15}^{(6)} + 2 n_{19}^{(6)} \bigg\} \\
		&\quad + e g Q_{ji} t^c \gamma_\mu \gamma_5 p_c^\gamma g^{\mu\delta} \bigg\{ 2 n_6^{(6)} + 2 n_7^{(6)} + 2 n_{14}^{(6)} + 2 n_{19}^{(6)} \bigg\} \\
		&\quad + e g Q_{ji} t^c \gamma_\mu \gamma_5 p_d^\gamma g^{\mu\delta} \bigg\{ - 4 n_2^{(6)} + 2 n_6^{(6)} + 2 n_7^{(6)} + 2 n_{19}^{(6)} \bigg\} \\
		&\quad + e g Q_{ji} t^c \gamma_\mu \gamma_5 p_c^\delta g^{\mu\gamma} \bigg\{ - 4 n_1^{(6)} + 2 n_6^{(6)} + 2 n_7^{(6)} - 2 n_{16}^{(6)} - 2 n_{20}^{(6)} \bigg\} \\
		&\quad + e g Q_{ji} t^c \gamma_\mu \gamma_5 p_d^\delta g^{\mu\gamma} \bigg\{ 2 n_6^{(6)} + 2 n_7^{(6)} - 2 n_{20}^{(6)} \bigg\} \\
		&\quad + e g Q_{ji} t^c \gamma_\mu \gamma_5 g^{\gamma\delta} (p_a-p_b)^\mu \bigg\{ - 4 n_7^{(6)} - n_{15}^{(6)} + 2 n_{20}^{(6)} \bigg\} \\
		&\quad + e g Q_{ji} t^c \gamma_\mu \gamma_5 g^{\gamma\delta} p_c^\mu \bigg\{ 4 n_1^{(6)} - 4 n_7^{(6)} + 2 n_{16}^{(6)} + 2 n_{20}^{(6)} \bigg\} \\
		&\quad + e g Q_{ji} t^c \gamma_\mu \gamma_5 g^{\gamma\delta} p_d^\mu \bigg\{ 4 n_2^{(6)} - 4 n_7^{(6)} + 2 n_{20}^{(6)} \bigg\}
		\, .
	\mytag
\end{align*}
The 8 conditions that could be obtained from the quark-gluon-photon four-point function are linearly dependent with the previously listed ones.

\paragraph{Quark-gluon-photon five-point function}

The quark-gluon-photon five-point function is given by
\begin{align*}
	\T_{ji,cd} &= \epsilon_\gamma(p_c) \epsilon_\delta(p_d) \epsilon_\epsilon(p_e) \bar u_j(p_b) \Gamma_{ji,cd}^{\gamma\delta\epsilon} u_i(p_a) \, , \\
	\Gamma_{ji,cd}^{\gamma\delta\epsilon} &= i e g^2 Q_{ji} f^{cda} t^a \Big( \gamma^\delta g^{\gamma\epsilon} - \gamma^\gamma g^{\delta\epsilon} \Big) \gamma_5  \bigg\{ - 4 n_1^{(6)} - n_{15}^{(6)} \bigg\}
		\, .
	\mytag
\end{align*}
It only gives a condition that is linearly dependent with previously listed ones.

\subsection{Projection of scalar structures}
\label{sec:Projections}

In order to renormalize the $CP$-odd three-gluon operator, we need to impose $3+1+30$ linearly independent renormalization conditions on Green's functions, corresponding to the counter\-terms from dimension-four, -five, and -six operators. The number of available structures in the Green's functions is larger---we choose to use the lower $n$-point functions as far as possible, which leads to the set of structures listed in Table~\ref{tab:nPointFunctionsConditions}. We use structures from two-, three-, and four-point functions with additional momentum insertion into the operator. The five-point functions are not needed. The photonic Green's function $\bar q q A$ is required to provide 3 conditions that fix the photonic counterterms $\O_3^{(6)}$, $\O_8^{(6)}$, and $\N_2^{(6)}$.

\begin{table}[t]
	\centering
	\begin{tabular}{c c l l l }
		\toprule
		operator dimension &	$n$	&	gluons only			& quarks \& gluons		& photonic \\
		\midrule
		4	&	2	&	$g^2$ : 1 / 1		&	$\bar q q$ : 2 / 2							\\
			&	3	&	$g^3$ : 0 / 1		&						&					\\
		\midrule
		5	&	2	&					&	$\bar q q$ : 1 / 1							\\
		\midrule
		6	&	2	&	$g^2$ : 3 / 3		&	$\bar q q$ : 10 / 10							\\
			&	3	&	$g^3$ : 3 / 6		&	$\bar qqg$ : 8 / 10		&	$\bar qqA$ : 3 / 7	\\
			&	4	&					&	$\bar qqg^2$ : 3 / 11		&	$\bar qqgA$ : 0 / 8	\\
			&	5	&					&	$\bar qqg^3$ : 0 / 1		&	$\bar qqg^2A$ : 0 / 1	\\
		\bottomrule
	\end{tabular}
	\caption{Overview of the available $n$-point functions with up to one photon. $a$ / $b$ indicates that $a$ constraints are used out of $b$ linearly independent ones provided by the $n$-point function at $\O(e)$.}
	\label{tab:nPointFunctionsConditions}
\end{table}

In the following, we define $3+1+30$ projections in Lorentz, Dirac, color, and flavor space out of the Green's functions. The explicit renormalization conditions will be formulated in Sect.~\ref{sec:RenormalizationConditions} by requiring these projections to agree with their tree-level expressions.
We remark that all the Lorentz contractions in the projections are performed in $D=4$ dimensions, see Sect.~\ref{sec:DimRegAndRenormalization}.

The quark-mass and charge matrices take the values
\begin{align}
	\M = \mathrm{diag}(m_u, m_d, m_s) \, , \quad Q = \frac{1}{3} \mathrm{diag}(2,-1,-1) \, .
\end{align}
However, after taking possible derivatives with respect to the quark masses, all conditions will be understood in the chiral limit, $\M\to0$. This leads to a mass-independent renormalization scheme.

\paragraph{\boldmath Gluon two-point function $g^2$}

We evaluate the gluon two-point function with momentum insertion into the operator $\O_i$ as a function of the three Lorentz invariants $p_a^2$, $p_b^2$, and $q^2$. The conditions are imposed on one Lorentz contraction of the amputated two-point function and on its partial derivatives with respect to $q^2$, $p_a^2$, and the $s$-quark mass at the symmetric point in the chiral limit defined by
\begin{align}\label{S2}
	S_2 : \quad p_a^2 = p_b^2 = q^2 = - \Lambda^2 \, , \quad m_u = m_d = m_s = 0 \, ,
\end{align}
i.e., all invariants take large space-like values.
Denoting by   $\lambda(a,b,c) = a^2 + b^2 + c^2 - 2(ab+bc+ca)$  the K\"all\'en triangle function, the projections are
\begin{align}\label{g2p}
	R_{1}[\O_i] &:= \frac{1}{\lambda(p_a^2,p_b^2,q^2)} i \epsilon_{\alpha\beta\mu\nu} p_a^\mu p_b^\nu \delta^{ab} \Pi_{ab}^{\alpha\beta} \bigg|_{S_2} \, , \nn
	R_{2}[\O_i] &:= \frac{\p}{\p q^2} \left[ \frac{1}{\lambda(p_a^2,p_b^2,q^2)} i \epsilon_{\alpha\beta\mu\nu} p_a^\mu p_b^\nu \delta^{ab} \Pi_{ab}^{\alpha\beta} \right] \bigg|_{S_2} \, , \nn
	R_{3}[\O_i] &:= \frac{\p}{\p p_a^2} \left[ \frac{1}{\lambda(p_a^2,p_b^2,q^2)} i \epsilon_{\alpha\beta\mu\nu} p_a^\mu p_b^\nu \delta^{ab} \Pi_{ab}^{\alpha\beta} \right] \bigg|_{S_2} \, , \nn
	R_{4}[\O_i] &:= \frac{\p^2}{\p m_s^2} \left[ \frac{1}{\lambda(p_a^2,p_b^2,q^2)} i \epsilon_{\alpha\beta\mu\nu} p_a^\mu p_b^\nu \delta^{ab} \Pi_{ab}^{\alpha\beta} \right] \bigg|_{S_2} \, .
\end{align}
In $R_{4}$, the derivative is taken with respect to the renormalized \msbar{} $s$-quark mass $m_s^\text{\msbar}(\mu=\Lambda)$. On the lattice, this can be implemented as a derivative with respect to the bare mass times the appropriate renormalization factor connecting the bare lattice mass to the \msbar{} mass.

\paragraph{\boldmath Gluon three-point function $g^3$}

The three-point function with momentum insertion effectively has four-point kinematics and depends on six Lorentz invariants, e.g., $p_a^2$, $p_b^2$, $p_c^2$, as well as the Mandelstam variables
\begin{align}
	 s=(p_a+p_b)^2 \, , \quad t=(p_a+p_c)^2 \, , \quad u=(p_b+p_c)^2 \, .
\end{align}
The conditions are imposed on three different Lorentz contractions of the amputated three-point function at the non-symmetric point defined by
\begin{align}
	\tilde S_3: \quad p_a^2 &= p_b^2 = p_c^2 = s = u = - \Lambda^2 \, , \quad t=-2\Lambda^2 \, , \quad q^2 = s + t + u - p_a^2 - p_b^2 - p_c^2 = - \Lambda^2 \, , \nn
		m_u &= m_d = m_s = 0 \, .
\end{align}
The conditions will be imposed on the following contractions:
\begin{align}\label{g3p}
	R_{5}[\O_i] &:= \epsilon_{\alpha\beta\gamma\mu} (p_a+p_b+p_c)^\mu f^{abc} \Pi_{abc}^{\alpha\beta\gamma} \Big|_{\tilde S_3} \, , \nn
	R_{6}[\O_i] &:= \epsilon_{\alpha\beta\gamma\mu} (p_a^\mu p_b \cdot p_c + p_b^\mu p_c \cdot p_a + p_c^\mu p_a \cdot p_b) f^{abc} \Pi_{abc}^{\alpha\beta\gamma} \Big|_{\tilde S_3} \, , \nn
	R_{7}[\O_i] &:= \epsilon^{\mu\nu\lambda\sigma} (\bar g_{\alpha\beta} \bar g_{\gamma\sigma} + \bar g_{\beta\gamma} \bar g_{\alpha\sigma} + \bar g_{\gamma\alpha} \bar g_{\beta\sigma} ) {p_a}_\mu {p_b}_\nu {p_c}_\lambda f^{abc} \Pi_{abc}^{\alpha\beta\gamma} \Big|_{\tilde S_3} \, ,
\end{align}
where $\bar g_{\mu\nu}$ denotes the metric tensor in $D=4$ spacetime dimensions.

\paragraph{\boldmath Quark two-point function $\bar qq$}

The quark two-point function is evaluated at the non-symmetric point defined by
\begin{align}
	\tilde S_2 : \quad p_a^2 = q^2 = - \Lambda^2 \, , \quad p_b^2 = - 2 \Lambda^2 \, , \quad m_u = m_d = m_s = 0 \, .
\end{align}
The renormalization conditions will be imposed on suitable contractions of the two-point function and derivatives with respect to $q^2$, $p_a^2$, and the \msbar{} masses (again evaluated at the scale $\mu=\Lambda$):
\begin{align}\label{q2p}
	R_{8}[\O_i] &:= \frac{\p}{\p m_s} \tr\left[ \gamma_5 \delta^{ij} \Gamma_{ji} \right] \Big|_{\tilde S_2} \, , \nn
	R_{9}[\O_i] &:= \frac{\p^2}{\p m_s \p q^2} \tr\left[ \gamma_5 \delta^{ij} \Gamma_{ji} \right] \Big|_{\tilde S_2} \, , \nn
	R_{10}[\O_i] &:= \frac{\p^2}{\p m_s \p p_a^2} \tr\left[ \gamma_5 \delta^{ij} \Gamma_{ji} \right] \Big|_{\tilde S_2} \, , \nn
	R_{11}[\O_i] &:= \frac{\p^2}{\p m_u \p m_s} \tr\left[ \gamma_5 \delta^{ij} \Gamma_{ji} \right] \Big|_{\tilde S_2} \, , \nn
	R_{12}[\O_i] &:= \frac{\p^3}{\p m_u \p m_s^2} \tr\left[ \gamma_5 \delta^{ij} \Gamma_{ji} \right] \Big|_{\tilde S_2} \, , \nn
	R_{13}[\O_i] &:= \frac{\p^3}{\p m_s^3} \tr\left[ \gamma_5 \delta^{ij} \Gamma_{ji} \right] \Big|_{\tilde S_2} \, , \nn
	R_{14}[\O_i] &:= \frac{\p}{\p m_s} \tr\left[ i p_a^\mu p_b^\nu \tilde\sigma_{\mu\nu} Q^{ij} \Gamma_{ji} \right] \Big|_{\tilde S_2} \, , \nn
	R_{15}[\O_i] &:= \frac{\p^2}{\p m_s^2} \tr\left[ (p_a-p_b)^\mu \tilde\gamma_\mu Q^{ij} \Gamma_{ji} \right] \Big|_{\tilde S_2} \, , \nn
	R_{16}[\O_i] &:= \tr\left[ \frac{(p_a+p_b)^2 (p_a-p_b)^\mu - (p_a^2-p_b^2) (p_a+p_b)^\mu}{\lambda(p_a^2,p_b^2,q^2)} \tilde\gamma_\mu \delta^{ij} \Gamma_{ji} \right] \bigg|_{\tilde S_2} \, , \nn
	R_{17}[\O_i] &:= \frac{\p}{\p q^2} \tr\left[ \frac{(p_a+p_b)^2 (p_a-p_b)^\mu - (p_a^2-p_b^2) (p_a+p_b)^\mu}{\lambda(p_a^2,p_b^2,q^2)} \tilde\gamma_\mu \delta^{ij} \Gamma_{ji} \right] \bigg|_{\tilde S_2} \, , \nn
	R_{18}[\O_i] &:= \frac{\p}{\p p_a^2} \tr\left[ \frac{(p_a+p_b)^2 (p_a-p_b)^\mu - (p_a^2-p_b^2) (p_a+p_b)^\mu}{\lambda(p_a^2,p_b^2,q^2)} \tilde\gamma_\mu \delta^{ij} \Gamma_{ji} \right] \bigg|_{\tilde S_2} \, , \nn
	R_{19}[\O_i] &:= \frac{\p^2}{\p m_s^2} \tr\left[ \frac{(p_a+p_b)^2 (p_a-p_b)^\mu - (p_a^2-p_b^2) (p_a+p_b)^\mu}{\lambda(p_a^2,p_b^2,q^2)} \tilde\gamma_\mu \delta^{ij} \Gamma_{ji} \right] \bigg|_{\tilde S_2} \, , \nn
	R_{20}[\O_i] &:= \tr\left[ [ (p_a^2-p_b^2) (p_a-p_b)^\mu - (p_a-p_b)^2 (p_a+p_b)^\mu ] \tilde\gamma_\mu \delta^{ij} \Gamma_{ji} \right] \Big|_{\tilde S_2} \, .
\end{align}
$\tr$ stands for the trace in Dirac space. The traces in flavor space are written explicitly with summed indices.

\paragraph{\boldmath Quark-gluon three-point function $\bar qqg$}

The quark-gluon three-point function with momentum insertion again depends on six Lorentz invariants, e.g., $p_a^2$, $p_b^2$, $p_c^2$, as well as the Mandelstam variables
\begin{align}
	 s=(p_a-p_b)^2 \, , \quad t=(p_a+p_c)^2 \, , \quad u=(p_b-p_c)^2 \, .
\end{align}
At the second non-symmetric point defined by
\begin{align}
	\hat S_3: \quad p_a^2 &= p_c^2 = s = - \Lambda^2 \, , \quad p_b^2 = t = u =-2\Lambda^2 \, , \quad q^2 = s + t + u - p_a^2 - p_b^2 - p_c^2 = - \Lambda^2 \, , \nn
		m_u &= m_d = m_s = 0 \, ,
\end{align}
the renormalization conditions will be imposed on eight different contraction of the amputated three-point function:
\begin{align}\label{qg3p}
	R_{21}[\O_i] &:= \frac{\p}{\p m_s} \tr\left[ i p_c^\mu \tilde\sigma_{\gamma\mu} t^c  Q^{ij} \Gamma_{ji,c}^{\gamma} \right] \Big|_{\hat S_3} \, , \nn
	R_{22}[\O_i] &:= \frac{\p}{\p m_s} \tr\left[ i (p_a-p_b)^\mu \tilde\sigma_{\gamma\mu} t^c  Q^{ij} \Gamma_{ji,c}^{\gamma} \right] \Big|_{\hat S_3} \, , \nn
	R_{23}[\O_i] &:= \frac{\p}{\p m_s} \tr\left[ (p_a+p_b)_\gamma \gamma_5 t^c Q^{ij} \Gamma_{ji,c}^{\gamma} \right] \Big|_{\hat S_3} \, , \nn
	R_{24}[\O_i] &:= \tr\left[ \tilde\gamma_\gamma t^c \delta^{ij} \Gamma_{ji,c}^{\gamma} \right] \Big|_{\hat S_3} \, , \nn
	R_{25}[\O_i] &:= \tr\left[ {p_a}_\mu {p_a}_\gamma \tilde\gamma^\mu t^c \delta^{ij} \Gamma_{ji,c}^{\gamma} \right] \Big|_{\hat S_3} \, , \nn
	R_{26}[\O_i] &:= \tr\left[ (p_a+p_b)_\mu {p_c}_\gamma \tilde\gamma^\mu t^c \delta^{ij} \Gamma_{ji,c}^{\gamma} \right] \Big|_{\hat S_3} \, , \nn
	R_{27}[\O_i] &:= \tr\left[ {p_c}_\mu (p_a+p_b)_\gamma \tilde\gamma^\mu t^c \delta^{ij} \Gamma_{ji,c}^{\gamma} \right] \Big|_{\hat S_3} \, , \nn
	R_{28}[\O_i] &:= \tr\left[ i \epsilon_{\gamma\mu\nu\lambda} p_b^\nu p_c^\lambda \gamma^\mu t^c \delta^{ij} \Gamma_{ji,c}^{\gamma} \right] \Big|_{\hat S_3} \, ,
\end{align}
where $\tr$ stands for the trace both in Dirac space and in color space.

\paragraph{\boldmath Quark-gluon four-point function $\bar qqg^2$}

Due to momentum insertion, the quark-gluon four-point function has five-point kinematics, i.e., there are 10 independent Lorentz invariants that can be chosen as $p_a^2$, $p_b^2$, $p_c^2$, $p_d^2$, as well as the Mandelstam variables
\begin{align}
	s_{ab} &= (p_a - p_b)^2 \, , \quad s_{ac} = (p_a+p_c)^2 \, , \quad s_{ad} = (p_a+p_d)^2 \, , \nn
	s_{bc} &= (p_b - p_c)^2 \, , \quad s_{bd} = (p_b - p_d)^2 \, , \quad s_{cd} = (p_c + p_d)^2 \, .
\end{align}
At the non-symmetric kinematical point
\begin{align}
	\tilde S_4 : \quad p_a^2 &= p_b^2 = p_c^2 = - \Lambda^2 \, , \quad p_d^2 = s_{ab} = s_{ac} = s_{ad} = s_{bc} = s_{bd} = s_{cd} = - 2 \Lambda^2 \, , \nn
		q^2 &= s_{ab} + s_{ac} + s_{ad} + s_{bc} + s_{bd} + s_{cd} - 2 p_a^2 - 2 p_b^2 - 2p_c^2 - 2p_d^2 = - 2 \Lambda^2 \, , \nn
		m_u &= m_d = m_s = 0
\end{align}
we use the following projections for the renormalization:
\begin{align}
	\label{eq:qg4p}
	R_{29}[\O_i] &:= \frac{\p}{\p m_s} \tr\left[  {p_c}_\gamma {p_d}_\delta p_c^\mu p_d^\nu \tilde\sigma_{\mu\nu} f^{cda} t^a Q^{ij} \Gamma_{ji,cd}^{\gamma\delta} \right] \Big|_{\tilde S_4} \, , \nn
	R_{30}[\O_i] &:= \frac{\p}{\p m_s} \tr\left[ {p_c}_\gamma {p_d}_\delta \gamma_5 \delta^{cd} Q^{ij} \Gamma_{ji,cd}^{\gamma\delta} \right] \Big|_{\tilde S_4} \, , \nn
	R_{31}[\O_i] &:= \frac{\p}{\p m_s} \tr\left[ {p_c}_\gamma {p_d}_\delta \gamma_5 d^{cda} t^a Q^{ij} \Gamma_{ji,cd}^{\gamma\delta} \right] \Big|_{\tilde S_4} \, .
\end{align}
$\tr$ again stands for the trace both in Dirac and color space.

\paragraph{\boldmath Quark-photon three-point function $\bar qqA$}

For the quark-photon three-point function, we choose the same kinematical configuration as for the quark-gluon three-point function. The projections are:
\begin{align}\label{qgamma3p}
	R_{32}[\O_i] &:= \frac{\p}{\p m_s} \tr\left[ i p_c^\mu \tilde\sigma_{\gamma\mu} \delta^{ij} \Gamma_{ji}^{\gamma} \right] \Big|_{\hat S_3} \, , \nn
	R_{33}[\O_i] &:= \tr\left[ \tilde\gamma_\gamma Q^{ij} \Gamma_{ji}^{\gamma} \right] \Big|_{\hat S_3} \, , \nn
	R_{34}[\O_i] &:= \tr\left[ i \epsilon_{\gamma\mu\nu\lambda} (p_a+p_b)^\nu p_c^\lambda \gamma^\mu Q^{ij} \Gamma_{ji}^{\gamma} \right] \Big|_{\hat S_3} \, ,
\end{align}
where $\tr$ denotes the trace in Dirac space.

\subsection{Renormalization conditions in the \rismom{} scheme}
\label{sec:RenormalizationConditions}

The physical matrix element of the $CP$-odd three-gluon operator in the \msbar{} scheme, defined 
in~\eqref{eq:physicalME}, depends on the matching coefficients $C_{ij}$ and on the matrix elements of the
gauge-invariant \rismom{} operators. One needs to provide renormalization conditions
to define \rismom{} operators on the lattice, but, as can be seen from~\eqref{eq:physicalME},  
at $\O(\alpha_s)$ it is not necessary to give the entries of the renormalization matrices $Z^\text{\msbar}_{ij}$ and $Z^{\rm RI}_{ij}$ with $i \geq 2$. 

In the following, we define the \rismom{} scheme by providing explicit renormalization conditions for all physical operators. We impose $3+1+30$ conditions on Green's functions with insertions of the gCEDM by requiring that the projections of two-, three-, and four-point functions defined in Sect.~\ref{sec:Projections} agree with their tree-level values. For the renormalization of the additional physical operators, only a subset of these conditions is needed to fix all possible mixings.

The gauge-invariant operators at dimension six~\eqref{eq:FinalOperatorsDim6} include the qCEDM, $\O^{(6)}_2$, and the qEDM, $\O^{(6)}_3$. The five operators $\O_{4,6,7,9,10}^{(6)}$ are related to the divergence of the axial current and the QCD $\bar\theta$ term,
with additional powers of the external momentum or of the quark masses, which have little influence on the renormalization.
As an alternative to the conditions provided below, these additional operators could be renormalized by using the conditions given in~\cite{Bhattacharya:2015rsa}, where the case of generic flavor structure was discussed. With minor modifications the renormalization conditions of~\cite{Bhattacharya:2015rsa} could be adjusted to the case considered in this paper, where the flavor structure is determined by the mass $\M$ and charge $Q$ matrices.

\subsubsection{Conditions for the gCEDM}

In order to renormalize the gCEDM, we need to impose $3+1+30$ renormalization conditions. They are given as
\begin{align}
	\label{eq:ConditionsO1}
	R_k\big[\O_1^{(6)}\big] &= 0 \, , \quad k \in \{ 1, \ldots, 4, 8, \ldots, 34 \} \, , \nn
	R_{5}\big[\O_1^{(6)}\big] &= - 3 g N_c^2 C_F \Lambda^4 \, , \nn
	R_{6}\big[\O_1^{(6)}\big] &= - 3 g N_c^2 C_F \Lambda^6 \, , \nn
	R_{7}\big[\O_1^{(6)}\big] &= 9 g N_c^2 C_F \Lambda^6 \, ,
\end{align}
where $C_F = \frac{N_c^2-1}{2N_c}$. The coupling $g$ on the RHS of these equations could be chosen as the renormalized coupling in any scheme. In order to simplify the matching between the \msbar{} and \rismom{} schemes, we choose $g^\text{\msbar{}}(\mu = \Lambda,\varepsilon=0)$. Due to the contractions chosen in~\eqref{eq:qg4p}, no gluon-exchange diagrams survive in the projections of the quark-gluon four-point function and only the contact terms contribute at tree level in~\eqref{eq:ConditionsO1}.

\subsubsection{Conditions for the qCEDM}

As shown in Table~\ref{tab:MixingStructure}, the qCEDM operator $\O_2^{(6)}$ in total mixes with 13 operators: it mixes with one dimension-four operator, $\O_2^{(4)} - \N_1^{(4)}$, which corresponds to the pseudoscalar density, see~\eqref{eq:PseudoscalarDensityRelations}.
At dimension five, it mixes into $\O_1^{(5)}$. At dimension six, $\O_2^{(6)}$ is renormalized by five gauge-invariant operators that do not vanish by EOMs, $\O^{(6)}_{2,3,4,6,7}$, four gauge-invariant nuisance operators 
$\N^{(6)}_{3,4,5,9}$, and one gauge-variant nuisance operator, $\N^{(6)}_{17}$. In addition, in the basis of~\eqref{eq:FinalOperatorsDim6}
and~\eqref{eq:FinalOperatorsDim6IIa}, $\O^{(6)}_2$ can mix into the combination $\O_{10}^{(6)} - \N_{10}^{(6)}$, see~\eqref{eq:PseudoscalarDensityRelations}.

Therefore, we need to impose 13 renormalization conditions on Green's functions with insertions of $\O_2^{(6)}$:
\begin{align}
	\label{eq:ConditionsO2}
	R_k\big[\O_2^{(6)}\big] &= 0 \, , \quad k \in \{ 4, 8, 9, 10, 11, 12, 13, 14, 15, 19, 32 \} \, , \nn
	R_{21}\big[\O_2^{(6)}\big] &= 8 g N_c C_F \Lambda^2 \, , \nn
	R_{22}\big[\O_2^{(6)}\big] &= -4 g N_c C_F \Lambda^2 \, .
\end{align}
Again, the coupling on the RHS is chosen as $g^\text{\msbar{}}(\mu = \Lambda,\varepsilon=0)$.

\subsubsection{Condition for the qEDM}

The qEDM operator renormalizes diagonally. Hence, it suffices to impose a single renormalization condition on the quark-photon three-point function:
\begin{align}
	\label{eq:ConditionsO3}
	R_{32}\big[\O_3^{(6)}\big] &= 8 e \Lambda^2 \, .
\end{align}
The coupling $e$ to the external electromagnetic field is not renormalized in QCD.

\subsubsection{Conditions for the remaining operators}

The remaining physical operators that mix with the gCEDM are the dimension-four QCD $\theta$ term, $\O_1^{(4)}$, the dimension-five operator $\O_1^{(5)}$, and the dimension-six operator $\O_4^{(6)}$, a mass correction to the $\theta$ term. The remaining dimension-six operators are total derivatives and do not contribute for vanishing momentum insertion into the physical matrix element.

The operator $\O_4^{(6)}$ mixes into $\O_{4,7}^{(6)}$ and $\N_5^{(6)}$. We need three renormalization conditions, one condition on the gluon two-point function and two conditions on the quark two-point function:
\begin{align}
	\label{eq:ConditionsO4}
	R_{4}\big[\O_4^{(6)}\big] &= -4 N_c C_F \, , \nn
	R_k\big[\O_4^{(6)}\big] &= 0 \, , \quad k \in \{ 12, 19 \} \, .
\end{align}
The operator $\O_1^{(5)}$ renormalizes diagonally. We impose the single condition
\begin{align}
	\label{eq:ConditionsO1D5}
	R_{11}\big[\O_1^{(5)}\big] &= -8 \, .
\end{align}
The operator $\O_1^{(4)}$ mixes into $\O_{1,2}^{(4)}$ and $\N_1^{(4)}$. We need three renormalization conditions, one condition on the gluon two-point function and two conditions on the quark two-point function:
\begin{align}
	\label{eq:ConditionsO1D4}
	R_{1}\big[\O_1^{(4)}\big] &= -2 N_c C_F \, , \nn
	R_k\big[\O_1^{(4)}\big] &= 0 \, , \quad k \in \{ 8, 16 \} \, .
\end{align}
As discussed in~\cite{Bhattacharya:2015rsa}, these conditions define a renormalized  $G \tilde G$ operator that does not satisfy the singlet Ward identity.
The Ward identity can be restored by a finite renormalization, as done in~\cite{Bhattacharya:2015rsa}.


\begin{fmffile}{diags/oneloop}	

\section{Matching at one loop}
\label{sec:OneLoopMatching}

In this section we calculate the matching coefficients $C_{1j}$, defined in~\eqref{eq:Cij}, at one loop in QCD.
Since the \rismom{} operators are independent of the chosen regulator, we can obtain the matching coefficients $C_{1j}$ by calculating the $n$-point functions
in dimensional regularization, and then imposing the \msbar{} and \rismom{} renormalization conditions.
Together with the nonperturbative definition 
of the \rismom{} operators, ensured by the renormalization conditions discussed in Sect.~\ref{sec:Scheme}, this will allow to convert lattice-QCD calculations of the nucleon 
EDM induced by the gCEDM to the \msbar{} scheme, up to $\O(\alpha_s^2)$ corrections.

In Sect.~\ref{sec:GaugeFixing}, we discuss two different gauge fixing procedures: conventional covariant gauge and background-field gauge. In Sect.~\ref{sec:DimRegAndRenormalization}, we define our dimensional \msbar{} scheme. We present the results for the matching coefficients at one loop in Sect.~\ref{sec:Results}.

\subsection{Gauge fixing}
\label{sec:GaugeFixing}

We provide results with two gauge-fixing choices. First, we
work in a generic covariant gauge, where the QCD Lagrangian in~\eqref{eq:LQCD} is complemented by the gauge-fixing term 
\begin{equation}
	\mathcal L_{\rm GF} = - \frac{1}{2\xi} \left( \partial^\mu G^a_\mu\right)^2
\end{equation}
and by the ghost Lagrangian given in~\eqref{eq:LGFgh}. This family includes the Landau gauge $\xi=0$  that can be easily implemented on the lattice. 

Second, we will employ the background-field method~\cite{Abbott:1980hw,Abbott:1983zw}, which greatly simplifies the mixing structure. In the background-field method, all fields are split into a classical background field $\hat F$ and a quantum field $F$,
\begin{align}
	F \mapsto F + \hat F \, .
\end{align}
The quantum fields are the integration variables in the functional integral. External fields and tree-level propagators are background fields, while internal loop propagators are quantum fields. For fermion fields, quantum and background fields need not be distinguished. The gauge of the background and quantum fields can be fixed independently. The background-field method manifestly preserves gauge invariance with respect to the background fields, hence one only has to consider mixing with gauge-invariant operators in the classes I and IIa defined in Sect.~\ref{sec:Mixing}, whereas no counterterms of class IIb are required.

The gauge-fixing term for the quantum fields is given by
\begin{align}
	\label{eq:GFback}
	\L_{\rm GF} = - \frac{1}{2\xi} \left( \hat{D}^\mu G^a_\mu\right)^2 \, ,
\end{align}
where $\hat{D}^\mu$ denotes the covariant derivative with respect to the background field $\hat G^a_\mu$,
\begin{align}
	\hat{D}^\mu G^a_\mu  = \p^\mu G^{a}_\mu + g f^{a b c} \hat G^{b\, \mu} G^{c}_\mu \, ,
\end{align}
while we retain the symbol $G^a_\mu$ for the quantum gluon field. The ghost Lagrangian reads~\cite{Abbott:1980hw}
\begin{align}
	\L_{\rm gh} = - \bar c^a \left[ \Box \delta^{ab} - g \overleftarrow\p_\mu f^{abc}( \hat G^\mu_c + G^\mu_c ) + g f^{acb} \hat G^c_\mu \p^\mu + g^2 f^{ace} f^{edb} \hat G_\mu^c( \hat G^\mu_d + G^\mu_d) \right] c^b \, .
\end{align}
The background-field gauge-fixing term is simply given by
\begin{align}
	\mathcal L_{\rm GF}^\mathrm{BG} = - \frac{1}{2\hat\xi} \left( \partial^\mu \hat G^a_\mu\right)^2 \, ,
\end{align}
where the gauge-fixing parameter $\hat\xi$ is independent of $\xi$. As the background fields only appear at tree level, ghost terms can be ignored.

\subsection{Dimensional regularization and renormalization}
\label{sec:DimRegAndRenormalization}

In dimensional regularization, we employ the 't\ Hooft--Veltman (HV) scheme~\cite{tHooft:1972tcz,Breitenlohner:1977hr}. The definition of $\gamma_5$ is
\begin{align}
	\label{eq:g5}
	\gamma_5 = - \frac{i}{4!}  \epsilon_{\mu\nu\lambda\sigma} \gamma^{\mu} \gamma^{\nu} \gamma^{\lambda} \gamma^{\sigma} \, ,
\end{align}
where the Levi-Civita symbol $\epsilon_{\mu\nu\lambda\sigma}$ with $\epsilon^{0123} = +1$ strictly remains in four space-time dimensions. The commutation relations read
\begin{align}
	\{ \gamma_\mu, \gamma_5 \} &= 0 \, , \quad \text{for } \mu = 0, 1, 2, 3 \, , \nn{}
	[ \gamma_\mu, \gamma_5 ] &= 0 \, , \quad \text{else} \, .
\end{align}
In general, this scheme leads to spurious anomalies that break chiral invariance and require the introduction of symmetry-restoring counterterms~\cite{Breitenlohner:1977hr,Ferrari:1994ct}. The spurious anomalies can be traced back to higher powers of the anticommutator $\{\gamma_\mu, \gamma_5\}$, which are matrices of rank $D-4$~\cite{Jegerlehner:2000dz}. In the present case, we do not encounter these problems because QCD is a vector theory and we only consider single-operator insertions. For this reason, we do not work with chiral fields and use the $D$-dimensional Dirac matrix $\gamma^\mu$ both in the QCD quark-gluon vertex and the quark propagator. External momenta and polarization vectors in $S$-matrix elements are treated in the HV scheme as having components only in $D=4$. The same applies to the projectors that we introduced in Sect.~\ref{sec:Projections} to define the renormalization conditions.

We define the renormalization constants for the fields, coupling, and the quark masses by
\begin{align}
	\label{eq:RenormalizationConstants}
	q^{(0)}   &= \sqrt{Z_q} \, q \, , \quad
	G_\mu ^{(0)}  = \sqrt{Z_G}  \, G_\mu \, , \quad
	g^{(0)} = Z_g \, g  \, \bar\mu^\varepsilon \, , \quad
	m^{(0)} = Z_m \, m \, , \nn
	\bar\mu &:= \mu \ \frac{e^{\gamma_E/2}}{(4 \pi)^{1/2}} \, ,
\end{align}
where $\gamma_E$ is the Euler--Mascheroni constant and $\mu$ denotes an arbitrary parameter with dimensions of mass,  introduced 
to keep the renormalized coupling $g$ dimensionless ($[g]=0$) in $D = 4 - 2 \varepsilon$ spacetime dimensions, while $[m]=1$, $[q] = 3/2 - \varepsilon$, and $[G_\mu] = 1- \varepsilon$. 
Note that $g$  and $\alpha_s := g^2/(4 \pi)$   depend on both $\mu$ and $\varepsilon$, so that 
$d \alpha_s/ d (\log \mu) = - 2 \varepsilon \alpha_s + \O(\alpha_s^2)$.  

In dimensional regularization with the \msbar{} scheme, the renormalization prescription 
is to subtract poles proportional to
\begin{align}
	\label{eq:MSbarPole}
	\Lambda_\varepsilon = - \frac{1}{2} \bar\mu^{D-4} \left( \frac{1}{\varepsilon} + \log(4\pi) - \gamma_E \right)  \, .
\end{align}
This is conveniently done by using the redefined scale $\mu$ and subsequently subtracting poles in $\varepsilon = (4-D)/2$.

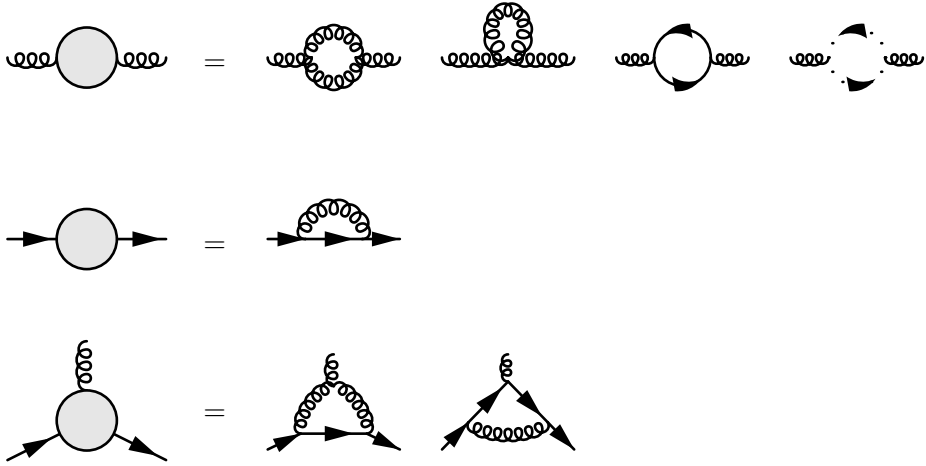
\begin{figure}[t]
	\begin{align*}
		\begin{gathered}
			\begin{fmfgraph}(60,60)
				\fmfset{curly_len}{1.75mm}
				\fmfleft{l1} \fmfright{r1}
				\fmf{gluon}{l1,v1}
				\fmf{gluon}{v1,r1}
				\fmfv{decor.shape=circle,decor.filled=10, decor.size=8mm}{v1}
			\end{fmfgraph}
		\end{gathered} \quad &= \quad
		\begin{gathered}
			\begin{fmfgraph}(50,50)
				\fmfset{curly_len}{1.5mm}
				\fmfleft{l1} \fmfright{r1}
				\fmf{gluon,tension=2}{l1,v1}
				\fmf{gluon,tension=2}{v2,r1}
				\fmf{gluon,right}{v1,v2,v1}
			\end{fmfgraph}
		\end{gathered} \quad\;
		\begin{gathered}
			\begin{fmfgraph}(50,50)
				\fmfset{curly_len}{1.5mm}
				\fmfleft{l1} \fmfright{r1}
				\fmf{gluon,tension=2}{l1,v1}
				\fmf{gluon,tension=2}{v1,r1}
				\fmf{gluon,right}{v1,v1}
			\end{fmfgraph}
		\end{gathered} \quad\;
		\begin{gathered}
			\begin{fmfgraph}(50,50)
				\fmfset{curly_len}{1.5mm}
				\fmfleft{l1} \fmfright{r1}
				\fmf{gluon,tension=3}{l1,v1}
				\fmf{gluon,tension=3}{v2,r1}
				\fmf{quark,right}{v1,v2,v1}
			\end{fmfgraph}
		\end{gathered} \quad\;
		\begin{gathered}
			\begin{fmfgraph}(50,50)
				\fmfset{curly_len}{1.5mm}
				\fmfleft{l1} \fmfright{r1}
				\fmf{gluon,tension=3}{l1,v1}
				\fmf{gluon,tension=3}{v2,r1}
				\fmf{ghost,right}{v1,v2,v1}
			\end{fmfgraph}
		\end{gathered} \\
		\begin{gathered}
			\begin{fmfgraph}(60,60)
				\fmfset{curly_len}{1.75mm}
				\fmfleft{l1} \fmfright{r1}
				\fmf{quark}{l1,v1}
				\fmf{quark}{v1,r1}
				\fmfv{decor.shape=circle,decor.filled=10, decor.size=8mm}{v1}
			\end{fmfgraph}
		\end{gathered} \quad &= \quad
		\begin{gathered}
			\begin{fmfgraph}(50,50)
				\fmfset{curly_len}{1.5mm}
				\fmfleft{l1} \fmfright{r1}
				\fmf{quark,tension=3}{l1,v1}
				\fmf{quark,tension=3}{v2,r1}
				\fmf{quark}{v1,v2}
				\fmf{gluon,right}{v2,v1}
			\end{fmfgraph}
		\end{gathered} \\
		\begin{gathered}
			\begin{fmfgraph}(60,50)
				\fmfset{curly_len}{2mm}
				\fmftop{t1} \fmfbottom{b1,b2}
				\fmf{gluon,tension=3}{t1,v1}
				\fmf{quark,tension=3}{b1,v1}
				\fmf{quark,tension=3}{v1,b2}
				\fmfv{decor.shape=circle,decor.filled=10, decor.size=8mm}{v1}
			\end{fmfgraph}
		\end{gathered} \quad &= \quad
		\begin{gathered}
			\begin{fmfgraph}(50,40)
				\fmfset{curly_len}{1.5mm}
				\fmftop{t1} \fmfbottom{b1,b2}
				\fmf{gluon,tension=3}{t1,v1}
				\fmf{gluon,right=0.25}{v3,v1,v2}
				\fmf{quark}{v2,v3}
				\fmf{quark,tension=3}{b1,v2}
				\fmf{quark,tension=3}{v3,b2}
			\end{fmfgraph}
		\end{gathered} \quad\;
		\begin{gathered}
			\begin{fmfgraph}(50,40)
				\fmfset{curly_len}{1.5mm}
				\fmftop{t1} \fmfbottom{b1,b2}
				\fmf{gluon,tension=3}{t1,v1}
				\fmf{quark,tension=1.5}{b1,v2}
				\fmf{quark,tension=1.5}{v3,b2}
				\fmf{quark}{v2,v1,v3}
				\fmffreeze
				\fmf{gluon,right=0.25}{v2,v3}
			\end{fmfgraph}
		\end{gathered} \\[-1.5cm]
	\end{align*}
	\caption{One-loop diagrams needed for the renormalization of the fields and the coupling.}
	\label{img:QCD}
\end{figure}

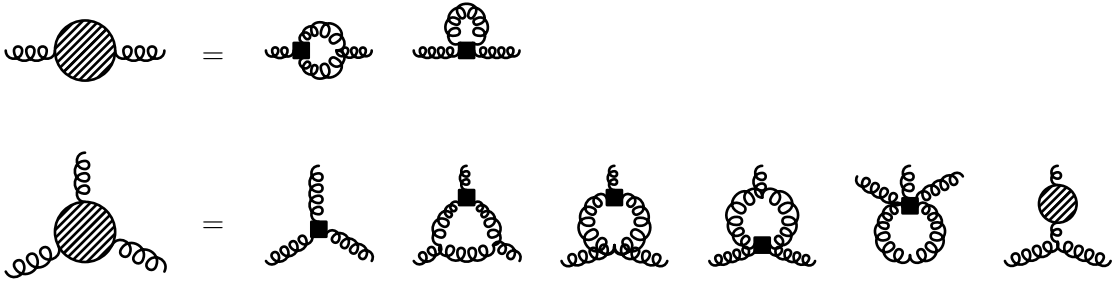
\begin{figure}[t]
	\begin{align*}
		\begin{gathered}
			\begin{fmfgraph}(60,60)
				\fmfset{curly_len}{1.75mm}
				\fmfleft{l1} \fmfright{r1}
				\fmf{gluon}{l1,v1}
				\fmf{gluon}{v1,r1}
				\fmfblob{8mm}{v1}
			\end{fmfgraph}
		\end{gathered} \quad &= \quad
		\begin{gathered}
			\begin{fmfgraph}(40,40)
				\fmfset{curly_len}{1.5mm}
				\fmfleft{l1} \fmfright{r1}
				\fmf{gluon,tension=2}{l1,v1}
				\fmf{gluon,tension=2}{v2,r1}
				\fmf{gluon,right}{v1,v2,v1}
				\fmfv{decoration.shape=square,decoration.size=2mm}{v1}
			\end{fmfgraph}
		\end{gathered} \quad\;
		\begin{gathered}
			\begin{fmfgraph}(40,40)
				\fmfset{curly_len}{1.5mm}
				\fmfleft{l1} \fmfright{r1}
				\fmf{gluon,tension=2}{l1,v1}
				\fmf{gluon,tension=2}{v1,r1}
				\fmf{gluon,right}{v1,v1}
				\fmfv{decoration.shape=square,decoration.size=2mm}{v1}
			\end{fmfgraph}
		\end{gathered} \\
		\begin{gathered}
			\begin{fmfgraph}(60,50)
				\fmfset{curly_len}{2mm}
				\fmftop{t1} \fmfbottom{b1,b2}
				\fmf{gluon,tension=3}{t1,v1}
				\fmf{gluon,tension=3}{b1,v1}
				\fmf{gluon,tension=3}{b2,v1}
				\fmfblob{8mm}{v1}
			\end{fmfgraph}
		\end{gathered} \quad &= \quad
		\begin{gathered}
			\begin{fmfgraph}(40,40)
				\fmfset{curly_len}{1.5mm}
				\fmftop{t1} \fmfbottom{b1,b2}
				\fmf{gluon,tension=3}{t1,v1}
				\fmf{gluon,tension=3}{b1,v1}
				\fmf{gluon,tension=3}{b2,v1}
				\fmfv{decoration.shape=square,decoration.size=2mm}{v1}
			\end{fmfgraph}
		\end{gathered} \quad\;
		\begin{gathered}
			\begin{fmfgraph}(40,40)
				\fmfset{curly_len}{1.5mm}
				\fmftop{t1} \fmfbottom{b1,b2}
				\fmf{gluon,tension=3}{t1,v1}
				\fmf{gluon,right=0.25}{v1,v2,v3,v1}
				\fmf{gluon,tension=3}{b1,v2}
				\fmf{gluon,tension=3}{b2,v3}
				\fmfv{decoration.shape=square,decoration.size=2mm}{v1}
			\end{fmfgraph}
		\end{gathered} \quad\;
		\begin{gathered}
			\begin{fmfgraph}(40,40)
				\fmfset{curly_len}{1.5mm}
				\fmftop{t1} \fmfbottom{b1,b2}
				\fmf{gluon,tension=3}{t1,v1}
				\fmf{gluon,right}{v1,v2,v1}
				\fmf{gluon,tension=3}{b1,v2,b2}
				\fmfv{decoration.shape=square,decoration.size=2mm}{v1}
			\end{fmfgraph}
		\end{gathered} \quad\;
		\begin{gathered}
			\begin{fmfgraph}(40,40)
				\fmfset{curly_len}{1.5mm}
				\fmftop{t1} \fmfbottom{b1,b2}
				\fmf{gluon,tension=3}{t1,v1}
				\fmf{gluon,right}{v1,v2,v1}
				\fmf{gluon,tension=3}{b1,v2,b2}
				\fmfv{decoration.shape=square,decoration.size=2mm}{v2}
			\end{fmfgraph}
		\end{gathered} \quad\;
		\begin{gathered}
			\begin{fmfgraph}(40,40)
				\fmfset{curly_len}{1.5mm}
				\fmftop{t1,t2,t3} \fmfbottom{b1}
				\fmf{gluon}{t1,v1}
				\fmf{gluon}{t2,v1}
				\fmf{gluon}{t3,v1}
				\fmf{gluon,right}{v1,v2,v1}
				\fmf{phantom,tension=6}{b1,v2}
				\fmfv{decoration.shape=square,decoration.size=2mm}{v1}
			\end{fmfgraph}
		\end{gathered} \quad\;
		\begin{gathered}
			\begin{fmfgraph}(40,40)
				\fmfset{curly_len}{1.5mm}
				\fmftop{t1} \fmfbottom{b1,b2}
				\fmf{gluon}{t1,v1}
				\fmf{gluon}{v1,v2}
				\fmf{gluon}{b1,v2,b2}
				\fmfblob{5mm}{v1}
			\end{fmfgraph}
		\end{gathered} \\[-1.5cm]
	\end{align*}
	\caption{Gluon two-point function $\Pi^{\alpha\beta}_{ab}$ and three-point function $\Gamma^{\alpha\beta\gamma}_{abc}$ at one-loop, with the insertion of the gCEDM operator $\O^{(1)}_6$ denoted by a square. Only one possible insertion of the gCEDM is shown.}
	\label{img:Glue}
\end{figure}

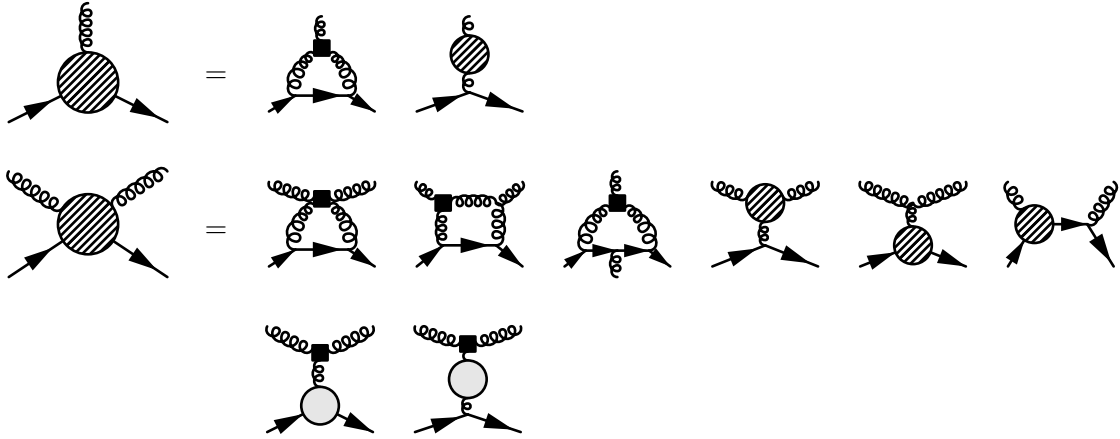
\begin{figure}[t]
	\begin{align*}
		\begin{gathered}
			\begin{fmfgraph}(60,50)
				\fmfset{curly_len}{1.5mm}
				\fmftop{t1} \fmfbottom{b1,b2}
				\fmf{gluon}{t1,v1}
				\fmf{quark}{b1,v1,b2}
				\fmfblob{8mm}{v1}
			\end{fmfgraph}
		\end{gathered} \quad &= \quad
		\begin{gathered}
			\begin{fmfgraph}(40,40)
				\fmfset{curly_len}{1.5mm}
				\fmftop{t1} \fmfbottom{b1,b2}
				\fmf{quark,tension=3}{b1,v1}
				\fmf{quark}{v1,v2}
				\fmf{quark,tension=3}{v2,b2}
				\fmf{gluon,tension=3}{t1,v3}
				\fmf{gluon,right=0.25}{v2,v3,v1}
				\fmfv{decoration.shape=square,decoration.size=2mm}{v3}
			\end{fmfgraph}
		\end{gathered} \quad\;
		\begin{gathered}
			\begin{fmfgraph}(40,40)
				\fmfset{curly_len}{1.5mm}
				\fmftop{t1} \fmfbottom{b1,b2}
				\fmf{gluon}{t1,v1}
				\fmf{gluon}{v1,v2}
				\fmf{quark}{b1,v2,b2}
				\fmfblob{5mm}{v1}
			\end{fmfgraph}
		\end{gathered} \\
		\begin{gathered}
			\begin{fmfgraph}(60,50)
				\fmfset{curly_len}{1.5mm}
				\fmftop{t1,t2} \fmfbottom{b1,b2}
				\fmf{gluon}{t1,v1}
				\fmf{gluon}{t2,v1}
				\fmf{quark}{b1,v1,b2}
				\fmfblob{8mm}{v1}
			\end{fmfgraph}
		\end{gathered} \quad &= \quad
		\begin{gathered}
			\begin{fmfgraph}(40,40)
				\fmfset{curly_len}{1.5mm}
				\fmftop{t1,t2} \fmfbottom{b1,b2}
				\fmf{quark,tension=3}{b1,v1}
				\fmf{quark}{v1,v2}
				\fmf{quark,tension=3}{v2,b2}
				\fmf{gluon,tension=3}{t1,v3,t2}
				\fmf{gluon,right=0.25}{v2,v3,v1}
				\fmfv{decoration.shape=square,decoration.size=2mm}{v3}
			\end{fmfgraph}
		\end{gathered} \quad\;
		\begin{gathered}
			\begin{fmfgraph}(40,40)
				\fmfset{curly_len}{1.5mm}
				\fmftop{t1,t2} \fmfbottom{b1,b2}
				\fmf{quark,tension=2}{b1,v1}
				\fmf{quark}{v1,v2}
				\fmf{quark,tension=2}{v2,b2}
				\fmf{gluon,tension=2}{t1,v3}
				\fmf{gluon,tension=2}{v4,t2}
				\fmf{gluon}{v2,v4,v3,v1}
				\fmfv{decoration.shape=square,decoration.size=2mm}{v3}
			\end{fmfgraph}
		\end{gathered} \quad\;
		\begin{gathered}
			\begin{fmfgraph}(40,40)
				\fmfset{curly_len}{1.5mm}
				\fmftop{t1} \fmfbottom{b1,b2,b3}
				\fmf{quark,tension=3}{b1,v1}
				\fmf{quark}{v1,v2,v3}
				\fmf{quark,tension=3}{v3,b3}
				\fmf{gluon,tension=3}{t1,v4}
				\fmf{gluon,right=0.25}{v3,v4,v1}
				\fmffreeze
				\fmf{gluon}{v2,b2}
				\fmfv{decoration.shape=square,decoration.size=2mm}{v4}
			\end{fmfgraph}
		\end{gathered} \quad\;
		\begin{gathered}
			\begin{fmfgraph}(40,40)
				\fmfset{curly_len}{1.5mm}
				\fmftop{t1,t2} \fmfbottom{b1,b2}
				\fmf{gluon}{t1,v1,t2}
				\fmf{gluon}{v1,v2}
				\fmf{quark}{b1,v2,b2}
				\fmfblob{5mm}{v1}
			\end{fmfgraph}
		\end{gathered} \quad\;
		\begin{gathered}
			\begin{fmfgraph}(40,40)
				\fmfset{curly_len}{1.5mm}
				\fmftop{t1,t2} \fmfbottom{b1,b2}
				\fmf{gluon}{t1,v1,t2}
				\fmf{gluon}{v1,v2}
				\fmf{quark}{b1,v2,b2}
				\fmfblob{5mm}{v2}
			\end{fmfgraph}
		\end{gathered} \quad\;
		\begin{gathered}
			\begin{fmfgraph}(40,40)
				\fmfset{curly_len}{1.5mm}
				\fmftop{t1,t2} \fmfbottom{b1,b2}
				\fmf{gluon}{t1,v1}
				\fmf{gluon}{v2,t2}
				\fmf{quark}{b1,v1,v2,b2}
				\fmfblob{5mm}{v1}
			\end{fmfgraph}
		\end{gathered} \\
		&\quad\quad\; \begin{gathered}
			\begin{fmfgraph}(40,50)
				\fmfset{curly_len}{1.5mm}
				\fmftop{t1,t2} \fmfbottom{b1,b2}
				\fmf{gluon}{t1,v1,t2}
				\fmf{gluon}{v1,v2}
				\fmf{quark}{b1,v2,b2}
				\fmfv{decor.shape=circle,decor.filled=10, decor.size=5mm}{v2}
				\fmfv{decoration.shape=square,decoration.size=2mm}{v1}
			\end{fmfgraph}
		\end{gathered} \quad\;
		\begin{gathered}
			\begin{fmfgraph}(40,50)
				\fmfset{curly_len}{1.5mm}
				\fmftop{t1,t2} \fmfbottom{b1,b2}
				\fmf{gluon}{t1,v1,t2}
				\fmf{gluon}{v1,v2,v3}
				\fmf{quark}{b1,v3,b2}
				\fmfv{decor.shape=circle,decor.filled=10, decor.size=5mm}{v2}
				\fmfv{decoration.shape=square,decoration.size=2mm}{v1}
			\end{fmfgraph}
		\end{gathered} \\[-1.75cm]
	\end{align*}
	\caption{Quark-gluon three-point and four-point functions at one-loop, with the insertion of the gCEDM operator $\O^{(1)}_6$ denoted by a square. Only one possible insertion of the gCEDM is shown. The hatched blobs denote one-loop subdiagrams with a gCEDM insertion, while the gray blobs denote QCD one-loop corrections.}
	\label{img:quark}
\end{figure}

The renormalization of the gCEDM operator in the \rismom{} scheme is accomplished by imposing the 34 renormalization conditions~\eqref{eq:ConditionsO1} on the 
gluon two- and three-point functions, on the quark two-point function, on the quark-gluon three and four-point functions, and on the quark-photon three-point function. The one-loop diagrams that need to be calculated are shown in Figs.~\ref{img:QCD}, \ref{img:Glue}, and~\ref{img:quark}. Note that the quark two-point function and the quark-photon three-point function with operator insertion only start at two-loop level.

\subsection{Results}
\label{sec:Results}

At one loop, the constants   $\Delta_{ij}$ introduced in~\eqref{eq:Zdef} can be defined as
\begin{align}
	\Delta_{ij}^{\overline{\rm{MS}}} &= \frac{\alpha_s}{4\pi} \frac{z_{ij}}{\varepsilon} \, , \nn*
	\Delta_{ij}^{\rm RI} &= \frac{\alpha_s}{4\pi} \left[  \left(\frac{1}{\varepsilon} + \log \frac{\mu^2}{\Lambda^2} \right)z_{ij} + c_{ij} \right] \, ,
\end{align}
which implies that the matching coefficients are given by
\begin{align}
	C_{ij} = \mathds{1}_{ij} +  \frac{\alpha_s}{4\pi} \left( z_{i j} \log \frac{\mu^2}{\Lambda^2} +  c_{ij} \right) \, .
\end{align}
As can be seen from~\eqref{eq:npoint}, the determination of $Z_{ij}$ requires the knowledge of the field renormalization. Furthermore, the renormalization conditions~\eqref{eq:ConditionsO1} involve the \msbar{} renormalized coupling $g^\text{\msbar}$ and quark masses $m^\text{\msbar}$.

\subsubsection{Covariant gauge}
\label{sec:CovariantGauge}

We start by calculating the renormalization of the gCEDM operator in a generic covariant gauge.
At one loop, it is sufficient to know the renormalization of the gluon field and of the strong coupling,
\begin{align}
	Z_G &= 1 + \frac{\alpha_s}{4\pi} \left(\frac{z_G}{\varepsilon}  +r_G \right) = 1 + \frac{\alpha_s}{4\pi} \left\{ \frac{1}{\varepsilon} \left[ C_A \left(  \frac{13}{6} -\frac{\xi}{2}  \right) - \frac{4}{3} N_f T_F \right] + r_G \right\} \, , \nn
	Z_g &= 1 + \frac{\alpha_s}{4\pi} \left( \frac{z_g}{\varepsilon}  + r_g\right) = 1 + \frac{\alpha_s}{4\pi} \left\{ -\frac{1}{\varepsilon} \frac{\beta_0}{2} + r_g \right\} \, ,
\end{align}
where $\beta_0$ is the lowest order $\beta$ function
\begin{align}
	\beta_0 = \frac{11}{3} C_A - \frac{4}{3} N_f T_F \, .
\end{align}
$C_A$ is the Casimir factor of the adjoint representation of $SU(3)$, $C_A = N_c$, while $T_F = 1/2$.
In the \msbar{} scheme, $r_G^\text{\msbar} = r_g^\text{\msbar} = 0$. In the \rismom{} scheme, we define the residue of the gluon propagator at $p^2 = -\Lambda^2$ to be equal to one, which determines the finite part of $Z_G$ in the chiral limit~\cite{Braaten:1981dv}:
\begin{align}
	r_G^{\rm RI} = C_A \left( \frac{97}{36} + \frac{\xi}{2} + \frac{\xi^2}{4} \right) - \frac{20}{9} N_f T_F + z_G \log \frac{\mu^2}{\Lambda^2} \, .
\end{align}
Since the renormalization conditions~\eqref{eq:ConditionsO1} are expressed in terms of the \msbar{} coupling $g^{\overline{\rm MS}}$, we do not need to define an \rismom{}
coupling constant, as $r_g$ drops out of the equations.

The gluon two- and three-point functions are shown in Fig.~\ref{img:Glue} and they determine the mixing of the gCEDM with the gluonic operators $\O^{(6)}_4$,
$\O^{(6)}_5$, $\O^{(6)}_9$, $\N^{(6)}_8$, $\N^{(6)}_{11}$, and $\N^{(6)}_{18}$ up to one linear combination of $\O^{(6)}_5$ and $\N^{(6)}_8$. In dimensional regularization, $\O^{(6)}_1$ does not induce divergences proportional to
the QCD $\bar\theta$ term $\O_1^{(4)}$. At the kinematic point $\tilde S_3$, one one-particle-reducible (1PR) diagram with the same topology as the last diagram in Fig.~\ref{img:Glue} contributes to the
three-point function. The 1PR contributions to the other two legs vanish due to the renormalization condition imposed on the two-point function at the kinematic point $S_2$.

The quark two-point function is zero at one loop. The quark-gluon three- and four-point functions are shown in Fig.~\ref{img:quark}.
The 1PR contributions to the gluon leg of the three-point function vanish at one loop due to the renormalization conditions $R_1$-$R_4$. The 1PR contributions to the incoming quark leg vanish due to the renormalization conditions imposed on the quark two-point function at the non-symmetric kinematic point $\tilde S_2$. The 1PR contributions to the other quark leg contain no loop corrections, which start at two-loop order, but a counterterm contribution has to be taken into account since the kinematic configuration does not correspond to $\tilde S_2$.

Imposing the conditions $R_{29}$-$R_{31}$ greatly simplifies the calculation of the four-point function at one loop. Since the three-gluon vertex from the operator
$\mathcal O^{(6)}_1$ vanishes when contracted with the gluon momentum, it is easy to see that the box diagrams in the second line of Fig.~\ref{img:quark}
do not contribute to $R_{29}$-$R_{31}$, leaving only the simpler triangle diagram. The same argument applies to the 1PR diagrams in the third line of Fig.~\ref{img:quark}.
Of the other 1PR loop contributions to the four-point function, the gluon two- and three-point function with insertion of the gCEDM do not
contribute to the projections $R_{29}$-$R_{31}$. The last topology shown in the second line of Fig.~\ref{img:quark}
does not receive a contribution from the loop, leaving only the contribution of the quark-gluon three-point function, contracted with the QCD three-gluon vertex. Several 1PR counterterm contributions need to be taken into account, as their kinematic configuration does not correspond to the renormalization point of the sub-amplitude.

We find the coefficients of the poles in $\varepsilon$ to be   
\begin{align}
	\label{eq:z1j}
	z_{1,1}^{(6)} &= - C_A \left(3 + \frac{3}{4} (1-\xi) \right) + z_g + \frac{3}{2} z_G =  - \frac{1}{2} \left( C_A + 2 N_f + \beta_0\right) \,, \nn
	z_{1,2}^{(6)} &= - \frac{3}{8} C_A \,, \nn
	z_{1,n_8}^{(6)} &= \frac{3}{16}C_A \,, \nn
	z_{1,n_{11}}^{(6)} &= \frac{3}{16} C_A \,,
\end{align}
while all other coefficients $z_{1j}$ vanish. Here, for clarity we introduced the notation $z_{i,j}^{(n)}$ for the mixing into the operator $\O_j^{(n)}$ and $z_{i,n_j}^{(n)}$ for the mixing into $\N_j^{(n)}$.

The finite pieces $c_{1j}$ are 
\begin{align}
	c_{1,1}^{(6)} &= - C_A \begin{aligned}[t] & \Bigg[ \left( \frac{17}{6} + \frac{25}{12} K + \frac{25}{18} \psi - \frac{1}{12} \log 2\right) -\frac{1}{12} (1-\xi) \left(21 + 10 K + 7 \psi + \log 2\right) \\
		& + \frac{3}{4} (1-\xi)^2 \Bigg] + \frac{3}{2} r_G - \left(  \frac{3}{2} z_G + z_g  \right)\log \frac{\mu^2}{\Lambda^2} \, , \end{aligned} \nn
	c_{1,2}^{(6)} &= C_A \left[ \frac{K}{2} + \frac{\log 2}{8} - \frac{811}{912} \right] \, , \nn
	c_{1,5}^{(6)} &= C_A \left[ \frac{13 K}{8} + \frac{11 \log 2}{8} - \frac{1285}{912} \right] \, , \nn
	c_{1,n_1}^{(6)} &= - \frac{583}{7296} C_A \, , \nn
	c_{1,n_8}^{(6)} &= C_A \left[ \left( \frac{7 K}{32} - \frac{\psi}{12}+\frac{41 \log 2}{32}-\frac{2819}{3648} \right) + (1-\xi ) \left(-\frac{K}{4}-\frac{\psi }{8}+\frac{11 \log 2}{16}+\frac{3}{16}\right) \right] \, , \nn
	c_{1,n_{11}}^{(6)} &=  C_A \left[ \left( -\frac{7 K}{4}+\frac{\psi }{12} - \frac{25 \log 2}{8}+\frac{55}{16} \right) + (1-\xi ) \left(\frac{K}{2}+\frac{5 \psi }{24}-\frac{3 \log 2}{2}-\frac{1}{3}\right) \right] \, , \nn
	c_{1,n_{12}}^{(6)} &= C_A \left[ \frac{K}{24}-\frac{\log 2}{24}-\frac{319}{24320}\right] \, , \nn
	c_{1,n_{13}}^{(6)} &= C_A \left[ \frac{K}{8}-\frac{\log 2}{8}-\frac{957}{24320}\right] \, , \nn
	c_{1,n_{14}}^{(6)} &= C_A \left[ \frac{K}{2}-\frac{\log 2}{2}-\frac{11}{95} \right] \, , \nn
	c_{1,n_{15}}^{(6)} &= \frac{21}{304} C_A \, , \nn
	c_{1,n_{16}}^{(6)} &= \frac{59}{1216} C_A \, , \nn
	c_{1,n_{17}}^{(6)} &= -\frac{21}{304} C_A \, , \nn
	c_{1,n_{18}}^{(6)} &= C_A \left[ \left( \frac{9 K}{4} + 3 \log 2 -\frac{29}{8} \right) + (1-\xi ) \left(-\frac{K}{2}-\frac{\psi}{6}+\frac{13 \log 2}{8}+\frac{7}{24}\right) \right] \, , \nn
	c_{1,n_{19}}^{(6)} &= C_A \left[ -K+\log 2 + \frac{197}{760}\right] \, , \nn
	c_{1,n_{20}}^{(6)} &= \frac{59}{608} C_A \, , \nn
	c_{1,1}^{(4)} &= \frac{3}{4} C_A \Lambda^2 \, ,
\end{align}
with analogous notation as for the $z_{1j}$. All other $c_{1j}$ vanish at one loop.
The additional logarithmic term in $c_{1,1}^{(6)}$ is an artifact of using $g^\text{\msbar}$ in the renormalization conditions and disappears if the matching is performed at $\mu = \Lambda$. Note also that a finite renormalization with the dimension-four operator $\O_1^{(4)}$ is present.

The triangle integrals in the three-point function depend on the two constants $\psi$ and $K$, which are defined as
\begin{equation}
\psi = \frac{2}{3} \left( \psi^{(1)} \left(\frac{1}{3}\right) - \frac{2}{3} \pi^2\right) = 2.34\ldots, \qquad K = \frac{1}{8}\left( \psi^{(1)} \left(\frac{1}{4}\right) -  \pi^2\right) = 0.916\ldots,
\end{equation}
where $\psi^{(1)}$ is the first derivative of the Digamma function.
To assess the numerical impact of the conversion between the \rismom{} and \msbar{}  scheme, we can evaluate the coefficient
$C_{11}$. At the scale $\mu = \Lambda = 3$ GeV, $C_{11} = 0.87$  in Landau gauge and $C_{11}= 0.76$ in Feynman gauge, indicating a
10\% - 25\% correction, as to be expected at one loop.

\subsubsection{Background-field method}
\label{sec:BackgroundMethod}

We can avoid mixing with gauge-variant operators by working with the background-field method~\cite{Abbott:1980hw},
with a gauge-fixing Lagrangian as specified in~\eqref{eq:GFback}. In this case, the class-IIb nuisance operators in~\eqref{eq:FinalOperatorsDim6IIb} can be disregarded and at dimension six, the operator basis reduces to the 10 operators in~\eqref{eq:FinalOperatorsDim6} and the 10 gauge-invariant EOM operators in~\eqref{eq:FinalOperatorsDim6IIa}.

We can define the  \rismom{} scheme by selecting a subset of 24 conditions $R_k[\O_1^{(6)}]$, with the understanding that these conditions
are imposed on Green's functions of the background field $\hat G^a$, not of the quantum field $G^a$. In the background-field method, we therefore replace the set of conditions~\eqref{eq:ConditionsO1} by
\begin{align}
	\label{eq:ConditionsO1BFM}
	R_k\big[\O_1^{(6)}\big] &= 0 \, , \quad k \in \{ 1, \ldots, 4, 8, \ldots, 21, 24, 28, 32, \ldots, 34 \} \, , \nn*
	R_{6}\big[\O_1^{(6)}\big] &= - 3 g N_c^2 C_F \Lambda^6 \, .
\end{align}
In particular, the quark-gluon four-point function is no longer required.

The field renormalization of the background field, which we still denote by $Z_G$, is given by
\begin{align}
	z_G &=   \beta_0 \, , \nn*
	r_G &=   C_A \left( \frac{67}{9} -2 (1-\xi) + \frac{(1-\xi)^2}{4}  \right) - \frac{20}{9} N_f T_F + z_G \log \frac{\mu^2}{\Lambda^2} \, .
\end{align}
Because explicit gauge invariance is preserved, the divergent part of  $Z_G$ and $Z_g$ satisfy $z_G = -2 z_g$. The same relation should be retained for the finite pieces, $r_G = -2 r_g$ in order to preserve the Ward identity. However, we again remark that $r_g$ does not enter our matching relations, as we use the \msbar{} coupling $g^\text{\msbar}(\mu=\Lambda,\epsilon=0)$ on the RHS of the conditions~\eqref{eq:ConditionsO1BFM}.

The result for the divergent pieces of the matching relations in the background-field gauge are given by
\begin{align}
	\label{eq:z1jBFM}
	z_{1,1}^{(6)} &= - 6 C_A + z_g + \frac{3}{2} z_G =  - \frac{1}{2} \left( C_A + 2 N_f + \beta_0\right) \,, \nn
	z_{1,2}^{(6)} &= - \frac{3}{8} C_A \,, \nn
	z_{1,n_8}^{(6)} &= \frac{3}{8}C_A \,,
\end{align}
while the finite pieces are
\begin{align}
	c_{1,1}^{(6)} &= - C_A \begin{aligned}[t] & \Bigg[ \left( \frac{47}{6} + \frac{8 \psi}{3} - 3 \log 2\right) -\frac{1}{12} (1-\xi) \left(6K + 3\psi + 6\log 2 + 44 \right) \\
		& + \frac{3}{4} (1-\xi)^2 \Bigg] + \frac{3}{2} r_G - \left(  \frac{3}{2} z_G + z_g  \right)\log \frac{\mu^2}{\Lambda^2} \, , \end{aligned} \nn
	c_{1,2}^{(6)} &= C_A \left[-\frac{89}{96}+\frac{5 \log 2}{8}\right] \, , \nn
	c_{1,5}^{(6)} &= C_A \left[ \frac{21 K}{8}+\frac{3 \log 2}{8}-\frac{61}{48}\right] \, , \nn
	c_{1,n_1}^{(6)} &= C_A \left[-\frac{K}{4}+\frac{\log 2}{4}-\frac{1}{24}\right] \, , \nn
	c_{1,n_8}^{(6)} &= C_A \left[-\frac{21 K}{32}-\frac{3 \log 2}{32}+\frac{157}{192}\right] \, , \nn
	c_{1,1}^{(4)} &= \frac{3}{4} C_A \Lambda^2 \, .
\end{align}
It can be checked that the background-gluon two- and three-point functions 
respect the Ward identities implied by gauge invariance at one loop for any value of the quantum gauge parameter $\xi$. However, even in the background-field method, off-shell Green's functions are unphysical quantities and they are gauge-parameter dependent. Therefore, our matching relations depend on the quantum gauge parameter $\xi$ (but of course not on the background gauge parameter $\hat\xi$), since the \rismom{} conditions themselves are gauge dependent.

Finally, we note that the results for $z_{1j}$ agree in the background-field gauge~\eqref{eq:z1jBFM} and the conventional gauge~\eqref{eq:z1j} for the case of mixing into physical (class-I) operators $\O_j$, as is required by gauge invariance~\cite{Deans:1978wn}. For the mixing into nuisance operators, gauge invariance does not provide a similar constraint: in the case of class-IIb operators, the mixing vanishes in the background-field method but not in conventional gauge, whereas for class-IIa nuisance operators the result in conventional gauge depends on the choice of basis for the class-IIb operators. We observe that with the basis change
\begin{align}
	\N_{11}^{(6)}{}' = \N_{11}^{(6)} - \N_8^{(6)}
\end{align}
the results for the mixing transform as
\begin{align}
	z_{1,n_8}^{(6)} \N_8^{(6)} + z_{1,n_{11}}^{(6)} \N_{11}^{(6)} = (z_{1,n_8}^{(6)} + z_{1,n_{11}}^{(6)} ) \N_8^{(6)} + z_{1,n_{11}}^{(6)} \N_{11}^{(6)}{}' \, .
\end{align}
The transformed operator $\N_{11}^{(6)}{}'$ still belongs to class-IIb. In this particular basis, the divergent pieces $z_{1,n_j}$ of the mixing into class-IIa operators $\N_j$ in conventional gauge agree with the results obtained in the background-field method.

\end{fmffile}


\section{Conclusions}
\label{sec:Conclusions}

The $CP$-odd three-gluon operator gives the main contribution to the nucleon EDM 
in several beyond the Standard Model scenarios, especially  when  $CP$ is violated in the interactions of heavy particles, such as the Higgs~\cite{Weinberg:1989dx,Cirigliano:2019vfc}
or the Higgs and the top quark~\cite{Cirigliano:2016nyn}. First-principle calculations, with controlled theoretical uncertainties, of the matrix elements of the gCEDM
on the nucleon are necessary  to derive the constraints of EDM experiments on this operator, and the implications for BSM physics. At the moment, the best estimates of the nucleon
EDM from the gCEDM have been obtained with QCD sum rule calculations~\cite{Demir:2002gg,Haisch:2019bml}, which are however affected by large theoretical uncertainties,
at the level of $50\%$-$100\%$. While for this operator lattice QCD calculations are still in their infancy~\cite{Dragos:2017wms,Rizik:2018lrz,Rizik:2020naq}, this method can in principle
provide fully nonperturbative results, in  which  all  sources  of  systematic  uncertainty can be quantified, controlled, and improved. 
LQCD and continuum calculations are interfaced via the definition of a renormalization scheme. In this paper, 
we have defined an \rismom{} scheme for the renormalization of the gCEDM, and we have provided the conversion matrix to the \msbar{} scheme at $\O(\alpha_s)$.
The derived operator basis will be of relevance also for matching calculations in other schemes, e.g., using the gradient flow~\cite{Rizik:2018lrz,Rizik:2020naq}.

As a dimension-six, flavor-singlet operator, the gCEDM has a complicated mixing pattern in an off-shell scheme. On the lattice, insertions of $\mathcal O_1^{(6)}$ induce power
divergences, which under the assumption of good chiral symmetry can be absorbed by three dimension-four and one dimension-five operator, defined in~\eqref{eq:FinalOperatorsDim4} and~\eqref{eq:FinalOperatorDim5}.
Both on the lattice and in the continuum, the gCEDM mixes into 10 dimension-six gauge-invariant operators that do not vanish by EOM, given in~\eqref{eq:FinalOperatorsDim6},
10 gauge-invariant nuisance operators~\eqref{eq:FinalOperatorsDim6IIa}, and 10 gauge-variant nuisance operators~\eqref{eq:FinalOperatorsDim6IIb}.
In this work, we have provided 34 renormalization conditions that define our \rismom{} scheme. In order to obtain enough independent 
conditions, it is necessary to compute the  gluon two- and three-point functions, the quark two-point function,
the  quark-gluon and quark-photon three-point functions, and some projections of the quark-gluon four-point function. 
We have imposed the renormalization conditions at one loop and computed the conversion matrix between the \rismom{} and \msbar{} schemes,
both in a conventional covariant gauge and in background-field gauge.

The number of operators and renormalization conditions make the lattice implementation of the \rismom{} renormalization scheme challenging, even though 
calculations of comparable complexity have been carried out for $\Delta S=1$ operators that contribute to $K \rightarrow \pi \pi$ decays~\cite{Lehner:2011fz,Bai:2015nea}.
For this reason, we explored the definition of the \rismom{} scheme in the background-field gauge~\cite{Abbott:1980hw}, which allows
to discard gauge-variant operators. Using the background-field method, the definition of the \rismom{} scheme involves only two- and three-point functions,
a very noticeable simplification. While background-field methods have not extensively been used to study higher-dimensional operators on the lattice, there are 
no particular technical problems for the implementation of the background-field condition~\cite{Cucchieri:2012ii}, and thus of the renormalization conditions
enumerated in Sect.~\ref{sec:BackgroundMethod}. It will be interesting to further explore the use of background-field methods in actual numerical simulations.

\section*{Acknowledgements}
\addcontentsline{toc}{section}{Acknowledgements}

We thank T.~Bhattacharya, W.~Dekens, J.~de Vries, A.~Kobach, A.~Manohar, S.~Pal, \linebreak A.~Shindler, and J.~Song for useful discussions.
Support by the DOE (Grant No.\ DE-SC0009919) and  
the Swiss National Science Foundation (Project No.\ P300P2\_167751) 
is gratefully acknowledged.
This work was supported by the US Department of Energy through the Los Alamos National Laboratory LDRD  program  under project number 20190041DR.
P.\,S.\ thanks the LANL for its hospitality at various stages of this work.

\appendix


\section{Construction of gauge-invariant operator basis}
\label{sec:BasisConstructionI}

In this appendix, we provide details on the construction of the basis of gauge-invariant operators. In App.~\ref{sec:SymmetriesBuildingBlocks}, we describe the symmetries of the building blocks. In App.~\ref{sec:PureGauge}, \ref{sec:TwoQuark}, and~\ref{sec:FourQuark}, we construct a complete list of pure gauge operators, two-quark, and four-quark operators, respectively, which we summarize in App.~\ref{sec:IntermediateSummaryOperators}. Here, we disregard evanescent operators away from $D=4$ dimensions, which will be discussed in App.~\ref{sec:Evanescent}.

\subsection{Symmetries and building blocks}
\label{sec:SymmetriesBuildingBlocks}

\begin{table}[t]
	\centering\small
	\setlength{\tabcolsep}{4.3pt}
	\begin{tabular}{c c c c c c c c c}
		\toprule
		field 	&				comm. &		mass dim. &	Lorentz & 			$SU(3)_c$ &		$\chi$ &				$\dagger$ &					$P$ &					$CP$ \\
		\midrule
		$q_L$ &				$-$ &		$\frac{3}{2}$ &	$(2,1)$ &			$3$ &			$(3,1)$ &				$\bar q_L \gamma^0$ &			$\gamma^0 q_R$ &			$\gamma^0 C \bar q_L^T$	 \\[0.2cm]
		$q_R$ &				$-$ &		$\frac{3}{2}$ &	$(1,2)$ &			$3$ &			$(1,3)$ &				$\bar q_R \gamma^0$ &			$\gamma^0 q_L$ &			$\gamma^0 C \bar q_R^T$	 \\[0.2cm]
		$\bar q_L$ &			$-$ &		$\frac{3}{2}$ &	$(1,2)$ &			$\bar 3$ &			$(\bar 3, 1)$ &			$\gamma^0 q_L$ &				$\bar q_R \gamma^0$ &		$q_L^T C \gamma^0$		 \\[0.2cm]
		$\bar q_R$ &			$-$ &		$\frac{3}{2}$ &	$(2,1)$ &			$\bar 3$ &			$(1, \bar 3)$ &			$\gamma^0 q_R$ &				$\bar q_L \gamma^0$ &		$q_R^T C \gamma^0$		 \\[0.2cm]
		$G_{\mu\nu}^a$ &		$+$ &		2 &			$(3,1)\oplus(1,3)$ &			$8$ &			$(1,1)$ &				$G_{\mu\nu}^a$ &				$G^{\mu\nu}_a$ &			$-\eta(a) G^{\mu\nu}_a$		 \\[0.1cm]
		\midrule
		$F_{\mu\nu}^L$ &		$+$ &		2 &			$(3,1)\oplus(1,3)$ &			$1$ &			$(8,1)$ &				$F_{\mu\nu}^L$ &					$F^{\mu\nu}_R$ &				$-{F^{\mu\nu}_L}^T$				 \\[0.2cm]
		$F_{\mu\nu}^R$ &		$+$ &		2 &			$(3,1)\oplus(1,3)$ &			$1$ &			$(1,8)$ &				$F_{\mu\nu}^R$ &					$F^{\mu\nu}_L$ &				$-{F^{\mu\nu}_R}^T$				 \\[0.2cm]
		$\M$ &				$+$ &		1 &			$(1,1)$ &			$1$ &			$(3,\bar 3)$ &			$\M^\dagger$ &				$\M^\dagger$ &			$\M^*$					 \\[0.2cm]
		$\M^\dagger$ &		$+$ &		1 &			$(1,1)$ &			$1$ &			$(\bar 3, 3)$ &			$\M$ &						$\M$ &					$\M^T$					 \\[0.1cm]
		\midrule
		$\p_\mu$ &			$+$ &		1 &			$(2,2)$ &			$1$ &			$(1,1)$ &				$\p_\mu$ &					$\p^\mu$ &				$\p^\mu$					 \\[0.2cm]
		$D_\mu(\cdot)$ &		$+$ &		1 &			$(2,2)$ &			$1\oplus8$ &			$\substack{(1,1)\oplus{} \\ (8,1)\oplus(1,8)}$ &				$(\cdot)\overleftarrow{D}_\mu$ &	$\gamma^0 D^\mu \gamma^0(\cdot)$ &			${D^{\mu}}^*(\cdot)$			 \\[0.1cm]
		\bottomrule
	\end{tabular}
	\caption{Properties of dynamical fields, spurion and external fields, and derivative operators. For simplicity, additional arbitrary phases in $P$- and $CP$-conjugation are neglected. $\eta(a)$ is defined in~\eqref{eq:etafactor}. The Lorentz group is locally isomorphic to $SU(2)_L\times SU(2)_R$.}
	\label{tab:FieldsAndSourcesGaugeInvariant}
\end{table}

The gauge-invariant class-I operators that are needed to renormalize the $CP$-odd three-gluon operator are constructed from the building blocks~\eqref{eq:BuildingBlocks}. The mass matrix has been promoted to a spurion field. The chiral transformations~\eqref{eq:SpurionTransformations} assigned to spurion and external fields allow us to take into account explicit chiral-symmetry breaking.

In order to renormalize the three-gluon operator, the operators have to be chirally invariant in the spurion sense, Lorentz scalars, $CP$-odd, and $P$-odd. Since we are working at leading order in the electromagnetic coupling and the external photon field always comes together with a charge matrix $Q$ and a coupling $e$, we only consider operators with at most one QED field-strength tensor.

The symmetry properties of the building blocks are listed in Table~\ref{tab:FieldsAndSourcesGaugeInvariant}. The charge-conjugation matrix fulfills
\begin{align}
	C \gamma_\mu^T C^{-1} = -\gamma_\mu
\end{align}
and can be written in the Dirac representation as $C = i \gamma^2 \gamma^0$, hence
\begin{align}
	C = C^* = - C^{-1} = - C^\dagger = - C^T \, .
\end{align}
The Dirac field transforms under charge conjugation as
\begin{align}
	\mathcal{C} \psi(x) \mathcal{C}^{-1} &= \eta_c C \bar\psi^T(x) \, , \nn
	\mathcal{C} \bar\psi(x) \mathcal{C}^{-1} &= \eta_c^* \psi^T(x) C \, ,
\end{align}
where $\eta_c$ is a phase factor, which we put equal to $1$ in the following. The electromagnetic gauge field transforms as
\begin{align}
	\mathcal{C} A_\mu(x) \mathcal{C}^{-1} = - A_\mu(x) \, ,
\end{align}
whereas the non-abelian gauge field transforms under charge conjugation as
\begin{align}
	\mathcal{C} G_\mu^a(x) \mathcal{C}^{-1} = - \eta(a) G_\mu^a(x) \, ,
\end{align}
with
\begin{align}
	\label{eq:etafactor}
	\eta(a) =  \left\{ \begin{matrix} 1 , & a=1, 3, 4, 6, 8,  \\
			-1, & a = 2, 5, 7 . \end{matrix} \right. 
\end{align}
We classify the operators according to the field content and mass dimension (up to $\O(e)$):
\begin{itemize}
	\item pure gauge operators:
		\begin{itemize}
			\item dimension 4: $G^2$,
			\item dimension 5: $G^2 D$, $G^2 \M$,
			\item dimension 6: $G^3$, $G^2 D^2$, $G^2 D \M$, $G^2 \M^2$,
		\end{itemize}
	\item two-quark operators:
		\begin{itemize}
			\item dimension 3: $\psi^2$,
			\item dimension 4: $\psi^2 \M$, $\psi^2 D$,
			\item dimension 5: $\psi^2 \M^2$, $\psi^2 D \M$, $\psi^2 D^2$, $\psi^2 G$, $\psi^2 F$,
			\item dimension 6: $\psi^2 \M^3$, $\psi^2 D \M^2$, $\psi^2 D^2 \M$, $\psi^2 D^3$, $\psi^2 G \M$, $\psi^2 G D$, $\psi^2 F \M$, $\psi^2 F D$,
		\end{itemize}
	\item four-quark operators:
		\begin{itemize}
			\item dimension 6: $\psi^4$,
		\end{itemize}
\end{itemize}
where $\psi^2$ denotes a quark bilinear. In this list, we have already excluded classes that obviously contain no gauge-invariant operators, e.g., $FG$, $FG^2$ classes. In the following, we construct the explicit operators by hand. We use the Hilbert series techniques~\cite{Lehman:2015via,Henning:2015daa,Lehman:2015coa,Henning:2015alf,Henning:2017fpj} as a cross-check to count the number of operators in each class, including total derivatives and EOM operators. As the known Hilbert series method does not include the discrete symmetries, even in this cross-check we select by hand the operators that are $P$-odd and $CP$-odd.

\subsection{Pure gauge operators}
\label{sec:PureGauge}

As a first class of operators, we consider the pure gauge operators. The building blocks are the quark-mass matrix, partial and covariant derivatives, and field-strength tensors. There are no operators at dimension two or three, hence we start at dimension four. Note that we need at least two field-strength tensors in order to have a non-vanishing trace.
As we are only interested in operators up to $\O(e)$, we disregard the electromagnetic field-strength tensor in this section.

Although in dimensional regularization mixing is only possible within operators of the same mass dimension (the mass matrix is treated as a spurion field), this is not necessarily true for other schemes. Therefore, we also look for $P$-odd, $CP$-odd operators of dimension smaller than six.

\paragraph{dim = 4}

At dimension four, we have two operators that consist only of gauge fields:
\begin{align}
	\tr[G_{\mu\nu} G^{\mu\nu}] \, , \quad \tr[G_{\mu\nu} \widetilde G^{\mu\nu}] \, .
\end{align}
The first term is the standard $CP$-even kinetic term for the gauge field, the second one is the $CP$-odd and $P$-odd QCD $\theta$-term. It belongs to our basis:
\begin{align}
	\tilde\O_1^{G,(4)} = \tr[G_{\mu\nu} \widetilde G^{\mu\nu}] \, .
\end{align}
Here, we use the tilde to distinguish a preliminary set of operators from the final ones after having removed redundancies.

\paragraph{dim = 5}

At dimension five, there are no chirally invariant pure gauge operators.

\paragraph{dim = 6}

We reach dimension six by adding either two mass matrices or two derivatives to a dimension-four operator (adding one mass matrix and one derivative does not give a Lorentz scalar). The only way to add two mass matrices is within a trace:
\begin{align}
	\tilde\O_1^{G,(6)} &= \tr[\M\M^\dagger] \tr[G_{\mu\nu} \widetilde G^{\mu\nu}] \, ,
\end{align}
where now the first trace is in flavor space, the second one in color space.

We consider the addition of two partial derivatives. The six Lorentz indices can be contracted either with $g^{\mu\nu}g^{\lambda\sigma}g^{\alpha\beta}$ or with $\epsilon^{\mu\nu\lambda\sigma}g^{\alpha\beta}$ to form a Lorentz scalar. In total, there are only four different contractions:
\begin{align}
	& \Box \, \tr[ G_{\mu\nu} G^{\mu\nu}] , \quad \Box \, \tr[ G_{\mu\nu} \widetilde G^{\mu\nu}] , \quad \p_\mu \p^\nu \tr[ G^{\mu\lambda} G_{\nu\lambda} ] , \quad \p_\mu \p^\nu \tr[ G^{\mu\lambda} \widetilde G_{\nu\lambda} ] \, .
\end{align}
Furthermore, the Schouten identity
\begin{align}
	\label{eq:SchoutenIdentity}
	g_{\alpha\rho}\epsilon_{\mu\nu\lambda\sigma} + g_{\alpha\mu}\epsilon_{\nu\lambda\sigma\rho} + g_{\alpha\nu}\epsilon_{\lambda\sigma\rho\mu} + g_{\alpha\lambda}\epsilon_{\sigma\rho\mu\nu} + g_{\alpha\sigma}\epsilon_{\rho\mu\nu\lambda} = 0
\end{align}
implies the relation
\begin{align}
	\Box \tr[ G_{\mu\nu} \widetilde G^{\mu\nu} ] = 4 \p_\mu \p^\nu \tr[ G^{\mu\lambda} \widetilde G_{\nu\lambda} ] \, .
\end{align}
This only leaves the following $P$-odd and $CP$-odd operator:
\begin{align}
	\tilde\O_2^{G,(6)} &= \Box \, \tr[G_{\mu\nu} \widetilde G^{\mu\nu}] \, .
\end{align}

Next, we consider the case where we add one partial and one covariant derivative. The possible contractions of the Lorentz indices are:
\begin{align}
	& \p^\mu \tr[ (D_\mu G_{\nu\lambda}) G^{\nu\lambda} ] \, , \quad \p^\mu \tr[ (D_\nu G_{\mu\lambda}) G^{\nu\lambda} ] \, , \quad \p^\mu \tr[ (D_\nu G^{\nu\lambda}) G_{\mu\lambda} ] \, , \nn
	& \p^\mu \tr[ (D_\mu G_{\nu\lambda}) \widetilde G^{\nu\lambda} ] \, , \quad \p^\mu \tr[ (D_\nu G_{\mu\lambda}) \widetilde G^{\nu\lambda} ] \, , \quad \p^\mu \tr[ (D_\nu \widetilde G^{\nu\lambda}) G_{\mu\lambda} ] \, , \nn
	& \p_\nu \tr[ (D^\mu G_{\mu\lambda}) \widetilde G^{\nu\lambda} ] \, , \quad \p_\nu \tr[ (D^\mu \widetilde G^{\nu\lambda}) G_{\mu\lambda} ] \, , \quad \p_\nu \tr[ (D_\lambda {G^\mu}_\alpha) G_{\mu\beta} ] \epsilon^{\nu\lambda\alpha\beta} \, .
\end{align}
However, we only have to consider contractions with the Levi-Civita tensor: they are $P$-odd and $CP$-odd, while the other contractions are even.

By applying the Bianchi identity~\eqref{eq:BianchiIdentity}, we remove redundancies. Furthermore, we note the Leibniz rule~\eqref{eq:Leibniz} for the covariant derivative in adjoint representation:
\begin{align}
	\label{eq:AdjointCovariantToPartialDerivative}
		\tr[ (D_\mu A) B ] + \tr[ A (D_\mu B) ] &= \tr[ (\p_\mu A) B ] + \tr[A (\p_\mu B)] - i g \tr[ [ G_\mu, A] B + A [G_\mu, B] ] \nn
			&= \p_\mu \tr[A B] \, .
\end{align}
We find the relations
\begin{align}
	\p^\mu \tr[ (D_\nu \widetilde G^{\nu\lambda}) G_{\mu\lambda} ] &= 0 \, , \nn
	\p^\mu \tr[ (D_\mu G_{\nu\lambda}) \widetilde G^{\nu\lambda} ] &= 2 \p^\mu \tr[ (D_\nu G_{\mu\lambda}) \widetilde G^{\nu\lambda} ]  \, , \nn
	\p_\nu \tr[ (D_\lambda {G^\mu}_\alpha ) G_{\mu\beta} ] \epsilon^{\nu\lambda\alpha\beta} &= \p_\nu \tr[ (D^\mu \widetilde G^{\nu\lambda}) G_{\mu\lambda} ] \, , \nn
	\p^\mu \tr[ (D_\nu G_{\mu\lambda}) \widetilde G^{\nu\lambda} ] + \p^\mu \tr[ (D_\nu \widetilde G^{\nu\lambda}) G_{\mu\lambda} ] &= \p^\mu \p_\nu \tr[ G_{\mu\lambda}  \widetilde G^{\nu\lambda} ] \, , \nn
	\p_\nu \tr[ (D^\mu G_{\mu\lambda}) \widetilde G^{\nu\lambda} ] + \p_\nu \tr[ (D^\mu \widetilde G^{\nu\lambda}) G_{\mu\lambda} ] &= \p^\mu \p_\nu \tr[ G_{\mu\lambda}  \widetilde G^{\nu\lambda} ] \, ,
\end{align}
hence, there is only one additional independent operator:
\begin{align}
	\tilde\O_3^{G,(6)} &= \p_\nu \tr[ (D^\mu G_{\mu\lambda}) \widetilde G^{\nu\lambda} ] \, .
\end{align}

Finally, we can build operators with two covariant derivatives and two field-strength tensors. The requirement that the operator be $P$- and $CP$-odd allows again only the contraction with $\epsilon^{\mu\nu\lambda\sigma}g^{\alpha\beta}$. If the two derivatives do not act on the same field-strength tensor, we can use~\eqref{eq:AdjointCovariantToPartialDerivative} and obtain a linear relation to an operator where both derivatives act on the same tensor and an operator involving a partial derivative.

The possible contractions are
\begin{align}
	& \tr[ (D_\mu D_\nu G_{\lambda\sigma}) {G^\lambda}_\beta ] \epsilon^{\mu\nu\sigma\beta} \, , \quad \tr[ (D_\mu D_\nu G_{\lambda\sigma}) {G^\nu}_\beta ] \epsilon^{\mu\lambda\sigma\beta} \, , \quad \tr[ (D_\nu D_\mu G_{\lambda\sigma}) {G^\nu}_\beta ] \epsilon^{\mu\lambda\sigma\beta} \, , \nn
	& \tr[ (D^\nu D_\mu G_{\nu\lambda}) G_{\alpha\beta} ] \epsilon^{\alpha\beta\mu\lambda} \, , \quad \tr[ (D_\mu D^\nu G_{\nu\lambda}) G_{\alpha\beta} ] \epsilon^{\alpha\beta\mu\lambda} \, , \quad \tr[ (D_\mu D^\mu G_{\lambda\sigma}) G_{\alpha\beta} ] \epsilon^{\alpha\beta\lambda\sigma} \, .
\end{align}
We take some linear combinations to replace this set by
\begin{align}
	& \tr[ ([D_\mu, D_\nu] G_{\lambda\sigma}) {G^\lambda}_\beta ] \epsilon^{\mu\nu\sigma\beta} \, , \;\; \tr[ ([ D_\mu ,  D_\nu] G_{\lambda\sigma}) {G^\nu}_\beta ] \epsilon^{\mu\lambda\sigma\beta} \, , \;\; \tr[ (D_\nu D_\mu G_{\lambda\sigma}) {G^\nu}_\beta ] \epsilon^{\mu\lambda\sigma\beta} \, , \nn
	& \tr[ ([ D^\nu,  D_\mu] G_{\nu\lambda}) G_{\alpha\beta} ] \epsilon^{\alpha\beta\mu\lambda} \, , \;\; \tr[ (D_\mu D^\nu G_{\nu\lambda}) G_{\alpha\beta} ] \epsilon^{\alpha\beta\mu\lambda} \, , \;\; \tr[ (D_\mu D^\mu G_{\lambda\sigma}) G_{\alpha\beta} ] \epsilon^{\alpha\beta\lambda\sigma} \, .
\end{align}
Using the Jacobi and Bianchi identities, we can eliminate three elements of the set:
\begin{align}
	\tr[ (D_\nu D_\mu G_{\lambda\sigma}) {G^\nu}_\beta ] \epsilon^{\mu\lambda\sigma\beta} &= 0 \, , \nn
	\tr[ (D^\nu D_\mu G_{\nu\lambda}) G_{\alpha\beta} ] \epsilon^{\alpha\beta\mu\lambda} &= \frac{1}{2} \tr[ (D_\mu D^\mu G_{\lambda\sigma}) G_{\alpha\beta} ] \epsilon^{\alpha\beta\lambda\sigma} \, , \nn
	\tr[ (D_\mu D^\nu G_{\nu\lambda}) G_{\alpha\beta} ] \epsilon^{\alpha\beta\mu\lambda} &= \p_\mu \tr[ (D^\nu G_{\nu\lambda}) G_{\alpha\beta} ] \epsilon^{\alpha\beta\mu\lambda} - \tr[ (D^\nu G_{\nu\lambda}) (D_\mu G_{\alpha\beta}) ] \epsilon^{\alpha\beta\mu\lambda} \nn
			&= \p_\mu \tr[ (D^\nu G_{\nu\lambda}) G_{\alpha\beta} ] \epsilon^{\alpha\beta\mu\lambda} \, .
\end{align}
The commutators of covariant derivatives in the adjoint representation can be expressed in terms of the field-strength according to
\begin{align}
	[D_\rho, D_\sigma] (\cdot) = -i g [ G_{\rho\sigma}, \;\cdot{}\; ] \, .
\end{align}
Therefore, the only additional operator is the $CP$-odd three-gluon operator itself,
\begin{align}
	\tilde\O_4^{G,(6)} &= i \tr[G_{\mu\nu} {G^\mu}_\lambda \widetilde G^{\nu\lambda}] \, ,
\end{align}
which is the only $P$-odd and $CP$-odd operator that can be constructed with three field-strength tensors. This completes the construction of the set of pure gauge operators.

\subsection{Two-quark operators}
\label{sec:TwoQuark}

We continue with two-quark operators, which at least have mass dimension three.

\paragraph{dim = 3}
There is no quark bilinear that is a Lorentz scalar and chirally invariant.

\paragraph{dim = 4}

In order to reach mass dimension four, we can add either one mass matrix or one derivative to a quark bilinear.

The chirally invariant operators obtained by adding a mass matrix to a quark bilinear are
\begin{align}
	\bar q_L \M q_R \, , \quad \bar q_R \M^\dagger q_L \, .
\end{align}
There is one Hermitian linear combination that is $P$- and $CP$-odd:
\begin{align}
	\tilde\O_1^{2q,(4)} &= \bar q_L i \M q_R - \bar q_R i \M^\dagger q_L \stackrel{\text{fixed spurion}}{\mapsto} \bar q i \gamma_5 \M q \, .
\end{align}

Next, we consider the insertion of a derivative, which has to be contracted with a Lorentz-vector, hence we need a vector quark bilinear. The possible gauge-invariant and chirally invariant contractions with a partial or covariant derivative are the following:
\begin{align}
	\p_\mu ( \bar q_L \gamma^\mu q_L ) \, , \quad \p_\mu (\bar q_R \gamma^\mu q_R ) \, , \quad \bar q_L \gamma^\mu i \overleftrightarrow D_\mu q_L \, , \quad \bar q_R \gamma^\mu i \overleftrightarrow D_\mu q_R \, .
\end{align}
The third and fourth operators are $CP$-even (they are the standard kinetic terms), while the first two are $CP$-odd.
There is only one linear combination that is also $P$-odd, the divergence of the axial current:
\begin{align}
	\tilde\O_2^{2q,(4)} &= \p_\mu (\bar q_R \gamma^\mu q_R - \bar q_L \gamma^\mu q_L) = \p_\mu (\bar q \tilde\gamma^\mu q)  \, ,
\end{align}
where $\tilde\gamma^\mu$ is defined in~\eqref{eq:GammaSigmaTilde}.

\paragraph{dim = 5}

We reach mass dimension five by inserting two masses, one mass and one derivative, two derivatives, or a field strength tensor into a quark bilinear.

There are two chirally invariant operators obtained from the insertion of two mass matrices into a quark bilinear:
\begin{align}
	\label{eq:Dim5ChirallyInvariant}
	\epsilon_{ijk} \epsilon_{lmn} \M^\dagger_{mj} \M^\dagger_{nk} \bar q_L^i q_R^l \, , \quad \epsilon_{ijk} \epsilon_{lmn} \M_{mj} \M_{nk} \bar q_R^i q_L^l \, ,
\end{align}
which are possible due to the fact that for $SU(3)$, the following tensor decompositions hold:
\begin{align}
	3 \otimes 3 \otimes 3 &= 3 \otimes ( \bar 3 \oplus 6 )  = 1 \oplus 8 \oplus 8 \oplus 10 \, , \nn
	\bar 3 \otimes \bar 3 \otimes \bar 3 &= \bar 3 \otimes ( 3 \oplus \bar 6 )  = 1 \oplus 8 \oplus 8 \oplus \overline{10} \, ,
\end{align}
hence a product of three (anti-)fundamental representations contains a singlet. If the spurions $\M$ and $\M^\dagger$ are fixed to a diagonal mass matrix, the above operators are flavor conserving as well and represent a correction to the mass terms themselves. We can again form one $P$-odd and $CP$-odd combination:
\begin{align}
	\tilde\O_1^{2q,(5)} &= i \epsilon_{ijk} \epsilon_{lmn} ( \M^\dagger_{mj} \M^\dagger_{nk} \bar q_L^i q_R^l - \M_{mj} \M_{nk} \bar q_R^i q_L^l ) \stackrel{\text{fixed spurion}}{\mapsto} \epsilon_{ijk} \epsilon_{lmn} \M_{mj} \M_{nk} \bar q^i i \gamma_5 q^l \, .
\end{align}
For a diagonal mass matrix, the following relation holds~\cite{Kaplan:1986ru,Leutwyler:1989pn}:
\begin{align}
	\tilde\O_1^{2q,(5)} = 2 \det(\M) \bar q i \gamma_5 \M^{-1} q \, .
\end{align}

Consider the insertion of a single mass matrix and a derivative into a quark bilinear. In order to contract the Lorentz index of the derivative, we need a vector bilinear. With an additional mass matrix, it is impossible to construct a chirally invariant operator.
Finally, we consider the case of two derivatives. We have to start either with a (pseudo-)scalar or with a tensor quark bilinear, and add two derivatives. Also here, we cannot construct a chirally invariant operator. The same is obviously true for the insertion of a field-strength tensor in a two-quark operator.

\paragraph{dim = 6}

We obtain operators of dimension six by inserting either three mass matrices, two mass matrices and one derivative, one mass matrix and two derivatives, or three derivatives into a quark bilinear. Furthermore, a field-strength tensor can take the role of two derivatives.

We start with the insertion of three mass matrices. A basis for the chirally invariant operators is given by:
\begin{align}
	\bar q_L \M \M^\dagger \M q_R \, , \quad \bar q_R \M^\dagger \M \M^\dagger q_L \, , \quad \tr[\M \M^\dagger] \bar q_L \M q_R \, , \quad \tr[\M \M^\dagger] \bar q_R \M^\dagger q_L \, .
\end{align}
The $SU(N_f)$ Fierz identity
\begin{align}
	t^A_{kl} t^A_{ij} &= \frac{1}{2} \delta_{kj} \delta_{il} - \frac{1}{2N_f} \delta_{kl} \delta_{ij}
\end{align}
implies
\begin{align}
	\tr[A t^A] \bar\psi t^A \chi = \frac{1}{2} \bar\psi A \chi - \frac{1}{2N_f} \tr[A] \bar\psi \chi \, ,
\end{align}
hence $t^A \otimes t^A$ operators are linearly dependent of $\mathds{1}\otimes\mathds{1}$ operators.

We can form the following Hermitian $P$-odd and $CP$-odd linear combinations:
\begin{align}
	\tilde\O_1^{2q,(6)} &= i \left( \bar q_L \M \M^\dagger \M q_R - \bar q_R \M^\dagger \M \M^\dagger q_L \right) \stackrel{\text{fixed spurion}}{\mapsto} \bar q i\gamma_5 \M^3 q \, , \nn
	\tilde\O_2^{2q,(6)} &= i \; \tr[\M \M^\dagger] \left( \bar q_L \M q_R - \bar q_R \M^\dagger q_L \right) \stackrel{\text{fixed spurion}}{\mapsto} \tr[\M^2] \bar q i \gamma_5 \M q \, .
\end{align}

Next, we insert two mass matrices and one derivative into a quark bilinear. We find the following chirally invariant Hermitian operators:
\begin{align}
	& \p_\mu \left( \bar q_L \gamma^\mu \M \M^\dagger q_L \right) \, , \quad \p_\mu \left( \bar q_R \gamma^\mu \M^\dagger \M q_R \right) \, , \quad \tr[ \M \M^\dagger] \p_\mu \left( \bar q_L \gamma^\mu q_L \right) \, , \nn
	& \tr[ \M \M^\dagger] \p_\mu \left( \bar q_R \gamma^\mu q_R \right) \, , \quad  \bar q_L \gamma^\mu i \overleftrightarrow D_\mu \M \M^\dagger q_L \, , \quad \bar q_R \gamma^\mu i \overleftrightarrow D_\mu \M^\dagger \M q_R \, , \nn
	& \tr[ \M \M^\dagger] \left( \bar q_L \gamma^\mu i \overleftrightarrow D_\mu q_L \right) \, , \quad \tr[ \M \M^\dagger] \left( \bar q_R \gamma^\mu i \overleftrightarrow D_\mu q_R \right) \, .
\end{align}
The four operators with covariant derivatives $i \overleftrightarrow D_\mu$ are $CP$-even. The following linear combinations are $P$- and $CP$-odd:
\begin{align}
	\tilde\O_3^{2q,(6)} &= \p_\mu \left( \bar q_R \gamma^\mu \M^\dagger \M q_R - \bar q_L \gamma^\mu \M \M^\dagger q_L \right) \stackrel{\text{fixed spurion}}{\mapsto} \p_\mu \left( \bar q \tilde\gamma^\mu \M^2 q \right) \, , \nn
	\tilde\O_4^{2q,(6)} &= \tr[ \M \M^\dagger] \p_\mu \left( \bar q_R \gamma^\mu q_R - \bar q_L \gamma^\mu q_L \right) \stackrel{\text{fixed spurion}}{\mapsto}  \tr[\M^2] \p_\mu \left( \bar q \tilde\gamma^\mu q \right) \, .
\end{align}

The next operator class consist of insertions of one mass matrix and two derivatives in a quark bilinear. We start with the following set of chirally invariant operators:
\begin{align}
	& \Box ( \bar q_L \M q_R ) \, , \quad \Box ( \bar q_R \M^\dagger q_L ) \, , \quad
		\p_\mu ( \bar q_L \overleftrightarrow D^\mu \M q_R ) \, , \quad \p_\mu ( \bar q_R \overleftrightarrow D^\mu \M^\dagger q_L ) \, , \nn
	& \p_\mu ( \bar q_L \sigma^{\mu\nu} \overleftrightarrow D_\nu \M q_R ) \, , \quad \p_\mu ( \bar q_R \sigma^{\mu\nu} \overleftrightarrow D_\nu \M^\dagger q_L ) \, , \quad
		\bar q_L ( \overleftarrow { D}^2 +  D^2 ) \M q_R \, , \nn
	& \bar q_R ( \overleftarrow { D}^2 +  D^2 ) \M^\dagger q_L \, , \quad \bar q_L \sigma^{\mu\nu} [D_\mu, D_\nu ] \M q_R \, , \quad \bar q_R \sigma^{\mu\nu} [D_\mu, D_\nu ] \M^\dagger q_L \, .
\end{align}
The following Hermitian linear combinations are $P$- and $CP$-odd:
\begin{align}
	\tilde\O_5^{2q,(6)} &= i \; \Box( \bar q_L \M q_R - \bar q_R \M^\dagger q_L ) \stackrel{\text{fixed spurion}}{\mapsto} \Box( \bar q i \gamma_5 \M q ) \, , \nn
	\tilde\O_6^{2q,(6)} &= \p_\mu ( \bar q_L \sigma^{\mu\nu} \overleftrightarrow{D}_\nu \M q_R - \bar q_R \sigma^{\mu\nu} \overleftrightarrow{D}_\nu \M^\dagger q_L ) \stackrel{\text{fixed spurion}}{\mapsto} \p_\mu ( \bar q \tilde\sigma^{\mu\nu} \overleftrightarrow{D}_\nu \M q ) \, , \nn
	\tilde\O_7^{2q,(6)} &= i \big( \bar q_L ( \overleftarrow { D}^2 +  D^2 ) \M q_R - \bar q_R ( \overleftarrow { D}^2 +  D^2 ) \M^\dagger q_L \big) \stackrel{\text{fixed spurion}}{\mapsto} \bar q i \gamma_5 ( \overleftarrow { D}^2 +  D^2 ) \M q \, , \nn
	\tilde\O_{8}^{2q,(6)} &= \bar q_L \sigma^{\mu\nu} [D_\mu, D_\nu] \M q_R - \bar q_R \sigma^{\mu\nu} [D_\mu,D_\nu] \M^\dagger q_L \stackrel{\text{fixed spurion}}{\mapsto} \bar q \tilde\sigma^{\mu\nu} [D_\mu,D_\nu] \M q \, ,
\end{align}
where $\tilde\sigma^{\mu\nu}$ is defined in~\eqref{eq:GammaSigmaTilde}.

The next class of two-quark operators contains insertions of three derivatives. The three Lorentz indices of the derivatives $\{\cdot\}_{\mu\nu\lambda}$ can either be contracted with $g^{\mu\nu} \gamma^\lambda$ or with $\epsilon^{\mu\nu\lambda\sigma}\gamma_\sigma$. For the moment, we disregard evanescent operators (see App.~\ref{sec:Evanescent}) and use the four-dimensional relation
\begin{align}
	\gamma^\mu \gamma^\nu \gamma^\lambda = g^{\mu\nu} \gamma^\lambda + g^{\nu\lambda} \gamma^\mu - g^{\mu\lambda} \gamma^\nu + i \epsilon^{\mu\nu\lambda\sigma} \gamma_\sigma \gamma_5 \, .
\end{align}
Due to the odd number of gamma matrices, all operators will be chirally invariant, hence we work directly in the parity basis. We start with the contractions with $g^{\mu\nu} \gamma^\lambda$. A $\gamma_5$ matrix is required for $P$-odd operators.
The derivatives can be either covariant derivatives or partial derivatives of a gauge singlet. Note that due to
\begin{align}
	[ D_\mu, D_\nu ] = - [\overleftarrow{D}_\mu, D_\nu ] = - [D_\mu, \overleftarrow{D}_\nu] = [\overleftarrow{D}_\mu, \overleftarrow{D}_\nu ] \, ,
\end{align}
the covariant derivatives acting on the left can always be put on the left-hand side of derivatives acting on the right. Furthermore, by using the relation
\begin{align}
	\bar A (\overleftarrow D_\mu + D_\mu) B = \p_\mu( \bar A B ) \, ,
\end{align}
left-acting covariant derivatives can be traded for partial derivatives of the gauge singlet. Hence, we find the following list of nine operators with three derivatives:
\begin{align}
	\begin{alignedat}{3}
		& \bar q \tilde\gamma^\nu D_\mu D^\mu D_\nu q \, , \quad
		&& \bar q \tilde\gamma^\nu D_\mu D_\nu D^\mu q \, , \quad
		&& \bar q \tilde\gamma^\nu D_\nu D^\mu D_\mu q \, , \\
		& \p_\mu( \bar q \tilde\gamma^\nu D^\mu D_\nu q ) \, , \quad
		&& \p_\mu( \bar q \tilde\gamma^\nu D_\nu D^\mu q ) \, , \quad
		&& \p_\nu( \bar q \tilde\gamma^\nu D^\mu D_\mu q ) \, , \\
		& \Box ( \bar q \tilde\gamma^\nu D_\nu q ) \, , \quad
		&& \p_\mu \p_\nu ( \bar q \tilde\gamma^\nu D^\mu q ) \, , \quad
		&& \Box \p_\nu( \bar q \tilde\gamma^\nu q ) \, .
	\end{alignedat}
\end{align}
By taking linear combinations, we make them manifestly Hermitian:
\begin{align}
	\begin{alignedat}{3}
		& \bar q \tilde\gamma^\nu (  \overleftarrow D^2 D_\nu +  \overleftarrow D_\nu D^2 ) q \, , \quad
		&& i\, \bar q \tilde\gamma^\nu \overleftarrow D_\mu \overleftrightarrow D_\nu D^\mu q \, , \quad
		&& i\, \bar q \tilde\gamma^\nu ( \overleftarrow D^2 D_\nu -  \overleftarrow D_\nu D^2 ) q \, , \\
		& \p_\mu \Big( \bar q \tilde\gamma^\nu (\overleftarrow D_\nu \overleftarrow D^\mu + D^\mu D_\nu) q \Big) \, , \quad
		&& i\, \p_\mu ( \bar q \tilde\gamma^\nu [ D^\mu , D_\nu ] q ) \, , \quad
		&& \p_\nu \Big( \bar q \tilde\gamma^\nu (\overleftarrow D^2 + D^2) q \Big) \, , \\
		& i\, \Box ( \bar q \tilde\gamma^\nu \overleftrightarrow D_\nu q ) \, , \quad
		&& i\, \p_\mu \p_\nu ( \bar q \tilde\gamma^\nu \overleftrightarrow D^\mu q ) \, , \quad
		&& \Box \p_\nu( \bar q \tilde\gamma^\nu q ) \, .
	\end{alignedat}
\end{align}
Four operators are $CP$-odd:
\begin{align}
	\tilde\O_{9}^{2q,(6)} &= \bar q \tilde\gamma^\nu (  \overleftarrow D^2 D_\nu +  \overleftarrow D_\nu D^2 ) q \, , \nn
	\tilde\O_{10}^{2q,(6)} &= \p_\mu \Big( \bar q \tilde\gamma^\nu (\overleftarrow D_\nu \overleftarrow D^\mu + D^\mu D_\nu) q \Big) \, , \nn
	\tilde\O_{11}^{2q,(6)} &= \p_\nu \Big( \bar q \tilde\gamma^\nu (\overleftarrow D^2 + D^2) q \Big) \, , \nn
	\tilde\O_{12}^{2q,(6)} &= \Box \p_\nu( \bar q \tilde\gamma^\nu q ) \, .
\end{align}

Next, we investigate the contractions of three derivatives with $\epsilon^{\mu\nu\lambda\sigma} \gamma_\sigma$. It is only possible to have three covariant derivatives or two covariant and one partial derivative: partial derivatives are commuting, hence two or three of them vanish upon contraction with the Levi-Civita tensor. Left-acting covariant derivatives can again be traded for right-acting and partial derivatives. If we choose two covariant and one partial derivative, we can immediately insert the commutator of the covariant derivatives. In order to have a $P$-odd operator, no $\gamma_5$ matrix is allowed. The only Hermitian operator with two covariant derivatives is therefore
\begin{align}
	\tilde\O_{13}^{2q,(6)} &= \p_\mu( \bar q i \gamma_\sigma [D_\nu, D_\lambda] q) \epsilon^{\mu\nu\lambda\sigma} \, ,
\end{align}
which indeed is $CP$-odd. 
Finally, consider the insertion of three covariant derivatives:
\begin{align}
	\bar q \gamma_\sigma D_\mu D_\nu D_\lambda q \epsilon^{\mu\nu\lambda\sigma} = \frac{1}{2} \bar q \gamma_\sigma D_\mu [ D_\nu , D_\lambda ] q \epsilon^{\mu\nu\lambda\sigma} \, .
\end{align}
Hermitian conjugation of this operator is identical to a $CP$ conjugation. The $CP$-odd Hermitian component is again identical to $\tilde\O_{13}^{2q,(6)}$, which is therefore the only $CP$-odd operator.

Finally, we consider the operator classes with field-strength tensors. Due to~\eqref{eq:CovariantDerivativeCommutator}, we only need to take into account the external (electromagnetic) field-strength tensor: the QCD field-strength tensor can be written as a linear combination of the commutator of covariant derivatives and the external field-strength tensors.
In the class $\psi^2 F \M$, the chirally invariant operators are
\begin{align}
	\bar q_L \sigma^{\mu\nu}  F_{\mu\nu}^L \M q_R \, , \quad \bar q_L \sigma^{\mu\nu} \M  F_{\mu\nu}^R q_R \, , \quad \bar q_R \sigma^{\mu\nu} F_{\mu\nu}^R \M^\dagger q_L \, \quad \bar q_R \sigma^{\mu\nu} \M^\dagger F_{\mu\nu}^L q_L  \, .
\end{align}
One Hermitian linear combination is both $P$-odd and $CP$-odd:
\begin{align}
	\tilde\O_{14}^{2q,(6)} &= i \big( \bar q_L \sigma^{\mu\nu} ( F_{\mu\nu}^L \M + \M F_{\mu\nu}^R ) q_R - \bar q_R \sigma^{\mu\nu} ( F_{\mu\nu}^R \M^\dagger + \M^\dagger F_{\mu\nu}^L ) q_L \big) \nn
			& \stackrel{\text{fixed spurion}}{\mapsto}  i e ( \bar q \tilde\sigma^{\mu\nu}  \{ \M , Q \} q ) F_{\mu\nu} \, .
\end{align}
The last class of two-quark operators is $\psi^2 F D$. Here, we find the $P$-odd operators
\begin{align}
	& ( \bar q_R \gamma^\mu \overleftarrow D^\nu F_{\mu\nu}^R q_R - \bar q_L \gamma^\mu \overleftarrow D^\nu F_{\mu\nu}^L q_L ) \, , \quad
		( \bar q_R \gamma^\mu F_{\mu\nu}^R D^\nu q_R - \bar q_L \gamma^\mu F_{\mu\nu}^L D^\nu q_L ) \, , \nn
	& \p^\nu ( \bar q_R \gamma^\mu F_{\mu\nu}^R q_R - \bar q_L \gamma^\mu F_{\mu\nu}^L q_L ) \, , \quad
	( \bar q_R \gamma_\sigma F_{\nu\lambda}^R D_\mu q_R + \bar q_L \gamma_\sigma F_{\nu\lambda}^L D_\mu q_L ) \epsilon^{\mu\nu\lambda\sigma} \, , \nn
	& \p_\mu ( \bar q_R \gamma_\sigma F_{\nu\lambda}^R q_R + \bar q_L \gamma_\sigma F_{\nu\lambda}^L q_L ) \epsilon^{\mu\nu\lambda\sigma} \, .
\end{align}
Two $CP$-odd Hermitian linear combinations exist:
\begin{align}
	\tilde\O_{15}^{2q,(6)} &= i ( \bar q_R \gamma^\mu F_{\mu\nu}^R D^\nu q_R - \bar q_L \gamma^\mu F_{\mu\nu}^L D^\nu q_L ) - i ( \bar q_R \gamma^\mu \overleftarrow D^\nu F_{\mu\nu}^R q_R - \bar q_L \gamma^\mu \overleftarrow D^\nu F_{\mu\nu}^L q_L ) \nn
		&\quad \stackrel{\text{fixed spurion}}{\mapsto} i e ( \bar q \tilde\gamma^\mu Q \overleftrightarrow D^\nu q ) F_{\mu\nu} \, , \nn
	\tilde\O_{16}^{2q,(6)} &= \p_\mu ( \bar q_R \gamma_\sigma F_{\nu\lambda}^R q_R + \bar q_L \gamma_\sigma F_{\nu\lambda}^L q_L ) \epsilon^{\mu\nu\lambda\sigma} \stackrel{\text{fixed spurion}}{\mapsto} e \p_\mu ( \bar q \gamma_\sigma Q q ) F_{\nu\lambda} \epsilon^{\mu\nu\lambda\sigma} \, .
\end{align}

\subsection{Four-quark operators}
\label{sec:FourQuark}

A basis for the chirally invariant four-quark operators is given by the following six operators:
\begin{align}
	\O^{V,LL}_{1_c 1_f} &= (\bar q_L \gamma^\mu q_L)(\bar q_L \gamma_\mu q_L) \, , \; &
	\O^{V,LL}_{8_c 1_f} &= (\bar q_L \gamma^\mu t^a q_L)(\bar q_L \gamma_\mu t^a q_L) \, , \nn
	\O^{V,RR}_{1_c 1_f} &= (\bar q_R \gamma^\mu q_R)(\bar q_R \gamma_\mu q_R) \, , \; &
	\O^{V,RR}_{8_c 1_f} &= (\bar q_R \gamma^\mu t^a q_R)(\bar q_R \gamma_\mu t^a q_R) \, , \nn
	\O^{V,LR}_{1_c 1_f} &= (\bar q_L \gamma^\mu q_L)(\bar q_R \gamma_\mu q_R) \, , \; &
	\O^{V,LR}_{8_c 1_f} &= (\bar q_L \gamma^\mu t^a q_L)(\bar q_R \gamma_\mu t^a q_R) \, ,
\end{align}
where $t^a$ are the $SU(3)$ generators in color space. Note that flavor-octet operators $\O^{V,LL}_{1_c 8_f}$, $\O^{V,LL}_{8_c 8_f}$, $\O^{V,RR}_{1_c 8_f}$, and $\O^{V,RR}_{8_c 8_f}$ are related to the above operators through Fierz identities in Dirac space
\begin{align}
	\label{eq:ChiralFierzIdentities}
		( \gamma^\mu P_L ) [ \gamma_\mu P_L ] &= - (\gamma^\mu P_L ] [ \gamma_\mu P_L ) \, , \nn
		( \gamma^\mu P_R ) [ \gamma_\mu P_R ] &= - (\gamma^\mu P_R ] [ \gamma_\mu P_R ) \, , \nn
		( \gamma^\mu P_L ) [ \gamma_\mu P_R ] &= 2 ( P_R ] [ P_L ) \, , \nn
		( \gamma^\mu P_R ) [ \gamma_\mu P_L ] &= 2 ( P_L ] [ P_R ) \, ,
\end{align}
as well as the $SU(N_f)$ and $SU(N_c)$ Fierz relations:
\begin{align}
	t^A_{ij} t^A_{kl} &= \frac{1}{2} \delta_{il} \delta_{kj} - \frac{1}{2N_f} \delta_{ij} \delta_{kl} \, , \quad
	t^a_{\alpha\beta} t^a_{\gamma\delta} = \frac{1}{2} \delta_{\alpha\delta} \delta_{\gamma\beta} - \frac{1}{2N_c} \delta_{\alpha\beta} \delta_{\gamma\delta} \, .
\end{align}
On the other hand, the flavor-octet $LR$ operators $\O^{V,LR}_{1_c 8_f}$, $\O^{V,LR}_{8_c 8_f}$ are not chirally invariant.

For the four-quark operators, Hermitian conjugation acts in the same way as a $CP$-transformation. All the operators $\O^{V,LL}$, $\O^{V,RR}$, and $\O^{V,LR}$ are Hermitian and $CP$-even.
We conclude that there is no four-quark operator that could mix with the $CP$-odd three-gluon operator.

Note that if $SU(2)$ instead of $SU(3)$ chiral symmetry is considered, there are additional chirally invariant operators~\cite{deVries:2012ab,Dekens:2013zca} due to the absence of the symmetric structure constants $d^{ABC}$. In $SU(3)$, these operators only appear at dimension 7 as structures similar to~\eqref{eq:Dim5ChirallyInvariant}:
\begin{align}
	&\epsilon_{ijk} \epsilon_{lmn} \M^\dagger_{mj} (\bar q_L^k q_R^n)(\bar q_L^i q_R^l) \, , \quad \epsilon_{ijk} \epsilon_{lmn} \M_{mj} (\bar q_R^k q_L^n)(\bar q_R^i q_L^l) \, , \nn
	&\epsilon_{ijk} \epsilon_{lmn} \M^\dagger_{mj} (\bar q_L^k t^a q_R^n)(\bar q_L^i t^a q_R^l) \, , \quad \epsilon_{ijk} \epsilon_{lmn} \M_{mj} (\bar q_R^k t^a q_L^n)(\bar q_R^i t^a q_L^l) \, ,
\end{align}
with two $P$-odd and $CP$-odd linear combinations. The $SU(3)$ analysis shows that the $SU(2)$ invariant operators
\begin{align}
	&(\bar u_L u_R) (\bar d_L d_R) - (\bar u_R u_L) (\bar d_R d_L) - (\bar d_L u_R) (\bar u_L d_R) + (\bar d_R u_L) (\bar u_R d_L) \, , \nn
	&(\bar u_L t^a u_R) (\bar d_L t^a d_R) - (\bar u_R t^a u_L) (\bar d_R t^a d_L) - (\bar d_L t^a u_R) (\bar u_L t^a d_R) + (\bar d_R t^a u_L) (\bar u_R t^a d_L) \, ,
\end{align}
always involve a factor $m_s$. Therefore, the gCEDM does not mix into these operators, which can only appear as power corrections. Here, we neglect any effects beyond dimension 6.

\subsection{Intermediate summary}
\label{sec:IntermediateSummaryOperators}

Here, we summarize the $P$-odd, $CP$-odd, Lorentz- and gauge-invariant, chirally invariant Hermitian operators up to dimension 6.

\paragraph{Pure gauge operators}

At dimension four, there is the QCD $\theta$-term:
\begin{align}
	\tilde\O_1^{G,(4)} &= \tr[G_{\mu\nu} \widetilde G^{\mu\nu}] \, ,
\end{align}
while at dimension six, we find four operators:
\begin{align}
		\tilde\O_1^{G,(6)} &= \tr[\M^2] \tr[G_{\mu\nu} \widetilde G^{\mu\nu}] \, , \nn
		\tilde\O_2^{G,(6)} &= \Box \, \tr[G_{\mu\nu} \widetilde G^{\mu\nu}] \, , \nn
		\tilde\O_3^{G,(6)} &= \p_\nu \tr[ (D^\mu G_{\mu\lambda}) \widetilde G^{\nu\lambda} ] \, , \nn
		\tilde\O_4^{G,(6)} &= i \tr[G_{\mu\nu} {G^\mu}_\lambda \widetilde G^{\nu\lambda}] \, .
\end{align}

\paragraph{Two-quark operators}

We find two operators at dimension four:
\begin{align}
	\tilde\O_1^{2q,(4)} &= \bar q i \gamma_5 \M q \, , \nn
	\tilde\O_2^{2q,(4)} &= \p_\mu (\bar q \tilde\gamma^\mu q) \, ,
\end{align}
one operator at dimension five:
\begin{align}
	\tilde\O_1^{2q,(5)} &= \epsilon_{ijk} \epsilon_{lmn} \M_{mj} \M_{nk} \bar q^i i \gamma_5 q^l \, ,
\end{align}
and 16 operators at dimension six:
\begin{align}
	\tilde\O_1^{2q,(6)} &= \bar q i\gamma_5 \M^3 q \, , \nn
	\tilde\O_2^{2q,(6)} &= \tr[\M^2] \bar q i \gamma_5 \M q \, , \nn
	\tilde\O_3^{2q,(6)} &= \p_\mu \left( \bar q \tilde\gamma^\mu \M^2 q \right) \, , \nn
	\tilde\O_4^{2q,(6)} &= \tr[\M^2] \p_\mu \left( \bar q \tilde\gamma^\mu q \right) , \nn
	\tilde\O_5^{2q,(6)} &= \Box \left( \bar q i \gamma_5 \M q \right) \, , \nn
	\tilde\O_6^{2q,(6)} &= \p_\mu ( \bar q \tilde\sigma^{\mu\nu} \overleftrightarrow{D}_\nu \M q ) \, , \nn
	\tilde\O_7^{2q,(6)} &= \bar q i \gamma_5 ( \overleftarrow { D}^2 +  D^2 ) \M q \, , \nn
	\tilde\O_8^{2q,(6)} &= \bar q \tilde\sigma^{\mu\nu} [D_\mu,D_\nu] \M q \, , \nn
	\tilde\O_9^{2q,(6)} &= \bar q \tilde\gamma^\mu (  \overleftarrow D^2 D_\mu +  \overleftarrow D_\mu D^2 ) q \, , \nn
	\tilde\O_{10}^{2q,(6)} &= \p_\mu \Big( \bar q \tilde\gamma^\nu (\overleftarrow D_\nu \overleftarrow D^\mu + D^\mu D_\nu) q \Big) \, , \nn
	\tilde\O_{11}^{2q,(6)} &= \p_\mu \Big( \bar q \tilde\gamma^\mu (\overleftarrow D^2 + D^2) q \Big) \, , \nn
	\tilde\O_{12}^{2q,(6)} &= \Box \p_\mu ( \bar q \tilde\gamma^\mu q ) \, , \nn
	\tilde\O_{13}^{2q,(6)} &= \p_\mu( \bar q i \gamma_\sigma [D_\nu, D_\lambda] q) \epsilon^{\mu\nu\lambda\sigma} \, , \nn
	\tilde\O_{14}^{2q,(6)} &= i e ( \bar q \tilde\sigma^{\mu\nu} \{ \M , Q \} q ) F_{\mu\nu} \, , \nn
	\tilde\O_{15}^{2q,(6)} &= i e ( \bar q \tilde\gamma^\mu Q \overleftrightarrow D^\nu q ) F_{\mu\nu} \, , \nn
	\tilde\O_{16}^{2q,(6)} &= e \p_\mu ( \bar q \gamma_\sigma Q q ) F_{\nu\lambda} \epsilon^{\mu\nu\lambda\sigma} \, .
\end{align}

\paragraph{Four-quark operators}
There are no four-quark operators that can mix with the gCEDM.


\section{BRST invariance and nuisance operators}
\label{sec:BasisConstructionII}

In this appendix, we provide details on the construction of the nuisance operators, which vanish by the EOM. We follow the method of~\cite{Deans:1978wn}.

In App.~\ref{sec:EoM}, we review the EOM. In App.~\ref{sec:STI}, we discuss the Slavnov--Taylor identities. The recipe for the construction of the nuisance operators is reviewed in App.~\ref{sec:NuisanceRecipe}. The symmetry properties of the building blocks are discussed in App.~\ref{sec:BuildingBlocksNuisance}. We construct the seed operators in App.~\ref{sec:SeedOperators} and list the resulting nuisance operators in App.~\ref{sec:ConstructionNuisanceOperators}. In App.~\ref{sec:NuisanceRedundancies}, we derive redundancies in the preliminary operator set, and we relate the operators to our final basis.

\subsection{Gauge fixing and equations of motion}
\label{sec:EoM}

The QCD Lagrangian including gauge fixing and Faddeev-Popov ghosts is given by
\begin{align}
	\label{eq:LGFgh}
	\L_\mathrm{QCD+GF+gh} &= - \frac{1}{4} G^{\mu\nu}_a G_{\mu\nu}^a + \bar q( i\slashed D - \M)q - \frac{1}{2\xi} (\p^\mu G^a_\mu)^2 + \p^\mu \bar c^a ( D_\mu^{ac} c^c ) \, ,
\end{align}
where $D_\mu^{ac} = \p_\mu \delta^{ac} + g f^{abc} G_\mu^b$ is the covariant derivative in the adjoint representation. The gauge-fixing term can also be written in terms of an auxiliary field $G^a$~\cite{Das:2008}:
\begin{align}
	\mathcal{L}_\mathrm{GF} = \frac{\xi}{2} G^a G^a + (\p^\mu G^a) G_\mu^a \, .
\end{align}
The EOM for $G^a$ is $\xi G^a = \p^\mu G_\mu^a$, which, inserted into $\mathcal{L}_\mathrm{GF}$, leads to the original gauge-fixing term plus a total derivative.

When the quark mass matrix is promoted to a spurion field, the mass term has to be replaced by
\begin{align}
	\mathcal{L}_m^\mathrm{QCD} = -\bar q_L \M q_R - \bar q_R \M^\dagger q_L .
\end{align}
The complete list of EOM reads:
\begin{align}
	( i \slashed D - \M^\dagger P_L - \M  P_R) q &= 0 \, , \quad \bar q ( i \overleftarrow{\slashed D} + \M^\dagger P_L + \M P_R ) = 0 \, , \nn
	D^\mu G_{\mu\nu}^a &= - g \bar q t^a \gamma_\nu q - \p_\nu G^a + g f^{abc} (\p_\nu \bar c^b) c^c \, , \nn
	\xi G^a &= \p^\mu G_\mu^a , \quad \p^\mu D_\mu^{ac} c^c = 0 \, , \quad D_\mu^{ac} \p^\mu \bar c^c = 0 \, .
\end{align}
For notational convenience, we define the EOM fields:
\begin{align}
	q_E &:= ( i \slashed D - \M^\dagger P_L - \M P_R ) q \, , \quad \bar q_E := - \bar q ( i \overleftarrow{\slashed D} + \M^\dagger P_L + \M P_R ) , \nn
	G_E^a &:= G^a - \frac{1}{\xi} \p^\mu G_\mu^a \, .
\end{align}
The definition of the EOM quark fields is chosen in such a way that they fulfil $q_E^\dagger = \bar q_E \gamma^0$.

\subsection{Slavnov--Taylor identities}
\label{sec:STI}

We add source terms for the fields and the (composite) BRST variations:
\begin{align}
	\label{eq:FaddeevPopovLagrangian}
		\L_0 &= - \frac{1}{4} G^{\mu\nu}_a G_{\mu\nu}^a + \bar q( i\slashed D - \M P_R - \M^\dagger P_L )q  + \p^\mu \bar c^a (D_\mu^{ac} c^c)  + \frac{\xi}{2} G^a G^a + (\p^\mu G^a) G_\mu^a \nn
			&\quad + H^{\mu,a} G_\mu^a + \bar L q + \bar q L + \bar N^a c^a + \bar c^a N^a \nn
			&\quad - J^{\mu,a} D_\mu^{ac} c^c + \bar K^a G^a + \frac{1}{2} g f^{abc} K^a c^b c^c  + g \bar M c^a t^a q + g \bar q c^a t^a M \, .
\end{align}
The action including a set of sources for gauge-invariant ghost-free operators $\O_i$ is defined as
\begin{align}
	S = \int d^4x \, \mathcal{L}_0 + \int d^4x \, \Phi_i(x) \O_i(x) =: S_0 + \Phi \cdot \O .
\end{align}
We introduce the BRST transformation:
\begin{align}
	\delta G_\mu^a &= - \frac{\delta S}{\delta J^{\mu,a}} \delta\lambda = D_\mu^{ac} c^c \delta\lambda \, , \nn
	\delta q &= -i \frac{\delta S}{\delta \bar M} \delta\lambda = - i g c^a t^a q \delta\lambda \, , \nn
	\delta \bar q &= -i \frac{\delta S}{\delta M} \delta\lambda = - i \bar q g c^a t^a \delta\lambda \, , \nn
	\delta c^a &= \frac{\delta S}{\delta K^a} \delta\lambda = \frac{1}{2} g f^{abc} c^b c^c \delta\lambda \, , \nn
	\delta \bar c^a &= \frac{\delta S}{\delta \bar K^a} \delta\lambda = G^a \delta\lambda \, , \nn
	\delta G^a &= 0 \, ,
\end{align}
where $\delta\lambda$ is an anticommuting infinitesimal parameter. With this transformation, we find
\begin{align}
	\delta( g c^a t^a q ) &= g \delta c^a t^a q + g c^a t^a \delta q = \frac{1}{2} g^2 f^{abc} c^b c^c \delta\lambda t^a q - i g^2 c^a c^b t^a t^b q \delta\lambda \nn
		&= -\frac{1}{2} g^2 f^{abc} c^b c^c t^a q \delta\lambda + \frac{1}{2} g^2 f^{cab} c^a c^b t^c q \delta\lambda = 0
\end{align}
and similarly
\begin{align}
	\delta( \bar q g c^a t^a ) &= 0 \, ,
\end{align}
as well as
\begin{align}
	\delta( D_\mu c^a ) &= \delta( \p_\mu c^a + g f^{abc} G_\mu^b c^c) = D_\mu \delta c^a + g f^{abc} \delta G_\mu^b c^c \nn
		&= \frac{1}{2} g D_\mu  (f^{abc} c^b c^c) \delta\lambda + g f^{abc} (D_\mu c^b) \delta\lambda c^c \nn
		&= \frac{1}{2} g D_\mu  (f^{abc} c^b c^c) \delta\lambda - g f^{abc} (D_\mu c^b) c^c \delta\lambda  = 0 \, , \nn
	\delta( f^{abc} c^b c^c ) &= f^{abc} ( \delta c^b c^c + c^b \delta c^c ) = \frac{1}{2} g f^{abc} \left( f^{bde} c^d c^e \delta\lambda c^c + c^b f^{cde} c^d c^e \delta\lambda \right) \nn
		&= g f^{abe} f^{cde} c^b c^c c^d \delta\lambda = \frac{g}{3} ( f^{abe} f^{cde} + f^{ace} f^{dbe} + f^{ade} f^{bce} ) c^b c^c c^d \delta\lambda = 0 \, ,
\end{align}
where in both relations we have used the Jacobi identity for the $SU(3)$ structure constants. Therefore, we see that for all fields $\phi \in \{ G_\mu^a, q, \bar q, c^a, \bar c^a, G^a \}$
\begin{align}
	\delta_1 \delta_2 \phi = 0 \, ,
\end{align}
i.e., the BRST transformation is nilpotent. Note that if the auxiliary field $G^a$ is not used, nilpotency for anti-ghosts only holds on-shell~\cite{Kugo:1979gm}.

For the fields $G_\mu^a$, $q$, $\bar q$, the BRST transformation corresponds to an infinitesimal gauge transformation with the parameter $\epsilon^a = c^a \delta\lambda$. Therefore, the physical part of the Lagrangian is invariant under BRST transformations. For the ghost and gauge-fixing part, one finds:
\begin{align}
	\delta \mathcal{L}_\mathrm{gh+GF} &= \delta\left( \p^\mu \bar c^a (D^{ac}_\mu c^c ) + \frac{\xi}{2} G^a G^a + (\p^\mu G^a) G_\mu^a \right) \nn*
		&= (\p^\mu G^a) \delta\lambda (D^{ac}_\mu c^c ) + (\p^\mu G^a) D^{ac}_\mu c^c \delta\lambda = 0 \, .
\end{align}
The source terms for the BRST variations are obviously invariant as well, hence the only variant part of the Lagrangian are the source terms for the fields:
\begin{align}
	\delta \mathcal{L}_0 &= H^{\mu,a} \delta G_\mu^a + \bar L \delta q + \delta \bar q L + \bar N^a \delta c^a + \delta \bar c^a N^a \nn
		&= \Big( H^{\mu,a} D_\mu^{ac} c^c  - i g \bar L c^a t^a q + i g \bar q c^a t^a L + \frac{1}{2} g f^{abc}  \bar N^a c^b c^c - G^a N^a \Big) \delta\lambda \, .
\end{align}
The generating functional is invariant under the variable transformation (see~\cite{Das:2008} for the invariance of the measure)
\begin{align}
	\phi \mapsto \phi + \delta\phi \, ,
\end{align}
which implies ($J$ generically denoting the sources)
\begin{align}
	0 &= \delta Z[J] = \int \mathcal{D}\phi \left( i \int d^4x \; \delta \mathcal{L}_0 \right) e^{i \int d^4x \;\mathcal{L}_0} \, , \nn
	0 &= \int d^4x \left( -H^{\mu,a} \frac{\delta Z}{\delta J^{\mu,a}} - i \bar L \frac{\delta Z}{\delta \bar M} + i \frac{\delta Z}{\delta M} L + \bar N^a \frac{\delta Z}{\delta K^a} - \frac{\delta Z}{\delta \bar K^a} N^a \right) \delta\lambda \, .
\end{align}
This implies for the generating functional of connected Green's functions, defined by $Z[J] = e^{i W[J]}$:
\begin{align}
	0 &= \int d^4x \left( -H^{\mu,a} \frac{\delta W}{\delta J^{\mu,a}}  - i \bar L \frac{\delta W}{\delta \bar M} + i \frac{\delta W}{\delta M} L + \bar N^a \frac{\delta W}{\delta K^a} - \frac{\delta W}{\delta \bar K^a} N^a \right) .
\end{align}
We introduce the effective action as the Legendre transform of $W$, which is the generating functional for one-particle-irreducible truncated Green's functions (we do not transform the auxiliary field, which is not propagating):
\begin{align}
	\Gamma[\phi^\text{(cl)}] := W[J] - \int d^4x & \bigg( H^{\mu,a} {G_\mu^a}^\text{(cl)} + \bar L q^\text{(cl)} + \bar q^\text{(cl)} L  + \bar N^a c^{a\text{(cl)}} + \bar c^{a\text{(cl)}} N^a  \bigg) \, ,
\end{align}
where the ``classical fields'' are expectation values (the functional derivatives are understood to act from the left):
\begin{align}
	{G_\mu^a}^\text{(cl)} &= \frac{\delta W}{\delta H_a^\mu} \, , \quad q^\text{(cl)} = \frac{\delta W}{\delta \bar L} \, , \quad \bar q^\text{(cl)} = - \frac{\delta W}{\delta L} \, , \quad c^{a\text{(cl)}} &= \frac{\delta W}{\delta \bar N^a} \, , \quad \bar c^{a\text{(cl)}} = - \frac{\delta W}{\delta N^a}  \, .
\end{align}
The variations of $\Gamma$ with respect to classical fields and sources are given by:
\begin{align}
		\frac{\delta \Gamma}{\delta {G_\mu^a}^\text{(cl)}} &= - H^{\mu,a} \, , \quad
		\frac{\delta \Gamma}{\delta q^\text{(cl)}} = \bar L \, , \quad
		\frac{\delta \Gamma}{\delta \bar q^\text{(cl)}} = - L \, , \quad
		\frac{\delta \Gamma}{\delta {c^a}^\text{(cl)}} = \bar N^a \, , \quad
		\frac{\delta \Gamma}{\delta \bar c^a{}^\text{(cl)}} = - N^a \, , \nn
		\frac{\delta \Gamma}{\delta J^{\mu,a}} &= \frac{\delta W}{\delta J^{\mu,a}} \, , \quad
		\frac{\delta \Gamma}{\delta M} = \frac{\delta W}{\delta M}  \, , \quad
		\frac{\delta \Gamma}{\delta \bar M} = \frac{\delta W}{\delta \bar M} \, , \quad
		\frac{\delta \Gamma}{\delta K^{a}} = \frac{\delta W}{\delta K^{a}} \, , \quad
		\frac{\delta \Gamma}{\delta \bar K^{a}} = \frac{\delta W}{\delta \bar K^{a}} \, ,
\end{align}
which leads to the Slavnov--Taylor identities:
\begin{align}
	0 = \int d^4x & \bigg(
			\frac{\delta \Gamma}{\delta {G_\mu^a}^\text{(cl)}} \frac{\delta \Gamma}{\delta J^{\mu,a}}
			- i \frac{\delta \Gamma}{\delta q^\text{(cl)}} \frac{\delta \Gamma}{\delta \bar M}
			- i \frac{\delta \Gamma}{\delta \bar q^\text{(cl)}} \frac{\delta \Gamma}{\delta M}
			+ \frac{\delta \Gamma}{\delta {c^a}^\text{(cl)}}\frac{\delta \Gamma}{\delta K^a}
			+ \frac{\delta \Gamma}{\delta \bar c^a{}^\text{(cl)}}  \frac{\delta \Gamma}{\delta \bar K^a}
			\bigg) \, .
\end{align}

In addition, we derive the (quantum) EOM for the ghost field. Consider the generating functional $Z[J]$, which must be invariant under the shift of the integration variable $\bar c^a \mapsto \bar c^a + \bar \epsilon$. Expanding to first order in $\bar\epsilon$ gives
\begin{align}
	0 = \int \mathcal{D}\phi ( \p^\mu D_\mu^{ab} c^b - N^a ) e^{i \int d^4x \, \mathcal{L}_0} \, .
\end{align}
This leads to
\begin{align}
	\label{eq:GhostQEOM}
	\frac{\delta\Gamma}{\delta\bar c^{a\text{(cl)}}} &= - N^a = - \frac{1}{Z} \int \mathcal{D}\phi ( \p^\mu D_\mu^{ab} c^b ) e^{i \int d^4x \, \mathcal{L}_0} = \p^\mu \frac{-i}{Z} \frac{\delta Z}{\delta J_a^\mu} = \p^\mu \frac{\delta W}{\delta J_a^\mu} = \p^\mu \frac{\delta \Gamma}{\delta J_a^\mu} \, .
\end{align}
Therefore, we can rewrite the Slavnov--Taylor identities as
\begin{align}
	\begin{split}
		0 = \int d^4x & \bigg(
			\bigg( \frac{\delta \Gamma}{\delta {G_\mu^a}^\text{(cl)}} - \p^\mu \frac{\delta \Gamma}{\delta \bar K^a} \bigg) \frac{\delta \Gamma}{\delta J^{\mu,a}}
			- i \frac{\delta \Gamma}{\delta q^\text{(cl)}} \frac{\delta \Gamma}{\delta \bar M}
			- i \frac{\delta \Gamma}{\delta \bar q^\text{(cl)}} \frac{\delta \Gamma}{\delta M}
			+ \frac{\delta \Gamma}{\delta {c^a}^\text{(cl)}}\frac{\delta \Gamma}{\delta K^a}
			\bigg) \, .
	\end{split}
\end{align}
The ghost quantum EOM~\eqref{eq:GhostQEOM} implies that the effective action depends on the anti-ghost only through the combination
\begin{align}
	J^{\mu,a} - \p^\mu \bar c^{a\text{(cl)}} \, .
\end{align}

\subsection{Construction of nuisance operators}
\label{sec:NuisanceRecipe}

To lowest order in the loop expansion, the Slavnov--Taylor identities become
\begin{align}
	0 = \int d^4x & \biggl( 
		\left( \frac{\delta S}{\delta G_\mu^a} - \p^\mu \frac{\delta S}{\delta \bar K^a} \right) \frac{\delta S}{\delta J^\mu_a}
		- i \frac{\delta S}{\delta q} \frac{\delta S}{\delta \bar M}
		- i \frac{\delta S}{\delta \bar q} \frac{\delta S}{\delta M}
		+ \frac{\delta S}{\delta c^a} \frac{\delta S}{\delta K^a}
		 \biggr) \, .
\end{align}
While $S = S_0 + \Phi \cdot \O$ satisfies the Ward identity, the general solution is given by
\begin{align}
	S + \hat\Phi \cdot \mathcal{N} \, ,
\end{align}
where $\mathcal{N}$ are additional nuisance operators. Working to first order in the external sources $\Phi$, $\hat\Phi$, one finds that the nuisance operators satisfy
\begin{align}
	\hat W( \hat\Phi \cdot \mathcal{N} ) = 0 \, ,
\end{align}
with the operator
\begin{align}
	\label{eq:WHatOperator}
	\hat W &=  \left( \frac{\delta S_0}{\delta G_\mu^a} - \p^\mu \frac{\delta S_0}{\delta \bar K^a} \right) \frac{\delta}{\delta J_a^\mu} + \frac{\delta S_0}{\delta J_a^\mu} \frac{\delta}{\delta G_\mu^a} + \left( \p^\mu \frac{\delta S_0}{\delta J_a^\mu} \right) \frac{\delta}{\delta \bar K^a} \nn
		&\quad -i \frac{\delta S_0}{\delta q} \frac{\delta}{\delta \bar M} -i \frac{\delta S_0}{\delta \bar M}\frac{\delta}{\delta q} 
			-i \frac{\delta S_0}{\delta \bar q}\frac{\delta}{\delta M} -i \frac{\delta S_0}{\delta M} \frac{\delta}{\delta \bar q} 
			+ \frac{\delta S_0}{\delta c^a}\frac{\delta}{\delta K^a} + \frac{\delta S_0}{\delta K^a} \frac{\delta}{\delta c^a} \, .
\end{align}
The BRST operator $\hat W$ is nilpotent, $\hat W^2 = 0$, and it carries ghost number $1$. In~\cite{Joglekar:1975nu} it was shown that the most general solution for the nuisance operators is given by
\begin{align}
	\hat\Phi \cdot \mathcal{N} = \int d^4x \, \hat W(x) ( \hat\Phi \cdot \F ) \, ,
\end{align}
where $\F$ is a set of anti-Hermitian ``seed operators'' with the same Lorentz, chiral, and global $SU(3)_c$ properties as $\O$, the same discrete symmetries and dimension, and ghost number $-1$. Hence, in the following we construct systematically the set $\F$ and derive from it the (gauge-variant) nuisance operators: we act with the operator $\hat W$ on the set $\mathcal{F}$ and afterwards we set the sources to zero.

Let us work out the explicit form of the operator $\hat W$. With sources already set to zero, we have
\begin{align}
			\frac{\delta S_0}{\delta G_\mu^a} &= D_\nu G^{\nu\mu,a} + g \bar q t^a \gamma^\mu q + \p^\mu G^a - g f^{abc} (\p^\mu \bar c^b) c^c \, , \quad
			\frac{\delta S_0}{\delta \bar K^a} = G^a \, , \quad
			\frac{\delta S_0}{\delta J_a^\mu} = - D_\mu^{ac} c^c \, , \nn
			\frac{\delta S_0}{\delta q} &= - \bar q_E \, , \quad
			\frac{\delta S_0}{\delta \bar q} = q_E \, , \quad
			\frac{\delta S_0}{\delta M} = g \bar q c^a t^a \, , \quad
			\frac{\delta S_0}{\delta \bar M} = g c^a t^a q \, , \nn
			\frac{\delta S_0}{\delta c^a} &= D_\nu^{ac} \p^\nu \bar c^c \, , \quad
			\frac{\delta S_0}{\delta K^a} = \frac{1}{2} g f^{abc} c^b c^c \, .
\end{align}
This leads to
\begin{align}
		\label{eq:WHatOperatorSimp}
		\hat W &=  \left( D_\nu G^{\nu\mu,a} + g \bar q t^a \gamma^\mu q - g f^{abc} (\p^\mu \bar c^b) c^c \right) \frac{\delta}{\delta J_a^\mu}  - \left( D_\mu^{ac} c^c \right) \frac{\delta}{\delta G_\mu^a} - \left( \p^\mu D_\mu^{ac} c^c \right) \frac{\delta}{\delta \bar K^a} \nn
			&\quad + i \bar q_E \frac{\delta}{\delta \bar M} - i g c^a t^a q \frac{\delta}{\delta q} 
			- i q_E \frac{\delta}{\delta M} - i g \bar q c^a t^a \frac{\delta}{\delta \bar q} \nn
			&\quad + \left( D_\nu^{ac} \p^\nu \bar c^c \right) \frac{\delta}{\delta K^a} + \left( \frac{1}{2} g f^{abc} c^b c^c \right) \frac{\delta}{\delta c^a} \, .
\end{align}
Note that after acting with the term
\begin{align}
	-i \frac{\delta S_0}{\delta \bar q} \frac{\delta}{\delta M} = -i q_E \frac{\delta}{\delta M}
\end{align}
on the seed operators, the EOM quark field $q_E$ needs to be anticommuted to the right-hand side, which produces an additional minus sign.

In case that the seed operator contains derivatives of the sources $J_a^\mu$, $\ldots$, we have to use partial integration, e.g.,
\begin{align}
		\int d^4x \, \hat W(x) & \int d^4y \, \hat\Phi(y) \mathcal{F}(y) = \int d^4x \, \hat W(x) \int d^4y \, \hat\Phi(y) \mathcal{\tilde F}(y)_b^{\nu\lambda} ( \p_\nu^y J_\lambda^b(y) ) \nn
			&= - \int d^4x \, \left( \frac{\delta S_0}{\delta G_\mu^a(x)} - \p^\mu \frac{\delta S_0}{\delta \bar K^a(x)} \right) \frac{\delta}{\delta J_a^\mu(x)} \int d^4y \, \p_\nu^y ( \hat\Phi(y) \mathcal{\tilde F}(y)_b^{\nu\lambda} ) J_\lambda^b(y) \nn
			&= - \int d^4x \, \left( \frac{\delta S_0}{\delta G_\mu^a(x)} - \p^\mu \frac{\delta S_0}{\delta \bar K^a(x)} \right) \p^\nu ( \hat\Phi(x) \mathcal{\tilde F}(x)^a_{\nu\mu} ) \nn
			&= \int d^4x \, \hat\Phi(x)  \p^\nu \left( \frac{\delta S_0}{\delta G_\mu^a(x)} - \p^\mu \frac{\delta S_0}{\delta \bar K^a(x)} \right) \mathcal{\tilde F}(x)^a_{\nu\mu} \, .
\end{align}

\subsection{Symmetry properties of sources and building blocks}
\label{sec:BuildingBlocksNuisance}

\begin{table}[t]
	\centering\small
	\setlength{\tabcolsep}{3.5pt}
	\begin{tabular}{c c c c c c c c c c}
		\toprule
		field 	&			comm. &		mass dim. &	ghost num. &		Lorentz & 			$SU(3)_c$ &	$\chi$ &			$\dagger$ &				$P$ &					$CP$ \\
		\midrule																																										
		$q_L$ &			$-$ &		$\frac{3}{2}$ &	$\phantom{+}0$ &	$(2,1)$ &			$3$ &		$(3,1)$ &			$\bar q_L \gamma^0$ &		$\gamma^0 q_R$ &			$\gamma^0 C \bar q_L^T$	 \\[0.2cm]
		$q_R$ &			$-$ &		$\frac{3}{2}$ &	$\phantom{+}0$ &	$(1,2)$ &			$3$ &		$(1,3)$ &			$\bar q_R \gamma^0$ &		$\gamma^0 q_L$ &			$\gamma^0 C \bar q_R^T$	 \\[0.2cm]
		$\bar q_L$ &		$-$ &		$\frac{3}{2}$ &	$\phantom{+}0$ &	$(1,2)$ &			$\bar 3$ &		$(\bar 3, 1)$ &		$\gamma^0 q_L$ &			$\bar q_R \gamma^0$ &		$q_L^T C \gamma^0$		 \\[0.2cm]
		$\bar q_R$ &		$-$ &		$\frac{3}{2}$ &	$\phantom{+}0$ &	$(2,1)$ &			$\bar 3$ &		$(1,\bar 3)$ &		$\gamma^0 q_R$ &			$\bar q_L \gamma^0$ &		$q_R^T C \gamma^0$		 \\[0.1cm]
		\midrule																																										
		$G_\mu^a$ &		$+$ &		1 &			$\phantom{+}0$ &	$(2,2)$ &			$8$ &		$(1,1)$ &			$G_\mu^a$ &				$G^\mu_a$ &				$-\eta(a) G^\mu_a$			 \\[0.2cm]
		$G^a$ &			$+$ &		2 &			$\phantom{+}0$ &	$(1,1)$ &			$8$ &		$(1,1)$ &			$G^a$ &					$G^a$ &					$-\eta(a) G^a$				 \\[0.1cm]
		\midrule																																										
		$c^a$ &			$-$ &		0 &			$\phantom{+}1$ &	$(1,1)$ &			$8$ &		$(1,1)$ &			$c^a$ &					$c^a$ &					$-\eta(a) c^a$				 \\[0.2cm]
		$\bar c^a$ &		$-$ &		2 &			$-1$ &			$(1,1)$ &			$8$ &		$(1,1)$ &			$-\bar c^a$ &				$\bar c^a$ &				$-\eta(a) \bar c^a$			 \\[0.1cm]
		\midrule																																										
		$J_\mu^a$ &		$-$ &		3 &			$-1$ &			$(2,2)$ &			$8$ &		$(1,1)$ &			$-J_\mu^a$ &				$J^\mu_a$ &				$-\eta(a) J^\mu_a$			 \\[0.2cm]
		$M_L$ &			$+$ &		$\frac{5}{2}$ &	$-1$ &			$(2,1)$ &			$3$ &		$(1,3)$ &			$\bar M_L \gamma^0$ &		$\gamma^0 M_R$ &			$-\gamma^0 C \bar M_L^T$	 \\[0.2cm]
		$M_R$ &			$+$ &		$\frac{5}{2}$ &	$-1$ &			$(1,2)$ &			$3$ &		$(3,1)$ &			$\bar M_R \gamma^0$ &		$\gamma^0 M_L$ &			$-\gamma^0 C \bar M_R^T$	 \\[0.2cm]
		$\bar M_L$ &		$+$ &		$\frac{5}{2}$ &	$-1$ &	 		$(1,2)$ &			$\bar 3$ &		$(1,\bar 3)$ &		$\gamma^0 M_L$ &			$\bar M_R \gamma^0$ &		$-M_L^T C\gamma^0$		 \\[0.2cm]
		$\bar M_R$ &		$+$ &		$\frac{5}{2}$ &	$-1$ &	 		$(2,1)$ &			$\bar 3$ &		$(\bar 3,1)$ &		$\gamma^0 M_R$ &			$\bar M_L \gamma^0$ &		$-M_R^T C\gamma^0$		 \\[0.2cm]
		$K^a$ &			$+$ &		4 &			$-2$ &			$(1,1)$ &			$8$ &		$(1,1)$ &			$-K^a$ &					$K^a$ &					$-\eta(a) K^a$				 \\[0.2cm]
		$\bar K^a$ &		$+$ &		2 &			$\phantom{+}0$ &	$(1,1)$ &			$8$ &		$(1,1)$ &			$\bar K^a$ &				$\bar K^a$ &				$-\eta(a) \bar K^a$			 \\[0.1cm]
		\midrule																																										
		$F_{\mu\nu}^L$ &	$+$ &		2 &			$\phantom{+}0$ &	$\!\!\!\!\!(3,1)\oplus(1,3)\!\!\!\!\!$ &			$1$ &			$(8,1)$ &				$F_{\mu\nu}^L$ &					$F^{\mu\nu}_R$ &				$-{F^{\mu\nu}_L}^T$				 \\[0.2cm]
		$F_{\mu\nu}^R$ &	$+$ &		2 &			$\phantom{+}0$ &	$\!\!\!\!\!(3,1)\oplus(1,3)\!\!\!\!\!$ &			$1$ &			$(1,8)$ &				$F_{\mu\nu}^R$ &					$F^{\mu\nu}_L$ &				$-{F^{\mu\nu}_R}^T$				 \\[0.2cm]
		$\M$ &			$+$ &		1 &			$\phantom{+}0$ &	$(1,1)$ &			$1$ &		$(3,\bar 3)$ &		$\M^\dagger$ &			$\M^\dagger$ &			$\M^*$					 \\[0.2cm]
		$\M^\dagger$ &	$+$ &		1 &			$\phantom{+}0$ &	$(1,1)$ &			$1$ &		$(\bar 3, 3)$ &		$\M$ &					$\M$ &					$\M^T$					 \\[0.1cm]
		\midrule																																										
		$\p_\mu$ &		$+$ &		1 &			$\phantom{+}0$ &	$(2,2)$ &			$1$ &		$(1,1)$ &			$\p_\mu$ &				$\p^\mu$ &				$\p^\mu$					 \\[0.2cm]
		$\nabla_\mu^L(\cdot)$ &	$+$ &	1 &			$\phantom{+}0$ &	$(2,2)$ &			$1$ &		{$(1,1)\oplus(8,1)$} &				$(\cdot)\overleftarrow{\nabla}_\mu^L$ &	$\nabla_R^\mu(\cdot)$ &			${\nabla_L^{\mu}}^*(\cdot)$			 \\[0.2cm]
		$\nabla_\mu^R(\cdot)$ &	$+$ &	1 &			$\phantom{+}0$ &	$(2,2)$ &			$1$ &		{$(1,1)\oplus(1,8)$} &				$(\cdot)\overleftarrow{\nabla}_\mu^R$ &	$\nabla_L^\mu(\cdot)$ &			${\nabla_R^{\mu}}^*(\cdot)$			 \\[0.1cm]
		\bottomrule
	\end{tabular}
	\caption{Properties of dynamical fields, sources, spurions, and derivative operators. For simplicity, additional arbitrary phases in $P$- and $CP$-conjugation are neglected. $\eta(a)$ is defined in~\eqref{eq:etafactor}.}
	\label{tab:FieldsAndSources}
\end{table}

In Table~\ref{tab:FieldsAndSources}, we list the transformation properties of the various fields and sources, which are the building blocks for the seed operators. In particular, the given transformation properties ensure that the leading-order Lagrangian is Hermitian,\footnote{See~\cite{Kugo:1979gm} for the Hermiticity properties of the ghost fields.} $P$-even, and $CP$-even.  Note that we define the complex conjugate of the product of Grassmann variables $c_{1,2}$ as $(c_1 c_2)^* = c_2^* c_1^* = - c_1^* c_2^*$.

We assign zero mass dimension to the ghost field and mass dimension 2 to the anti-ghost field. In this convention, the operator $\int d^4x \, \hat W(x)$ does not change the mass dimension of the seed operators. If a mass dimension 1 is assigned to both ghost and anti-ghost fields, the operator $\int d^4x \, \hat W(x)$ raises the mass dimension by one unit. The assignment of the mass dimensions is purely conventional and does not affect the results.

We assign the following chiral transformations:
\begin{align}
	M_{L,R} &\stackrel{\chi}{\mapsto} U_{R,L} M_{L,R} , \nn
	\bar M_{L,R} &\stackrel{\chi}{\mapsto} \bar M_{L,R} U_{R,L}^\dagger  ,
\end{align}
where
\begin{align}
	M_{L,R} := P_{L,R} M , \quad \bar M_{L,R} := \bar M P_{R,L} .
\end{align}

Due to gauge fixing, the seed operators need not be gauge invariant under $SU(3)_c$. Therefore, the gauge field $G_\mu^a$ is allowed as a separate building block and not only as part of the full covariant derivative $D_\mu$. However, the gauged $SU(3)_L\times SU(3)_R$ chiral symmetry, which also contains $U(1)_\mathrm{em}$ as a subgroup, remains intact. Therefore, we define covariant derivatives with respect to the external fields only,
\begin{align}
	\nabla_\mu^L &= \partial_\mu - i l_\mu \, , \qquad \overleftarrow{\nabla}_\mu^L = \overleftarrow{\p}_\mu + i l_\mu \, , \nn
	\nabla_\mu^R &= \partial_\mu - i r_\mu \, , \qquad \overleftarrow{\nabla}_\mu^R = \overleftarrow{\p}_\mu + i r_\mu \, ,
\end{align}
and impose on the seed operators invariance under the local chiral group.

\subsection{Seed operators}
\label{sec:SeedOperators}

Let us now systematically construct the gauge-variant seed operators using the building blocks in Table~\ref{tab:FieldsAndSources}.
Due to the ghost EOM~\eqref{eq:GhostQEOM}, the anti-ghost and the source $J_a^\mu$ only appear as a building block
\begin{align}
	\hat J^{\mu,a} :=  J^{\mu,a} - \p^\mu \bar c^a \, .
\end{align}

After applying the operator $\hat W$, we will set the sources $M_{L,R}$, $\bar M_{L,R}$, $K^a$, $\bar K^a$ to zero, hence we only need to take into account seed operators with at most one of these sources. The source $\hat J^{\mu,a}$ should be set to $-\p^\mu\bar c^a$.

In order to construct $SU(3)_c$ singlets, we contract open indices with the $SU(3)_c$ tensors $\delta^{ab}$ (two indices), $f^{abc}$, $d^{abc}$ (three indices), or
\begin{align}
	\delta^{ab} \delta^{cd} , \quad \delta^{ac} \delta^{bd} , \quad \delta^{ad} \delta^{bc} , \quad d^{abe} d^{cde}, \quad d^{ace} d^{bde} , \quad d^{abe} f^{cde}, \quad d^{ace} f^{bde}, \quad d^{ade} f^{bce}
\end{align}
in the case of four indices~\cite{Dittner:1971fy,Borodulin:1995xd}.

\paragraph{$K^a$ operators}

At dimension 4, the only operator
\begin{align*}
	c^a K^a
\end{align*}
is $P$- and $CP$-even. At dimension 5, there are no operators. At dimension 6, the following operators have to be considered:
\begin{align*}
	\text{seed operator}& \quad & &P \quad & &CP \\*[0.1cm]
	\hline \\*[-0.4cm]
	G^\mu_a G_\mu^a c^b K^b , \quad
	G^\mu_a G_\mu^b c^a K^b , \quad
	G^\mu_a G_\mu^b c^c K^d d^{ace} d^{bde} , 	 & & &\text{even},  & &\text{even}, \\*
	G^\mu_a G_\mu^b c^c K^d d^{ace} f^{bde} , 	 & & &\text{even},  & &\text{odd}, \\*
	(\p^\mu G_\mu^a) c^b K^c f^{abc} , \quad
	G_\mu^a (\p^\mu c^b) K^c f^{abc} , \quad
	G_\mu^a c^b (\p^\mu K^c) f^{abc} , 	 & & &\text{even},  & &\text{even}, \\*
	(\p^\mu G_\mu^a) c^b K^c d^{abc} , \quad
	G_\mu^a (\p^\mu c^b) K^c d^{abc} , \quad
	G_\mu^a c^b (\p^\mu K^c) d^{abc} , 	 & & &\text{even},  & &\text{odd}, \\
	\\
	c^a (\Box K^a) , 	 & & &\text{even},  & &\text{even}, \\
	(\Box c^a) K^a , 	 & & &\text{even},  & &\text{even}, \\
	(\p^\mu c^a) (\p_\mu K^a) , 	 & & &\text{even},  & &\text{even}, \\
	\tr[\M \M^\dagger] c^a K^a , 	 & & &\text{even},  & &\text{even} .
\end{align*}
Obviously, it is not possible to construct a $P$-odd operator.

\paragraph{$\bar K^a$ operators}

The source $\bar K^a$ has to come together with a source $\hat J_\mu^a$ in order to give ghost number $-1$. There are only dimension-six operators:
\begin{align*}
	\text{seed operator}& \quad & &P \quad & &CP \\[0.1cm]
	\hline \\[-0.4cm]
	G^\mu_a \hat J_\mu^b \bar K^c f^{abc} , 	 & & &\text{even},  & &\text{even}, \\
	G^\mu_a \hat J_\mu^b \bar K^c d^{abc} , 	 & & &\text{even},  & &\text{odd}, \\
	(\p^\mu \hat J_\mu^a) \bar K^a , \quad
	\hat J^\mu_a (\p_\mu \bar K^a) ,	 & & &\text{even},  & &\text{even} .
\end{align*}
No $P$-odd operators can be constructed.

\paragraph{$M$ and $\bar M$ operators}

At dimension 4, there are four operators:
\begin{align*}
	\text{seed operator}& \quad & &P \quad & &CP \\[0.1cm]
	\hline \\[-0.4cm]
	i( \bar q M + \bar M q ) , 	 						& & &\text{even},  & &\text{even}, \\
	\bar q M - \bar M q , 								& & &\text{even},  & &\text{odd}, \\
	i ( \bar q \gamma_5 M - \bar M \gamma_5 q ) , 	 		& & &\text{odd},  & &\text{even}, \\
	\bar q \gamma_5 M + \bar M \gamma_5 q , 	 		& & &\text{odd},  & &\text{odd}.
\end{align*}
Therefore, the only anti-Hermitian $P$- and $CP$-odd seed operator at dimension 4 is
\begin{align}
	\F_1^{(4)} = \bar q \gamma_5 M + \bar M \gamma_5 q \, .
\end{align}
No dimension-5 operator can be constructed. At dimension six, we find the following list of chirally invariant operators. Note that we already neglect operators that vanish when the spurions and external fields are fixed to their physical values:
\begin{align*}
	\text{seed operator}& \quad & &P \quad & &CP \\[0.1cm]
	\hline \\[-0.4cm]
	\tr[\M \M^\dagger] i ( \bar q_L M_R + \bar M_R q_L ), 											 & & & L \leftrightarrow R ,  & &\text{even}, \\
	\tr[\M \M^\dagger] ( \bar q_L M_R - \bar M_R q_L ), 												 & & & L \leftrightarrow R ,  & &\text{odd}, \\
	\tr[\M \M^\dagger] i ( \bar q_R M_L + \bar M_L q_R ), 											 & & & L \leftrightarrow R ,  & &\text{even}, \\
	\tr[\M \M^\dagger] ( \bar q_R M_L - \bar M_L q_R ), 												 & & & L \leftrightarrow R ,  & &\text{odd}, \\
	i ( \bar q_L \M \M^\dagger M_R + \bar M_R \M \M^\dagger q_L ) ,		 							 & & & L \leftrightarrow R, \; \M \leftrightarrow \M^\dagger ,  & &\text{even}, \\
	( \bar q_L \M \M^\dagger M_R - \bar M_R \M \M^\dagger q_L ) , 									 & & & L \leftrightarrow R, \; \M \leftrightarrow \M^\dagger ,  & &\text{odd}, \\
	i ( \bar q_R \M^\dagger \M M_L + \bar M_L \M^\dagger \M q_R ) ,		 							 & & & L \leftrightarrow R, \; \M \leftrightarrow \M^\dagger ,  & &\text{even}, \\
	( \bar q_R \M^\dagger \M M_L - \bar M_L \M^\dagger \M q_R ) , 									 & & & L \leftrightarrow R, \; \M \leftrightarrow \M^\dagger ,  & &\text{odd}, \\
	\\
	i ( \bar q_{L} \M \gamma_\mu t^a M_{L} + \bar M_{L} \M^\dagger \gamma_\mu t^a q_{L}) G_a^\mu , 			 & & & L \leftrightarrow R, \; \M \leftrightarrow \M^\dagger ,  & &\text{even}, \\
	( \bar q_{L} \M \gamma_\mu t^a M_{L} - \bar M_{L} \M^\dagger \gamma_\mu t^a q_{L}) G_a^\mu , 			 & & & L \leftrightarrow R, \; \M \leftrightarrow \M^\dagger ,  & &\text{odd}, \\
	i ( \bar q_{R} \M^\dagger \gamma_\mu t^a M_{R} + \bar M_{R} \M \gamma_\mu t^a q_{R}) G_a^\mu , 		 & & & L \leftrightarrow R, \; \M \leftrightarrow \M^\dagger ,  & &\text{even}, \\
	( \bar q_{R} \M^\dagger \gamma_\mu t^a M_{R} - \bar M_{R} \M \gamma_\mu t^a q_{R}) G_a^\mu , 			 & & & L \leftrightarrow R, \; \M \leftrightarrow \M^\dagger ,  & &\text{odd}, \\
	\\
	i ( \bar q_{L} \M \slashed \nabla_R M_{L} + \bar M_{L} \overleftarrow{\slashed \nabla}_R \M^\dagger q_{L}) , 			 & & & L \leftrightarrow R, \; \M \leftrightarrow \M^\dagger ,  & &\text{odd}, \\
	( \bar q_{L} \M \slashed \nabla_R M_{L} - \bar M_{L} \overleftarrow{\slashed \nabla}_R \M^\dagger q_{L}) , 				 & & & L \leftrightarrow R, \; \M \leftrightarrow \M^\dagger ,  & &\text{even}, \\
	i ( \bar q_{R} \M^\dagger \slashed \nabla_L M_{R} + \bar M_{R} \overleftarrow{\slashed \nabla}_L \M q_{R}) , 			 & & & L \leftrightarrow R, \; \M \leftrightarrow \M^\dagger ,  & &\text{odd}, \\
	( \bar q_{R} \M^\dagger \slashed \nabla_L M_{R} - \bar M_{R} \overleftarrow{\slashed \nabla}_L \M q_{R}) , 			 & & & L \leftrightarrow R, \; \M \leftrightarrow \M^\dagger ,  & &\text{even}, \\
	i ( \bar q_{L} \overleftarrow{\slashed \nabla}_L \M M_{L} + \bar M_{L} \M^\dagger \slashed \nabla_L q_{L}) , 			 & & & L \leftrightarrow R, \; \M \leftrightarrow \M^\dagger ,  & &\text{odd}, \\
	( \bar q_{L} \overleftarrow{\slashed \nabla}_L \M M_{L} - \bar M_{L} \M^\dagger \slashed \nabla_L q_{L}) , 				 & & & L \leftrightarrow R, \; \M \leftrightarrow \M^\dagger ,  & &\text{even}, \\
	i ( \bar q_{R} \overleftarrow{\slashed \nabla}_R \M^\dagger M_{R} + \bar M_{R} \M \slashed \nabla_R q_{R}) , 			 & & & L \leftrightarrow R, \; \M \leftrightarrow \M^\dagger ,  & &\text{odd}, \\
	( \bar q_{R} \overleftarrow{\slashed \nabla}_R \M^\dagger M_{R} - \bar M_{R} \M \slashed \nabla_R q_{R}) , 			 & & & L \leftrightarrow R, \; \M \leftrightarrow \M^\dagger ,  & &\text{even}, \\
	\\
	i ( \bar q_{L,R} M_{R,L} + \bar M_{R,L} q_{L,R}) G_\mu^a G^\mu_a , 								 & & & L \leftrightarrow R ,  & &\text{even}, \\
	( \bar q_{L,R} M_{R,L} - \bar M_{R,L} q_{L,R}) G_\mu^a G^\mu_a , 									 & & & L \leftrightarrow R ,  & &\text{odd}, \\
	i ( \bar q_{L,R} t^a M_{R,L} + \bar M_{R,L} t^a q_{L,R}) G_\mu^b G^\mu_c d^{abc} , 						 & & & L \leftrightarrow R ,  & &\text{even}, \\
	( \bar q_{L,R} t^a M_{R,L} - \bar M_{R,L} t^a q_{L,R}) G_\mu^b G^\mu_c d^{abc} , 						 & & & L \leftrightarrow R ,  & &\text{odd}, \\
	i ( \bar q_{L,R} \sigma^{\mu\nu} t^a M_{R,L} + \bar M_{R,L} \sigma^{\mu\nu} t^a q_{L,R}) G_\mu^b G_\nu^c f^{abc} , 							 & & & L \leftrightarrow R ,  & &\text{even}, \\
	( \bar q_{L,R} \sigma^{\mu\nu} t^a M_{R,L} - \bar M_{R,L} \sigma^{\mu\nu} t^a q_{L,R}) G_\mu^b G_\nu^c f^{abc} , 								 & & & L \leftrightarrow R ,  & &\text{odd}, \\
	\\
	i ( \bar q_{L,R} t^a M_{R,L} + \bar M_{R,L} t^a q_{L,R} ) \p_\mu G^\mu_a , 							 & & & L \leftrightarrow R ,  & &\text{odd}, \\
	( \bar q_{L,R} t^a M_{R,L} - \bar M_{R,L} t^a q_{L,R} ) \p_\mu G^\mu_a , 								 & & & L \leftrightarrow R ,  & &\text{even}, \\
	i ( \bar q_{L,R} \sigma^{\mu\nu} t^a M_{R,L} + \bar M_{R,L} \sigma^{\mu\nu} t^a q_{L,R}) \p_\mu G_\nu^a , 									 & & & L \leftrightarrow R ,  & &\text{even}, \\
	( \bar q_{L,R} \sigma^{\mu\nu} t^a M_{R,L} - \bar M_{R,L} \sigma^{\mu\nu} t^a q_{L,R}) \p_\mu G_\nu^a ,						 				 & & & L \leftrightarrow R ,  & &\text{odd}, \\
	\\
	i ( \bar q_{L,R} \overleftarrow{\nabla}_\mu^{L,R} t^a M_{R,L} + \bar M_{R,L} t^a \nabla_\mu^{L,R} q_{L,R} ) G^\mu_a , 			 & & & L \leftrightarrow R ,  & &\text{odd}, \\
	( \bar q_{L,R} \overleftarrow{\nabla}_\mu^{L,R} t^a M_{R,L} - \bar M_{R,L} t^a \nabla_\mu^{L,R} q_{L,R} ) G^\mu_a , 			 & & & L \leftrightarrow R ,  & &\text{even}, \\
	i ( \bar q_{L,R} \overleftarrow{\nabla}_\mu^{L,R} \sigma^{\mu\nu} t^a M_{R,L} + \bar M_{R,L} \sigma^{\mu\nu} t^a  \nabla_\mu^{L,R} q_{L,R}) G_\nu^a , 					 & & & L \leftrightarrow R ,  & &\text{even}, \\
	( \bar q_{L,R} \overleftarrow{\nabla}_\mu^{L,R} \sigma^{\mu\nu} t^a M_{R,L} - \bar M_{R,L} \sigma^{\mu\nu} t^a  \nabla_\mu^{L,R} q_{L,R}) G_\nu^a , 					 & & & L \leftrightarrow R ,  & &\text{odd}, \\
	\\
	i ( \bar q_{L,R}  t^a \nabla_\mu^{L,R} M_{R,L} + \bar M_{R,L} \overleftarrow{\nabla}_\mu^{L,R} t^a q_{L,R} ) G^\mu_a ,			 & & & L \leftrightarrow R ,  & &\text{odd}, \\
	( \bar q_{L,R}  t^a \nabla_\mu^{L,R} M_{R,L} - \bar M_{R,L} \overleftarrow{\nabla}_\mu^{L,R} t^a q_{L,R} ) G^\mu_a ,			 & & & L \leftrightarrow R ,  & &\text{even}, \\
	i ( \bar q_{L,R} \sigma^{\mu\nu} t^a \nabla_\mu^{L,R} M_{R,L} + \bar M_{R,L} \overleftarrow{\nabla}_\mu^{L,R} \sigma^{\mu\nu} t^a q_{L,R}) G_\nu^a , 					 & & & L \leftrightarrow R ,  & &\text{even}, \\
	( \bar q_{L,R} \sigma^{\mu\nu} t^a \nabla_\mu^{L,R} M_{R,L} - \bar M_{R,L} \overleftarrow{\nabla}_\mu^{L,R} \sigma^{\mu\nu} t^a q_{L,R}) G_\nu^a , 					 & & & L \leftrightarrow R ,  & &\text{odd}, \\
	\\
	i ( \bar q_{L,R} \overleftarrow \nabla_{L,R}^2 M_{R,L} + \bar M_{R,L} \nabla_{L,R}^2 q_{L,R}) ,								 & & & L \leftrightarrow R ,  & &\text{even}, \\
	( \bar q_{L,R} \overleftarrow \nabla_{L,R}^2 M_{R,L} - \bar M_{R,L} \nabla_{L,R}^2 q_{L,R}) ,								 & & & L \leftrightarrow R ,  & &\text{odd}, \\
	i ( \bar q_{L,R} \nabla_{L,R}^2 M_{R,L} + \bar M_{R,L} \overleftarrow \nabla_{L,R}^2 q_{L,R}) ,								 & & & L \leftrightarrow R ,  & &\text{even}, \\
	( \bar q_{L,R} \nabla_{L,R}^2 M_{R,L} - \bar M_{R,L} \overleftarrow \nabla_{L,R}^2 q_{L,R}) ,								 & & & L \leftrightarrow R ,  & &\text{odd}, \\
	i ( \bar q_{L,R} \overleftarrow{\nabla}_\mu^{L,R} \nabla^\mu_{L,R} M_{R,L} + \bar M_{R,L} \overleftarrow{\nabla}_\mu^{L,R} \nabla^\mu_{L,R} q_{L,R} ) , 	 & & & L \leftrightarrow R ,  & &\text{even}, \\
	( \bar q_{L,R} \overleftarrow{\nabla}_\mu^{L,R} \nabla^\mu_{L,R} M_{R,L} - \bar M_{R,L} \overleftarrow{\nabla}_\mu^{L,R} \nabla^\mu_{L,R} q_{L,R} ) , 	 & & & L \leftrightarrow R ,  & &\text{odd}, \\
	i ( \bar q_{L,R} \overleftarrow{\nabla}_\mu^{L,R} \sigma^{\mu\nu} \nabla^{L,R}_\nu M_{R,L} + \bar M_{R,L} \overleftarrow{\nabla}_\nu^{L,R} \sigma^{\mu\nu} \nabla^{L,R}_\mu q_{L,R}) , 		 & & & L \leftrightarrow R ,  & &\text{odd}, \\
	( \bar q_{L,R} \overleftarrow{\nabla}_\mu^{L,R} \sigma^{\mu\nu} \nabla^{L,R}_\nu M_{R,L} - \bar M_{R,L} \overleftarrow{\nabla}_\nu^{L,R} \sigma^{\mu\nu} \nabla^{L,R}_\mu q_{L,R}) , 			 & & & L \leftrightarrow R ,  & &\text{even} , \\
	\\
	i ( \bar q_{L,R} \sigma^{\mu\nu} F_{\mu\nu}^{L,R} M_{R,L} + \bar M_{R,L} \sigma^{\mu\nu} F_{\mu\nu}^{L,R} q_{L,R}) , 			 & & & L \leftrightarrow R ,  & &\text{even}, \\
	( \bar q_{L,R} \sigma^{\mu\nu} F_{\mu\nu}^{L,R} M_{R,L} - \bar M_{R,L} \sigma^{\mu\nu} F_{\mu\nu}^{L,R} q_{L,R}) ,			 & & & L \leftrightarrow R ,  & &\text{odd}.
\end{align*}
This results in the following list of 19 anti-Hermitian $P$- and $CP$-odd seed operators:
\begin{align}
	\F_1^{(6)} &= \tr[\M \M^\dagger] ( \bar q_L  M_R - \bar q_R M_L + \bar M_L q_R - \bar M_R q_L ) \, , \nn
	\F_2^{(6)} &= ( \bar q_L \M \M^\dagger M_R - \bar q_R \M^\dagger \M M_L + \bar M_L \M^\dagger \M q_R - \bar M_R \M \M^\dagger q_L ) \, , \nn
	\F_3^{(6)} &= ( \bar q_{L} \M \gamma_\mu t^a M_{L} -  \bar q_{R} \M^\dagger \gamma_\mu t^a M_{R} + \bar M_{R} \M \gamma_\mu t^a q_{R} - \bar M_{L} \M^\dagger \gamma_\mu t^a q_{L}) G_a^\mu \, , \nn
	\F_4^{(6)} &= i ( \bar q_{R} \M^\dagger \slashed\nabla_L M_{R} - \bar q_{L} \M \slashed\nabla_R M_{L} + \bar M_{R} \overleftarrow{\slashed\nabla}_L \M q_{R} - \bar M_{L} \overleftarrow{\slashed\nabla}_R \M^\dagger q_{L}) \, , \nn
	\F_5^{(6)} &= i ( \bar q_{R} \overleftarrow{\slashed \nabla}_R \M^\dagger M_{R} - \bar q_{L} \overleftarrow{\slashed\nabla}_L \M M_{L} + \bar M_{R} \M \slashed\nabla_R q_{R} - \bar M_{L} \M^\dagger  \slashed\nabla_L q_{L}) \, , \nn
	\F_6^{(6)} &= ( \bar q_L M_R - \bar q_R M_L + \bar M_L q_R  - \bar M_R q_L) G_\mu^a G^\mu_a \, , \nn
	\F_7^{(6)} &= ( \bar q_L t^a M_R - \bar q_R t^a M_L + \bar M_L t^a q_R - \bar M_R t^a q_L) G_\mu^b G^\mu_c d^{abc} \, , \nn
	\F_8^{(6)} &= ( \bar q_L \sigma^{\mu\nu} t^a M_R - \bar q_R \sigma^{\mu\nu} t^a M_L + \bar M_L \sigma^{\mu\nu} t^a q_R - \bar M_R \sigma^{\mu\nu} t^a q_L) G_\mu^b G_\nu^c f^{abc} \, , \nn
	\F_9^{(6)} &= i ( \bar q_L t^a M_R - \bar q_R t^a M_L + \bar M_R t^a q_L - \bar M_L t^a q_R ) \p_\mu G^\mu_a \, , \nn
	\F_{10}^{(6)} &= ( \bar q_L \sigma^{\mu\nu} t^a M_R - \bar q_R \sigma^{\mu\nu} t^a M_L + \bar M_L \sigma^{\mu\nu} t^a q_R - \bar M_R \sigma^{\mu\nu} t^a q_L ) \p_\mu G_\nu^a \, , \nn
	\F_{11}^{(6)} &= i ( \bar q_L \overleftarrow{\nabla}_\mu^L t^a M_R - \bar q_R \overleftarrow{\nabla}_\mu^R t^a M_L + \bar M_R t^a \nabla_\mu^L q_L - \bar M_L t^a \nabla_\mu^R q_R ) G^\mu_a \, , \nn
	\F_{12}^{(6)} &= ( \bar q_L \overleftarrow{\nabla}_\mu^L \sigma^{\mu\nu} t^a M_R - \bar q_R \overleftarrow{\nabla}_\mu^R \sigma^{\mu\nu} t^a M_L + \bar M_L \sigma^{\mu\nu} t^a  \nabla_\mu^R q_R - \bar M_R \sigma^{\mu\nu} t^a  \nabla_\mu^L q_L ) G_\nu^a \, , \nn
	\F_{13}^{(6)} &= i ( \bar q_L  t^a \nabla^L_\mu M_R - \bar q_R  t^a \nabla^R_\mu M_L + \bar M_R \overleftarrow{\nabla}_\mu^L t^a q_L - \bar M_L \overleftarrow{\nabla}_\mu^R t^a q_R ) G^\mu_a \, , \nn
	\F_{14}^{(6)} &= ( \bar q_L \sigma^{\mu\nu} t^a \nabla^L_\mu M_R - \bar q_R \sigma^{\mu\nu} t^a \nabla^R_\mu M_L + \bar M_L \overleftarrow{\nabla}_\mu^R \sigma^{\mu\nu} t^a q_R - \bar M_R \overleftarrow{\nabla}_\mu^L \sigma^{\mu\nu} t^a q_L) G_\nu^a \, , \nn
	\F_{15}^{(6)} &= ( \bar q_L \overleftarrow \nabla_L^2 M_R - \bar q_R \overleftarrow \nabla_R^2 M_L + \bar M_L \nabla_R^2 q_R - \bar M_R \nabla_L^2 q_L ) \, , \nn
	\F_{16}^{(6)} &= ( \bar q_L \nabla_L^2 M_R - \bar q_R \nabla_R^2 M_L + \bar M_L \overleftarrow \nabla_R^2 q_R - \bar M_R \overleftarrow \nabla_L^2 q_L ) \, , \nn
	\F_{17}^{(6)} &= ( \bar q_L \overleftarrow{\nabla}_\mu^L \nabla^\mu_L M_R - \bar q_R \overleftarrow{\nabla}_\mu^R \nabla^\mu_R M_L + \bar M_L \overleftarrow{\nabla}_\mu^R \nabla^\mu_R q_R - \bar M_R \overleftarrow{\nabla}_\mu^L \nabla^\mu_L q_L ) \, , \nn
	\F_{18}^{(6)} &= i ( \bar q_L \overleftarrow{\nabla}_\mu^L \sigma^{\mu\nu} \nabla_\nu^L M_R - \bar q_R \overleftarrow{\nabla}_\mu^R \sigma^{\mu\nu} \nabla_\nu^R M_L + \bar M_R \overleftarrow{\nabla}_\nu^L \sigma^{\mu\nu} \nabla_\mu^L q_L - \bar M_L \overleftarrow{\nabla}_\nu^R \sigma^{\mu\nu} \nabla_\mu^R q_R ) \, , \nn
	\F_{19}^{(6)} &= ( \bar q_L \sigma^{\mu\nu} F_{\mu\nu}^L M_R - \bar q_R \sigma^{\mu\nu} F_{\mu\nu}^R M_L + \bar M_L \sigma^{\mu\nu} F_{\mu\nu}^R q_R - \bar M_R \sigma^{\mu\nu} F_{\mu\nu}^L q_L ) \, .
\end{align}

\paragraph{$\hat J_\mu^a$ operators}

At dimension 4, there is the operator
\begin{align*}
	G^\mu_a \hat J_\mu^a \, ,
\end{align*}
which, however, is $P$-even.
At dimension 5, no operators can be constructed. At dimension 6, we find the following list of seed operators.
\begin{align*}
	\text{seed operator}& \quad & &P \quad & &CP \\[0.1cm]
	\hline \\[-0.4cm]
	\tr[\M \M^\dagger] G^\mu_a \hat J_\mu^a , 			 & & &	\text{even},  & &\text{even}, \\
	\\
	c^a \hat J_\mu^b \hat J^\mu_c f^{abc} , 				 & & &	\text{even},  & &\text{even}, \\
	\\
	G_\mu^a G_\nu^b G_\lambda^c \hat J_\sigma^d \epsilon^{\mu\nu\lambda\sigma} d^{ade} f^{bce} ,	 & & &\text{odd},  & &\text{even}, \\
	G^\mu_a G_\mu^a G^\nu_b \hat J_\nu^b , \quad
	G^\mu_a G_\mu^b G^\nu_a \hat J_\nu^b , \quad
	G^\mu_a G_\mu^b G^\nu_c \hat J_\nu^d d^{ace} d^{bde} , 	 & & &\text{even},  & &\text{even}, \\
	G^\mu_a G_\mu^b G^\nu_c \hat J_\nu^d d^{ade} f^{bce} ,	 & & &\text{even},  & &\text{odd}, \\
	\\
	(\p_\mu G_\nu^a) G_\lambda^b \hat J_\sigma^c  \epsilon^{\mu\nu\lambda\sigma} f^{abc} ,			 & & &\text{odd},  & &\text{odd}, \\
	(\p_\mu G_\nu^a) G_\lambda^b \hat J_\sigma^c  \epsilon^{\mu\nu\lambda\sigma} d^{abc} ,			 & & &\text{odd},  & &\text{even}, \\
	(\p^\mu G_\mu^a) G^\nu_b \hat J_\nu^c f^{abc} , \quad (\p^\mu G^\nu_a) G_\mu^b \hat J_\nu^c f^{abc} , \quad (\p^\mu G^\nu_a) G_\nu^b \hat J_\mu^c f^{abc} ,	 & & &\text{even},  & &\text{even}, \\
	(\p^\mu G_\mu^a) G^\nu_b \hat J_\nu^c d^{abc} , \quad (\p^\mu G^\nu_a) G_\mu^b \hat J_\nu^c d^{abc} , \quad (\p^\mu G^\nu_a) G_\nu^b \hat J_\mu^c d^{abc} , 	 & & &\text{even},  & &\text{odd}, \\
	G_\mu^a G_\nu^b (\p_\lambda \hat J_\sigma^c)  \epsilon^{\mu\nu\lambda\sigma} f^{abc} ,			 & & &\text{odd},  & &\text{odd}, \\
	G^\mu_a G^\nu_b (\p_\mu \hat J_\nu^c) f^{abc} , 	 & & &\text{even},  & &\text{even}, \\
	G^\mu_a G^\nu_b (\p_\mu \hat J_\nu^c) d^{abc} , \quad G^\mu_a G_\mu^b (\p^\nu \hat J_\nu^c) d^{abc} , 	 & & &\text{even},  & &\text{odd}, \\
	\\
	(\Box G^\mu_a) \hat J_\mu^a , \quad (\p^\mu \p^\nu G_\mu^a) \hat J_\nu^a  , 	 & & &\text{even},  & &\text{even}, \\
	G^\mu_a (\Box \hat J_\mu^a) , \quad G_\mu^a (\p^\mu \p^\nu \hat J_\nu^a)  , 	 & & &\text{even},  & &\text{even}, \\
	(\p_\mu G_\nu^a) (\p_\lambda \hat J_\sigma^a)  \epsilon^{\mu\nu\lambda\sigma} ,					 & & &\text{odd},  & &\text{odd}, \\
	(\p^\mu G_\mu^a) (\p^\nu \hat J_\nu^a) , \quad (\p^\mu G^\nu_a) (\p_\mu \hat J_\nu^a) , \quad (\p^\mu G^\nu_a) (\p_\nu \hat J_\mu^a)  , 	 & & &\text{even},  & &\text{even}, \\
	\\
	\bar q_L \gamma^\mu t^a q_L \hat J_\mu^a ,	\quad
	\bar q_R \gamma^\mu t^a q_R \hat J_\mu^a ,	 & & &L \leftrightarrow R,  & &\text{even} .
\end{align*}
Dropping all operators that are not both $P$- and $CP$-odd, we are left with three anti-Hermitian seed operators:
\begin{align}
	\label{eq:PureGaugeSeedOperators}
	\F_{20}^{(6)} &= (\p_\mu G_\nu^a) G_\lambda^b \hat J_\sigma^c  \epsilon^{\mu\nu\lambda\sigma} f^{abc} \, , \nn
	\F_{21}^{(6)} &= G_\mu^a G_\nu^b (\p_\lambda \hat J_\sigma^c)  \epsilon^{\mu\nu\lambda\sigma} f^{abc} \, , \nn
	\F_{22}^{(6)} &= (\p_\mu G_\nu^a) (\p_\lambda \hat J_\sigma^a)  \epsilon^{\mu\nu\lambda\sigma} \, .
\end{align}

\subsection{Nuisance operators}
\label{sec:ConstructionNuisanceOperators}

After acting with the $\hat W$ operator on the seed operators $\F_i$, we obtain the list of nuisance operators. Furthermore, we perform a basis change: we write as many nuisance operators as possible in a manifestly gauge-invariant form and as total derivatives. In order to write the operators in a more compact form, we fix the spurion and external fields to their physical value, $\M, \M^\dagger \mapsto \M$, $l_\mu, r_\mu \mapsto e Q A_\mu$, and write everything in the parity basis.

There is one nuisance operator at dimension four:
\begin{align}
	\tilde\N_1^{(4)} &= i ( \bar q_E \gamma_5 q + \bar q \gamma_5 q_E ) \, .
\end{align}
At dimension six, we find the following nuisance operators:
\begin{align*}
	\label{eq:IntermediateNuisanceOps}
	\tilde\N_{1}^{(6)} &= G_{\mu\nu}^a \left(\p_\lambda \Big( D^\rho G_{\rho\sigma}^a + g \bar q t^a \gamma_\sigma q - g f^{abc} (\p_\sigma \bar c^b) c^c \Big) \right)  \epsilon^{\mu\nu\lambda\sigma} \, , \\
	\tilde\N_{2}^{(6)} &= i ( \bar q_E \gamma_5 q + \bar q \gamma_5 q_E) G_\mu^a G^\mu_a \, , \\
	\tilde\N_{3}^{(6)} &= i ( \bar q_E \gamma_5 t^a q + \bar q \gamma_5 t^a q_E) G_\mu^b G^\mu_c d^{abc} \, , \\
	\tilde\N_{4}^{(6)} &= i ( \bar q_E \tilde\sigma^{\mu\nu} t^a q + \bar q \tilde\sigma^{\mu\nu} t^a q_E) G_{\mu\nu}^a \, , \\
	\tilde\N_{5}^{(6)} &= ( \bar q_E \gamma_5 t^a q - \bar q \gamma_5 t^a q_E ) \p_\mu G^\mu_a \, , \\
	\tilde\N_{6}^{(6)} &= ( \bar q_E \gamma_5 t^a D_\mu q - \bar q \overleftarrow{D}_\mu \gamma_5 t^a q_E ) G^\mu_a \, , \\
	\tilde\N_{7}^{(6)} &= i ( \bar q_E \tilde\sigma^{\mu\nu} t^a q + \bar q \tilde\sigma^{\mu\nu} t^a q_E ) \p_\mu G_\nu^a \, , \\
	\tilde\N_{8}^{(6)} &= i ( \bar q_E \tilde\sigma^{\mu\nu} t^a  D_\mu q + \bar q \overleftarrow{D}_\mu \tilde\sigma^{\mu\nu} t^a q_E ) G_\nu^a \, , \\
	\tilde\N_{9}^{(6)} &= i ( \bar q_E \gamma_5 D^2 q + \bar q \overleftarrow D^2 \gamma_5 q_E ) \, , \\
	\tilde\N_{10}^{(6)} &= i e ( \bar q_E \tilde\sigma^{\mu\nu} Q q + \bar q \tilde\sigma^{\mu\nu} Q q_E ) F_{\mu\nu} \, , \\
	\tilde\N_{11}^{(6)} &= ( \bar q_E \M \tilde\gamma^\mu D_\mu q + \bar q \overleftarrow{D}_\mu \tilde\gamma^\mu \M q_E ) \, , \\
	\tilde\N_{12}^{(6)} &= i ( \bar q_E \M \tilde\gamma_\mu t^a q - \bar q \M \tilde\gamma_\mu t^a q_E ) G_a^\mu \, , \\
	\tilde\N_{13}^{(6)} &= \tr[\M^2] i ( \bar q_E \gamma_5 q + \bar q \gamma_5  q_E ) \, , \\
	\tilde\N_{14}^{(6)} &= i ( \bar q_E \M^2 \gamma_5 q + \bar q \M^2 \gamma_5 q_E ) \, , \\
	\tilde\N_{15}^{(6)} &= \p_\lambda \left( G_{\mu\nu}^a \Big( D^\rho G_{\rho\sigma}^a + g \bar q t^a \gamma_\sigma q \Big) \right) \epsilon^{\mu\nu\lambda\sigma} \, , \\
	\tilde\N_{16}^{(6)} &= \p_\lambda \left( (\p_\mu G_\nu^a) \Big( D^\rho G_{\rho\sigma}^a + g \bar q t^a \gamma_\sigma q - g f^{abc} (\p_\sigma \bar c^b) c^c \Big) \right) \epsilon^{\mu\nu\lambda\sigma} \, , \\
	\tilde\N_{17}^{(6)} &= \p_\mu \Big( ( \bar q_E \gamma_5 t^a q - \bar q \gamma_5  t^a q_E ) G^\mu_a \Big) \, , \\
	\tilde\N_{18}^{(6)} &= i \p_\mu \Big( ( \bar q_E \tilde\sigma^{\mu\nu} t^a q + \bar q \tilde\sigma^{\mu\nu} t^a q_E ) G_\nu^a \Big) \, , \\
	\tilde\N_{19}^{(6)} &= i \p_\mu \Big( \bar q_E \gamma_5 D^\mu q + \bar q \overleftarrow{D}^\mu \gamma_5 q_E \Big) \, , \\
	\tilde\N_{20}^{(6)} &= \p_\mu ( \bar q_E \tilde\sigma^{\mu\nu} D_\nu q - \bar q \overleftarrow{D}_\nu \tilde\sigma^{\mu\nu} q_E ) \, , \\
	\tilde\N_{21}^{(6)} &= \p_\mu \Big( \bar q_E \M \tilde\gamma^\mu q + \bar q \M \tilde\gamma^\mu q_E \Big) \, , \\
	\tilde\N_{22}^{(6)} &= i \Box \Big( \bar q_E \gamma_5 q + \bar q \gamma_5 q_E \Big) \, .
	\mytag
\end{align*}

\subsection{Redundancies}
\label{sec:NuisanceRedundancies}

The nuisance operators are constructed as BRST variations of a complete set of linearly independent seed operators. However, it turns out that the nuisance operators themselves are redundant. This can be understood as follows: acting with the BRST operator on the seed operators replaces the BRST sources by EOM fields, i.e., this operation reduces the degrees of freedom. Therefore, linear independence of the seed operators does not imply linear independence of the resulting nuisance operators. The remaining redundancies are most easily identified by considering the vertex rules for all the operators, see Sect.~\ref{sec:VertexRules}, which leave two linear combinations of nuisance operators undetermined. An explicit calculation then confirms the following linear relations:
\begin{align}
	0 &= \tilde \N_6^{(6)} + \tilde \N_8^{(6)} - \tilde \N_{12}^{(6)} \, , \nn
	0 &= g \tilde \N_4^{(6)} - 2 \tilde \N_9^{(6)} + \tilde \N_{10}^{(6)} - 2 \tilde \N_{11}^{(6)} + \tilde \N_{19}^{(6)} + \tilde \N_{20}^{(6)} + \tilde \N_{21}^{(6)} \, .
\end{align}
This allows us to drop two nuisance operators from the set~\eqref{eq:IntermediateNuisanceOps}---we choose to drop the nuisance operators $\tilde\N_8^{(6)}$ and $\tilde\N_9^{(6)}$.

In a last step, we remove redundancies from the list of operators $\tilde\O$ in App.~\ref{sec:IntermediateSummaryOperators}: there are linear combinations that are identical to nuisance operators. Removing these redundancies leads to a minimal set of class-I operators $\O$, i.e., gauge-invariant operators that do not vanish by the EOM.

At dimension four, we find the linear relation
\begin{align}
	\tilde\N_1^{(4)} = \tilde\O_2^{2q,(4)} - 2 \tilde\O_1^{2q,(4)} .
\end{align}
At dimension six, the following relations hold (disregarding evanescent structures):
\begin{align}
	g \tilde\N_{4}^{(6)} + \tilde\N_{10}^{(6)} &= 2 \tilde\O_{8}^{2q,(6)} + 2 \tilde\O_{9}^{2q,(6)} + 2 \tilde\O_{10}^{2q,(6)} - \tilde\O_{11}^{2q,(6)} - \tilde\O_{12}^{2q,(6)} - \tilde\O_{13}^{2q,(6)} \, , \nn
	\tilde\N_{10}^{(6)} &= -\tilde\O_{14}^{2q,(6)} + 2 \tilde\O_{15}^{2q,(6)} - \tilde\O_{16}^{2q,(6)} \, , \nn
	\tilde\N_{11}^{(6)} &= -\tilde\O_{3}^{2q,(6)} - \tilde\O_{5}^{2q,(6)} - \tilde\O_{6}^{2q,(6)} + \tilde\O_{7}^{2q,(6)} + \tilde\O_{8}^{2q,(6)} \, , \nn
	\tilde\N_{13}^{(6)} &= \tilde\O_{4}^{2q,(6)} - 2 \tilde\O_{2}^{2q,(6)} \, , \nn
	\tilde\N_{14}^{(6)} &= \tilde\O_{3}^{2q,(6)} - 2 \tilde\O_{1}^{2q,(6)} \, , \nn
	\tilde\N_{15}^{(6)} &= 4 \tilde\O_{3}^{G,(6)} + \tilde\O_{13}^{2q,(6)} - \tilde\O_{16}^{2q,(6)} \, , \nn
	\tilde\N_{19}^{(6)} &= - \tilde\O_{5}^{2q,(6)} - \tilde\O_{10}^{2q,(6)} + \tilde\O_{12}^{2q,(6)} \, , \nn
	\tilde\N_{20}^{(6)} &= - \tilde\O_{6}^{2q,(6)} - \tilde\O_{10}^{2q,(6)} + \tilde\O_{11}^{2q,(6)} + \tilde\O_{13}^{2q,(6)} \, , \nn
	\tilde\N_{21}^{(6)} &= - 2 \tilde\O_{3}^{2q,(6)} - \tilde\O_{5}^{2q,(6)}  - \tilde\O_{6}^{2q,(6)} \, , \nn
	\tilde\N_{22}^{(6)} &= \tilde\O_{12}^{2q,(6)} - 2 \tilde\O_{5}^{2q,(6)} \, .
\end{align}
Finally, we arrive at an operator basis that is free of redundancies, presented in Sect.~\ref{sec:Operators}.
 
For the determination of the mixing structure, it is useful to express the intermediate redundant set of operators in terms of the final basis. At dimension four, the relations read:
\begin{align}
	\tilde\O_1^{G,(4)} &= \O_1^{(4)} \, , \quad
	\tilde\O_1^{2q,(4)} = \frac{1}{2} \left( \O_2^{(4)} - \N_1^{(4)} \right) \, , \quad
	\tilde\O_2^{2q,(4)} = \O_2^{(4)} \, , \quad
	\tilde\N_1^{(4)}  = \N_1^{(4)} \, .
\end{align}
At dimension five, there is only one operator:
\begin{align}
	\tilde\O_1^{2q,(5)} &= \O_1^{(5)} \, .
\end{align}
At dimension six, the gauge-invariant operators are given by
\begin{align}
	\tilde\O_1^{G,(6)} &= \O_4^{(6)} \, , \qquad
	\tilde\O_2^{G,(6)} = \O_9^{(6)} \, , \qquad
	\tilde\O_3^{G,(6)} = \O_5^{(6)} \, , \qquad
	\tilde\O_4^{G,(6)} = \frac{1}{g} \O_1^{(6)} \, , \nn
	\tilde\O_1^{2q,(6)} &= \frac{1}{2} \left( \O_6^{(6)} - \N_4^{(6)} + \frac{1}{N_f} \left( \O_7^{(6)} - \N_5^{(6)} \right) \right) \, , \qquad
	\tilde\O_2^{2q,(6)} = \frac{1}{2} \left( \O_7^{(6)} - \N_5^{(6)} \right) \, , \nn
	\tilde\O_3^{2q,(6)} &= \O_6^{(6)} + \frac{1}{N_f} \O_7^{(6)} , \qquad
	\tilde\O_4^{2q,(6)} = \O_7^{(6)} , \qquad
	\tilde\O_5^{2q,(6)} = \frac{1}{2} \left( \O_{10}^{(6)} - \N_{10}^{(6)} \right) \, , \nn
	\tilde\O_6^{2q,(6)} &= -2 \O_6^{(6)} - \frac{2}{N_f} \O_7^{(6)} - \frac{1}{2} \O_{10}^{(6)} - \N_{9}^{(6)} + \frac{1}{2} \N_{10}^{(6)} \, , \nn
	\tilde\O_7^{2q,(6)} &=  \O_2^{(6)} + \O_3^{(6)} - \O_6^{(6)} - \frac{1}{N_f} \O_7^{(6)} + \N_3^{(6)} - \N_{9}^{(6)} \, , \nn
	\tilde\O_8^{2q,(6)} &= - \O_2^{(6)} - \O_3^{(6)} \, , \nn
	\tilde\O_9^{2q,(6)} &= \O_2^{(6)} + \O_3^{(6)} - \O_6^{(6)} - \frac{1}{N_f} \O_7^{(6)} + \frac{1}{2} \left( \N_1^{(6)} + \N_2^{(6)} + \N_6^{(6)} + \N_7^{(6)} - \N_{9}^{(6)} \right) \, , \nn
	\tilde\O_{10}^{2q,(6)} &= \frac{1}{2} \O_{10}^{(6)} - \N_6^{(6)} + \frac{1}{2} \N_{10}^{(6)} \, , \nn
	\tilde\O_{11}^{2q,(6)} &= 4 \O_5^{(6)} - 2 \O_6^{(6)} - \frac{2}{N_f} \O_7^{(6)} - 2 \O_8^{(6)} - \N_6^{(6)} + \N_7^{(6)} - \N_8^{(6)} - \N_9^{(6)} + \N_{10}^{(6)} \, , \nn
	\tilde\O_{12}^{2q,(6)} &= \O_{10}^{(6)} \, , \qquad
	\tilde\O_{13}^{2q,(6)} = - 4 \O_5^{(6)} + 2 \O_8^{(6)} + \N_8^{(6)} \, , \qquad
	\tilde\O_{14}^{2q,(6)} = 2 \O_3^{(6)} \, , \nn
	\tilde\O_{15}^{2q,(6)} &= \O_3^{(6)} + \O_8^{(6)} + \frac{1}{2} \N_2^{(6)} \, , \qquad
	\tilde\O_{16}^{2q,(6)} = 2 \O_8^{(6)} \, .
\end{align}
For the nuisance operators, we have the relations
\begin{alignat}{4}
	\tilde\N_{1}^{(6)} &= \N_{11}^{(6)} \, , \qquad &
	\tilde\N_{2}^{(6)} &= \frac{1}{g^2} \N_{12}^{(6)} \, , \qquad &
	\tilde\N_{3}^{(6)} &= \frac{1}{g^2} \N_{13}^{(6)} \, , \qquad &
	\tilde\N_{4}^{(6)} &= \frac{1}{g} \N_1^{(6)} \, , \nn
	\tilde\N_{5}^{(6)} &= \frac{1}{g} \N_{14}^{(6)} \, , \qquad &
	\tilde\N_{6}^{(6)} &= \frac{1}{g} \N_{15}^{(6)} \, , \qquad &
	\tilde\N_{7}^{(6)} &= \frac{1}{g} \N_{16}^{(6)} \, , \qquad &
	\tilde\N_{8}^{(6)} &= \frac{1}{g} \left( - \N_{15}^{(6)} + \N_{17}^{(6)} \right) \, , \nn
	\tilde\N_{9}^{(6)} &= \frac{1}{2} \left( \N_1^{(6)} + \N_2^{(6)} - 2 \N_3^{(6)} + \N_6^{(6)} + \N_7^{(6)} + \N_9^{(6)} \right) \, , \span\span\span\span \qquad &
	\tilde\N_{10}^{(6)} &= \N_2^{(6)} \, , \nn
	\tilde\N_{11}^{(6)} &= \N_3^{(6)} \, , \qquad &
	\tilde\N_{12}^{(6)} &= \frac{1}{g} \N_{17}^{(6)} \, , \qquad &
	\tilde\N_{13}^{(6)} &= \N_5^{(6)} \, , \qquad &
	\tilde\N_{14}^{(6)} &= \N_4^{(6)} + \frac{1}{N_f} \N_5^{(6)} \, , \nn
	\tilde\N_{15}^{(6)} &= \N_8^{(6)} \, , \qquad &
	\tilde\N_{16}^{(6)} &= \N_{18}^{(6)} \, , \qquad &
	\tilde\N_{17}^{(6)} &= \frac{1}{g} \N_{19}^{(6)} \, , \qquad &
	\tilde\N_{18}^{(6)} &= \frac{1}{g} \N_{20}^{(6)} \, , \nn
	\tilde\N_{19}^{(6)} &= \N_6^{(6)} \, , \qquad &
	\tilde\N_{20}^{(6)} &= \N_7^{(6)} \, , \qquad &
	\tilde\N_{21}^{(6)} &= \N_{9}^{(6)} \, , \qquad &
	\tilde\N_{22}^{(6)} &= \N_{10}^{(6)} \, .
\end{alignat}


\section{Mixing with evanescent operators}
\label{sec:Evanescent}

\subsection{Generalities}

As is well known, dimensional regularization leads to the appearance of evanescent operators~\cite{Collins:1984xc,Buras:1989xd,Dugan:1990df,Herrlich:1994kh}. These operators are present in $D$ dimensions, but they vanish for $D=4$. If bare evanescent operators are inserted into loop diagrams, the combination of poles in $\epsilon$ with the evanescent structure can lead to finite contributions. Evanescent operators can be renormalized by finite counterterms so that the renormalized evanescent operators have vanishing matrix elements. As shown in~\cite{Dugan:1990df,Herrlich:1994kh}, the renormalized evanescent operators do not mix into physical operators. However, the counterterms affect the calculation of the anomalous dimension. Since the bare evanescent operators are ambiguous, their choice affects the anomalous dimension matrix of the physical operators and their definition is part of the scheme.

In dimensional regularization, the evanescent operators are present both in the \msbar{} scheme and in the MOM scheme. Let us denote the relation between bare and renormalized operators as
\begin{align}
	\begin{pmatrix} \O^\text{\msbar} \\ \E^\text{\msbar} \end{pmatrix} = \begin{pmatrix} \tilde Z^\text{\msbar}_{\O\O} & \tilde Z^\text{\msbar}_{\O\E} \\ \tilde Z^\text{\msbar}_{\E\O} & \tilde Z^\text{\msbar}_{\E\E} \end{pmatrix} \begin{pmatrix} \O^{(0)} \\ \E^{(0)} \end{pmatrix} \, , \quad 
	\begin{pmatrix} \O^\text{MOM} \\ \E^\text{MOM} \end{pmatrix} = \begin{pmatrix} \tilde Z^\text{MOM}_{\O\O} & \tilde Z^\text{MOM}_{\O\E} \\ \tilde Z^\text{MOM}_{\E\O} & \tilde Z^\text{MOM}_{\E\E} \end{pmatrix} \begin{pmatrix} \O^{(0)} \\ \E^{(0)} \end{pmatrix} \, , \quad 
\end{align}
where $\tilde Z_{ij}$ denote the entries of the inverse mixing matrices $Z^{-1}$, fulfilling $\tilde Z_{ij} = \delta_{ij} + \O(\alpha_s)$. The renormalized \msbar{} evanescent operators are defined to have vanishing matrix elements,
\begin{align}
	\< \E^\text{\msbar} \> = 0 \, .
\end{align}
This is achieved by adjusting the counterterms $\tilde Z_{\E\O}$, which at $\O(\alpha_s)$ are finite, i.e., $\O(\varepsilon^0)$. The minimal scheme defines $\tilde Z_{\O\O}^\text{\msbar}-1$, $\tilde Z_{\O\E}^\text{\msbar}$, and $\tilde Z_{\E\E}^\text{\msbar}-1$ to only contain poles in $\varepsilon$. The exact form of the evanescent operators $\E^\text{\msbar}$ is part of the scheme definition: suppose that we are choosing a different basis of bare operators
\begin{align}
	\begin{pmatrix} \O^{(0)\prime} \\ \E^{(0)\prime} \end{pmatrix} = \begin{pmatrix} 1 & 0 \\ a \varepsilon & 1 \end{pmatrix} \begin{pmatrix} \O^{(0)} \\ \E^{(0)} \end{pmatrix} \, ,
\end{align}
where with some arbitrary constant $a$ the operator $\E^{(0)\prime}$ still is evanescent, then the minimal scheme in the new basis reads
\begin{align}
	\begin{pmatrix} \O^{\text{\msbar}\prime} \\ \E^{\text{\msbar}\prime} \end{pmatrix} = \begin{pmatrix} \tilde Z^\text{\msbar}_{\O\O} & \tilde Z^\text{\msbar}_{\O\E} \\ \tilde Z^\text{\msbar}_{\E\O} + a \varepsilon \big( \tilde Z^\text{\msbar}_{\O\O} - \tilde Z^\text{\msbar}_{\E\E} \big) & \tilde Z^\text{\msbar}_{\E\E} \end{pmatrix} \begin{pmatrix} \O^{(0)\prime} \\ \E^{(0)\prime} \end{pmatrix} = \begin{pmatrix} \tilde Z^\text{\msbar}_{\O\O} + a \varepsilon \tilde Z^\text{\msbar}_{\O\E} & \tilde Z^\text{\msbar}_{\O\E} \\ \tilde Z^\text{\msbar}_{\E\O} + a \varepsilon \tilde Z^\text{\msbar}_{\O\O} & \tilde Z^\text{\msbar}_{\E\E} \end{pmatrix} \begin{pmatrix} \O^{(0)} \\ \E^{(0)} \end{pmatrix} \, ,
\end{align}
i.e., the alternative operator $\O^{\text{\msbar}\prime}$ differs from $\O^\text{\msbar}$ by a finite renormalization (and no longer looks minimally subtracted in the original basis), while the new renormalized evanescent still has vanishing matrix elements. Similarly, the choice of evanescent operators affects the anomalous-dimension matrix of the physical operators at two loops~\cite{Dugan:1990df,Herrlich:1994kh}.

Also in the MOM scheme, the renormalized evanescent operators are defined to have vanishing matrix element,
\begin{align}
	\label{eq:MOMEvan}
	 \< \E^\text{MOM} \> = 0 \, .
\end{align}
Therefore, the set of regularization-independent operators can be identified as the renormalized physical operators only, $\{ \O^\text{RI} \} = \{ \O^\text{MOM} \}$.

The conversion between the \msbar{} and MOM schemes is given by
\begin{align}
	\begin{pmatrix} \O^\text{\msbar} \\ \E^\text{\msbar} \end{pmatrix} = \begin{pmatrix} \tilde Z^\text{\msbar}_{\O\O} & \tilde Z^\text{\msbar}_{\O\E} \\ \tilde Z^\text{\msbar}_{\E\O} & \tilde Z^\text{\msbar}_{\E\E} \end{pmatrix} \begin{pmatrix} Z^\text{MOM}_{\O\O} & Z^\text{MOM}_{\O\E} \\ Z^\text{MOM}_{\E\O} & Z^\text{MOM}_{\E\E} \end{pmatrix}  \begin{pmatrix} \O^\text{MOM} \\ \E^\text{MOM} \end{pmatrix} \, ,
\end{align}
where $Z_{ij}$ are the entries of the mixing matrix $Z$.
In particular, the physical \msbar{} operators are given by
\begin{align}
	\O^\text{\msbar} = \left( \tilde Z^\text{\msbar}_{\O\O} Z^\text{MOM}_{\O\O} + \tilde Z^\text{\msbar}_{\O\E} Z^\text{MOM}_{\E\O}  \right) \O^\text{MOM} + \left( \tilde Z^\text{\msbar}_{\O\O} Z^\text{MOM}_{\O\E} + \tilde Z^\text{\msbar}_{\O\E} Z^\text{MOM}_{\E\E} \right) \E^\text{MOM} \, .
\end{align}
Due to~\eqref{eq:MOMEvan}, we only need to know the coefficient of $\O^\text{MOM}$ in order to determine $\< \O^\text{\msbar} \>$. Furthermore, in the matching at one loop, we have
\begin{align}
	\< \O^\text{\msbar} \> = \left( 1 - \Delta^\text{\msbar}_{\O\O} + \Delta^\text{MOM}_{\O\O} + \O(\alpha_s^2) \right) \< \O^\text{MOM} \> \, ,
\end{align}
where $Z = 1 + \Delta$, i.e.,~\eqref{eq:physicalME} is unaffected by evanescent operators.

Finally, we determine the conversion matrix $C = 1 - \Delta^\text{\msbar} + \Delta^\text{MOM}$ by imposing renormalization conditions
\begin{align}
	\text{const.} \stackrel{!}{=} R[ \O^\text{MOM} ]^\text{1-loop} = \tilde Z_{\O\O}^\text{MOM} R[ \O^{(0)} ]^\text{1-loop} + \tilde Z_{\O\E}^\text{MOM} R[ \E^{(0)} ]^\text{tree} \, ,
\end{align}
where
\begin{align}
	\tilde Z_{\O\O}^\text{MOM} = 1 - \frac{\alpha_s}{4\pi} \left( \frac{z_{\O\O}}{\varepsilon} + c_{\O\O} \right) + \O(\alpha_s^2) \, , \quad
	\tilde Z_{\O\E}^\text{MOM} = - \frac{\alpha_s}{4\pi} \left( \frac{z_{\O\E}}{\varepsilon} + c_{\O\E} \right) + \O(\alpha_s^2) \, .
\end{align}
Since in general $\< \E \>^\text{tree} = \O(\varepsilon)$, one expects that either the evanescent counterterms $z_{\O\E}$ need to be determined separately, or the number of conditions to be imposed must match the number of physical plus evanescent operators. However, this can be avoided if the renormalization conditions and the set of evanescent operators is chosen so that $R[\E^{(0)}]^\text{tree} = 0$ instead of $R[\E^{(0)}]^\text{tree} = \O(\varepsilon)$. This allows us to consider only as many conditions as physical operators are present and to solve the system for the coefficients $c_{\O\O}$ that determine the conversion matrix.

In the following, we define the set of relevant evanescent operators and show that their tree-level insertions into the renormalization conditions vanish identically. Throughout, we use the HV scheme~\cite{tHooft:1972tcz,Breitenlohner:1977hr} 
to deal with the Levi-Civita symbol and the Dirac matrices in $D \neq 4$ spacetime dimensions. 
In defining our scheme, it is useful to divide the operators in our basis (up to and including dimension six)  in two categories:  (i) purely bosonic  operators ${\cal O}_B$, involving 
one  gluonic dual field strength (such as gCEDM);   (ii)  fermionic operators  ${\cal O}_F$ containing a quark bilinear with 
a Dirac structure involving one $\gamma_5$,  and  possibly gluonic structures, the external electromagnetic field, and derivatives 
(such as the pseudoscalar density and the  qCEDM).

\subsection{Definition of evanescent operators}

The bosonic operators at dimension four and six can be written schematically as
\begin{align}
	\O_B^{(4)} &= \epsilon_{\mu \nu \lambda \sigma}  \, O_B^{\mu \nu \lambda \sigma} \, , \nn
	\O_{B,\M}^{(6)} &= \epsilon_{\mu \nu \lambda \sigma} \,\tr[\M^2] \,O_B^{\mu \nu \lambda \sigma} \, , \nn
	\O_B^{(6)} &= \epsilon_{\mu \nu \lambda \sigma}   \, g_{\alpha \beta}  \,O_B^{\mu \nu \lambda \sigma  \alpha \beta} \, ,
\end{align}
where $O_B$ are Lorentz tensors of rank four and six, respectively,  built out of $\partial_\mu$, $G_\mu^a$, and $A_\mu$. In the HV scheme, the indices of the Levi-Civita symbol are restricted to $D=4$ dimensions. In addition, external momenta and polarization vectors in $S$-matrix elements are considered to be objects in $D=4$ dimensions. However, the restriction to $D=4$ of external momenta and polarizations can be performed after performing the loop calculation. As we are considering only QCD corrections, all vertices and propagators in loops are continued to $D$ dimensions. Therefore, any metric tensor that appears in a loop calculation (either from propagators, tensor reductions of loop integrals, or the Dirac trace of closed fermion loops) is $D$-dimensional. In particular, an evanescent structure
\begin{align}
	\E_B^{(6)} &= \epsilon_{\mu \nu \lambda \sigma}   \, g_{\hat\alpha \hat\beta}  \,O_B^{\mu \nu \lambda \sigma  \hat\alpha \hat\beta} \, ,
\end{align}
where the indices $\hat\alpha$ and $\hat\beta$ are restricted to $-2\varepsilon$ dimensions, cannot be independently generated in QCD with single insertions of the gCEDM operator. This implies that the only evanescent bosonic operator appears at dimension six due to the Schouten identity~\eqref{eq:SchoutenIdentity}:
\begin{align}
	\label{eq:SchoutenEvanescent}
	{\cal E}_S^{(6)} &= \partial_{\alpha} \partial^{\alpha} \, \tr[G_{\mu\nu} \widetilde G^{\mu\nu}] - 4  \,
	 \partial_{\alpha} \partial^{\mu} \, \tr[G^{\alpha \nu} \widetilde G_{\mu \nu}] \, ,
\end{align}
where the indices $\mu$, $\nu$ are in 4, and $\alpha$ in $D$ dimensions.

For the fermionic operators, we make use of the HV definition~\eqref{eq:g5} for $\gamma_5$
and the fact that in any spacetime dimension $D$
a string of $k$  Dirac matrices can be decomposed as a linear combination of 
the fully antisymmetric products
\begin{equation}
	\Gamma^{(n)}_{\alpha_1 ... \alpha_n} =  \frac{1}{n!}  \, \gamma_{[\alpha_1} ... \gamma_{\alpha_n]}~.
\end{equation}
In particular,  the product of $n$ gamma matrices is expressed as a combination of 
$\Gamma^{(n)}$ with $n=k, k-2, ...$, with the remaining Lorentz indices provided by 
appropriate powers of the $D$-dimensional metric tensor $g_{\mu \nu}$~\cite{Kennedy:1981kp}.
Note that all structures involving $\Gamma^{(n)}$ with $n>4$ are evanescent. 

The fermionic operators at dimension four, five, and six can be written schematically as
\begin{align}
	\label{eq:GenericFermionicOperator}
	\O_F^{(k,n)}  &=  \epsilon^{\mu_1 \mu_2 \mu_3 \mu_4} \,\bar{q} \, O_F^{\nu_1 \ldots \nu_k}  \Gamma_{\lambda_1 \ldots \lambda_{n}}^{(n)} \,  q \, g_{\mu_a \nu_b} {g_{\mu_c}}^{\lambda_d} g_{\nu_e \nu_f} {g_{\nu_g}}^{\lambda_h} \, ,   \nn
	k&=0,1,2,3 \, , \quad n = 4-k, 6-k, \ldots, 4+k \, ,
\end{align}
where $O_F$ is built out of $\partial_\mu$, $G_\mu^a$, $A_\mu$, color structures, and the charge and mass matrices. The highly symbolic product of metric tensors is needed 
to ensure the final result is a Lorentz scalar (note that $\mu_a$ and $\lambda_b$ indices 
cannot be contracted among themselves due to the antisymmetry of the Levi-Civita symbol and $\Gamma^{(n)}$).
As in the case of the bosonic operators, QCD loops only generate metric tensors in $D$ dimensions, since we only consider single insertions of the gCEDM. Therefore, the only possible evanescent operators arise either due to the evanescent structures $\Gamma^{(n)}$ with $n>4$, or due to the Schouten identity. An explicit list (with generic operators $O_F$) is given by
\begin{align}
	\E_F^{(1,5)} &= \epsilon^{\mu_1\ldots\mu_4} \, \bar q \, \Gamma^{(5)}_{\mu_1\ldots\mu_5} O_F^{\mu_5} q \, , \nn
	\E_F^{(2,4)} &= \epsilon^{\mu_1\ldots\mu_4} \, \bar q \, \Gamma^{(4)}_{\mu_1\ldots\mu_3\mu_5} O_F^{\mu_5\mu_6} q \, g_{\mu_4\mu_6} - \frac{1}{4}  \epsilon^{\mu_1\ldots\mu_4} \, \bar q \, \Gamma^{(4)}_{\mu_1\ldots\mu_4} O_F^{\mu_5\mu_6} q \, g_{\mu_5\mu_6} \, , \nn
	\E_F^{(2,6)} &= \epsilon^{\mu_1\ldots\mu_4} \, \bar q \, \Gamma^{(6)}_{\mu_1\ldots\mu_6} O_F^{\mu_5\mu_6} q \, , \nn
	\E_F^{(3,3)} &= \epsilon^{\mu_1\ldots\mu_4} \, \bar q \, \Gamma^{(3)}_{\mu_1\ldots\mu_3} O_F^{[\mu_5\mu_6]\mu_7} q \, g_{\mu_4 \mu_5} g_{\mu_6 \mu_7} + 3 \epsilon^{\mu_1\ldots\mu_4} \, \bar q \, \Gamma^{(3)}_{\mu_1\mu_2\mu_7} O_F^{\mu_5\mu_6\mu_7} q \, g_{\mu_3 \mu_5} g_{\mu_4 \mu_6} \, , \nn
	\E_F^{(3,5)} &= \epsilon^{\mu_1\ldots\mu_4} \, \bar q \, \Gamma^{(5)}_{\mu_1\ldots\mu_5} O_F^{\mu_5\mu_6\mu_7} q \, g_{\mu_6 \mu_7} \, , \nn
	\tilde \E_F^{(3,5)} &= \epsilon^{\mu_1\ldots\mu_4} \, \bar q \, \Gamma^{(5)}_{\mu_1\ldots\mu_3\mu_5\mu_6} O_F^{\mu_5\mu_6\mu_7} q \, g_{\mu_4 \mu_7} \, , \nn
	\E_F^{(3,7)} &= \epsilon^{\mu_1\ldots\mu_4} \, \bar q \, \Gamma^{(7)}_{\mu_1\ldots\mu_7} O_F^{\mu_5\mu_6\mu_7} q \, ,
\end{align}
and operators where the indices of $O_F^{\mu_5\mu_6\mu_7}$ are permuted. The operators $\E_F^{(k,n)}$ with $n>4$ are evanescent due to the evanescent structure $\Gamma^{(n)}$. The other two operators are evanescent due to the Schouten identity. When the Dirac algebra is worked out in the HV scheme replacing the Levi-Civita symbol with $\gamma_5$, all the fermionic evanescent operators involve contractions of the Lorentz indices of $O_F^{\nu_1\ldots\nu_k}$ with objects in $D-4 = -2\varepsilon$ dimensions.

Finally, our scheme requires that the tree-level insertions of the evanescent operators into the renormalization conditions vanish in $D$ dimensions. This follows immediately from the fact that the renormalization conditions are formulated in terms of projections in Sect.~\ref{sec:Projections} that contract all open Lorentz indices of the truncated vertex functions with four-dimensional objects, in the same way as it happens with $S$-matrix elements in the HV scheme. When the evanescent operators are inserted into the truncated vertex functions at tree level, the fields are removed and derivatives turn into four-dimensional external momenta. All indices are either contracted with the four-dimensional Levi-Civita symbol or the four-dimensional indices of the projectors, hence both the Dirac structures $\Gamma^{(n)}$ and the metric tensors in the evanescent operators are projected to $D=4$ dimensions. Therefore, both the four-dimensional Dirac algebra and the Schouten identity apply and the tree-level insertions of the evanescent operators vanish identically.

\bibliographystyle{utphysmod}
\bibliography{Literature}

\end{document}